\documentclass[12pt,a4paper]{article}
\pdfoutput=1

\usepackage{amsmath,amssymb}
\usepackage[bookmarks=true]{hyperref}
\usepackage[nosort]{cite}
\usepackage[bulletsep]{collect}
\def\gfxon{\usepackage[final]{graphicx}}

\gfxon

\makeatletter
\let\old@startsection=\@startsection
\renewcommand{\@startsection}[6]{\old@startsection{#1}{#2}{#3}{#4}{#5}{#6\mathversion{bold}}}
\makeatother

\newcommand{\dpod}[1]{\partial_{#1}}

\makeatletter \@addtoreset{equation}{section} \makeatother

\makeatletter
\let\old@makecaption=\@makecaption
\def\@makecaption{\small\old@makecaption}
\makeatother

\newcommand{\lagr}{\mathcal{L}}
\newcommand{\suppot}{\mathcal{W}}
\newcommand{\prepot}{\mathcal{F}}

\newcommand{\Reals}{\mathbb{R}}

\makeatletter
\newcommand{\ellSN}{\mathop{\operator@font sn}\nolimits}
\newcommand{\ellCN}{\mathop{\operator@font cn}\nolimits}
\newcommand{\ellDN}{\mathop{\operator@font dn}\nolimits}
\newcommand{\ellAM}{\mathop{\operator@font am}\nolimits}
\newcommand{\ellK}{\mathop{\smash{\operator@font K}\vphantom{a}}\nolimits}
\newcommand{\ellE}{\mathop{\smash{\operator@font E}\vphantom{a}}\nolimits}
\makeatother

\ifx\genfrac\sdflkaj

\else

\fi
\newcommand{\sfrac}[2]{{\textstyle\frac{#1}{#2}}}
\newcommand{\half}{\sfrac{1}{2}}

\newcommand{\quarter}{\sfrac{1}{4}}

\newcommand{\ssN}{\mathcal{N}}

\newcommand{\nln}{\nonumber\\}
\newcommand{\nl}[1][0pt]{\nonumber\\[#1]&\hspace{-4\arraycolsep}&\mathord{}}

\newcommand{\earel}[1]{\mathrel{}&\hspace{-2\arraycolsep}#1\hspace{-2\arraycolsep}&\mathrel{}}
\newcommand{\eq}{\earel{=}}
\newcommand{\beq}{\begin{equation}}
\newcommand{\eeq}{\end{equation}}

\def\[{\begin{equation}}
\def\]{\end{equation}}
\def\<{\begin{eqnarray}}
\def\>{\end{eqnarray}}

\makeatletter
\def\mr@ignsp#1 {\ifx\:#1\@empty\else #1\expandafter\mr@ignsp\fi}%
\newcommand{\multiref}[1]{\begingroup
\xdef\mr@no@sparg{\expandafter\mr@ignsp#1 \: }%
\def\mr@comma{}%
\@for\mr@refs:=\mr@no@sparg\do{\mr@comma\def\mr@comma{,}\ref{\mr@refs}}%
\endgroup}
\makeatother

\newcommand{\hypref}[2]{\ifx\href\asklfhas #2\else\href{#1}{#2}\fi}
\newcommand{\Secref}[1]{Section~\multiref{#1}}
\newcommand{\secref}[1]{Sec.~\multiref{#1}}

\newcommand{\appref}[1]{App.~\multiref{#1}}

\newcommand{\figref}[1]{Fig.~\multiref{#1}}
\renewcommand{\eqref}[1]{(\multiref{#1})}

\ifx\href\asklfhas\newcommand{\href}[2]{#2}\fi
\newcommand{\arxivno}[1]{\href{http://arxiv.org/abs/#1}{#1}}

\begin{document}

\begin{flushright}\footnotesize
\texttt{ArXiv:\arxivno{1207.0460}}\\
\texttt{UMN-TH-3108/12}\\
\texttt{FTPI-MINN-12/21} \\
\texttt{NSF-KITP-12-108}\\
\texttt{ITEP-TH-30/12}\\
\vspace{0.5cm}
\end{flushright}
\vspace{0.3cm}

\renewcommand{\thefootnote}{\arabic{footnote}}

\setcounter{footnote}{0}

\begin{center}%

\vspace{-.7cm}%

{\Large\textbf{\mathversion{bold}
BPS States in Omega Background \\and Integrability}
\par}

\vspace{.7cm}%

\textsc{Kseniya Bulycheva$^{0}$, Heng-yu Chen$^{3}$, \\
 Alexander Gorsky$^{1}$, and Peter Koroteev$^{2}$}

\vspace{4mm}

\textit{$^{0}$Moscow Institute of Physics and Technology\\%
Institutskii Per. 9a, Dolgoprudny 141701, Russia}

\vspace{3mm}%

\textit{$^{0,1}$Institute for Theoretical and Experimental Physics\\%
Bolshaya Cheremushkinskaya, 25, Moscow 117218, Russia}

\vspace{3mm}%

\textit{$^{1,2}$University of Minnesota, School of Physics and Astronomy\\%
116 Church Street S.E. Minneapolis, MN 55455, USA}

\vspace{3mm}%

\textit{$^{2}$Kavli Institute for Theoretical Physics\\%
Santa Barbara, CA 93106-4030, USA}

\vspace{3mm}%

\textit{$^{3}$Department of Physics and Center for Theoretical Sciences\\%
National Taiwan University, Taipei 10617, Taiwan}

\vspace{4mm}

\thispagestyle{empty}

\texttt{bulycheva@itep.ru, heng.yu.chen@phys.ntu.edu.tw,\\ gorsky@itep.ru, koroteev@physics.umn.edu}

\par\vspace{.5cm}

\vfill

\textbf{Abstract}\vspace{5mm}

\begin{minipage}{12.7cm}
We reconsider string and domain wall central charges in $\ssN=2$ supersymmetric gauge theories in four dimensions in presence of the Omega background in the Nekrasov-Shatashvili (NS) limit. Existence of these charges entails presence of the corresponding topological defects in the theory -- vortices and domain walls. In spirit of the 4d/2d duality we discuss the worldsheet  low energy effective theory living on the BPS vortex in $\ssN=2$ Supersymmetric Quantum Chromodynamics (SQCD). We discuss some aspects of the brane realization of the dualities between various quantum integrable models. A chain of such dualities enables us to check the AGT correspondence in the NS limit.
\end{minipage}

\vspace*{\fill}
\end{center}

\newpage

\tableofcontents

\newpage

\section{Introduction}\label{sec:Intro}

Recent work by Nekrasov and Shatashvili \cite{Nekrasov:2009rc} has initiated the program of quantization of integrable systems by deforming four dimensional supersymmetric (SUSY) theories these integrable systems are associated with. The relationship between classical integrable systems and $\ssN=2$ supersymmetric gauge theories has been extensively studied for a long time \cite{Gorsky:1995zq, Donagi:1995cf, Gorsky:1997jq}. Remarkably the low energy dynamics of $\ssN=2$ gauge theories is captured by finite dimensional integrable systems such that the phase space of the latter is related  to the instanton moduli space of the former. To be precise, there are two different classical  integrable systems involved here. The first one is a holomorphic integrable system of the Hitchin or the spin chain type giving a Seiberg-Witten curve whose Jacobian is mapped onto complex Liouville tori and action variables are identified with order parameters in the gauge theory. The second integrable system is of the Whitham type emerges from the renormalization group (RG) flows, and the very RG equation plays the role of the Hamilton-Jacobi equation written in proper variables. 

The quantum integrable systems we are considering in this paper can be extracted from four dimensional theories in Omega background \cite{Nekrasov:2002qd} with $\epsilon_1=\epsilon,\,\epsilon_2=0$ (the so called Nekrasov-Shatashvili (NS) limit \cite{Nekrasov:2009ui}). Given a prepotential of the 4d theory $\prepot(a,\epsilon_1,\epsilon_2)$ as a function of the Coulomb branch moduli parameters $\{a\}$ and the Omega deformation parameters $\epsilon_{1,2}$,  we can consider in the NS limit an effective 2d theory with the following effective exact twisted superpotential
\[\label{eq:EffTwistedExactSuperpot}
\widetilde{\suppot}(a,\epsilon) = \lim_{\epsilon_2\to 0}\frac{\prepot(a,\epsilon,\epsilon_2)}{\epsilon_2}\,,
\]
where $\epsilon_1$ is replaced by $\epsilon$ \cite{Nekrasov:2009rc}. The F-terms of the effective Lagrangian effectively become two dimensional in the NS limit and are described by $\widetilde{\suppot}(a,\epsilon)$. For small $\epsilon$ formula \eqref{eq:EffTwistedExactSuperpot} can be even further simplified
\[
\widetilde{\suppot}(a,\epsilon) = \frac{\prepot(a)}{\epsilon}+\dots\,,
\]
where the ellipses denote terms which are regular in $\epsilon$. The above twisted superpotential includes both perturbative and instantonic contributions.

Minimization of superpotential \eqref{eq:EffTwistedExactSuperpot} yields the supersymmetric vacua which, according to the same authors \cite{Nekrasov:2009uh, Nekrasov:2009ui} are intimately connected with quantum integrable systems. Indeed, according to Nekrasov and Shatashvili, supersymmetric vacua of an appropriate two (also three and four) dimensional $\ssN=2$ gauge theory are in one-to-one correspondence with the Bethe roots of a certain integrable system. It is useful
to make the Legendre transform and consider the dual superpotential $\widetilde{W_D}(a_D, \epsilon)$ depending on the dual variable $a_D$. Thus the equation
\[\label{eq:BAEInt}
\exp{\left(\frac{\partial\widetilde{W_D}(a_D,\epsilon)}{\partial a_D}\right)}=1\,,
\]
can be viewed as a Bethe ansatz equation for some integrable system. This is consistent with the interpretation of the prepotential as an action in the Whitham system. Since $a_D$ is the coordinate variable  in the Whitham dynamics, one can recognize  the exponent in (\ref{eq:BAEInt}) as the canonical conjugate momentum.

There also exists a different well known duality between four dimensional gauge theories and two dimensional gauge theories (linear sigma models) \cite{Dorey:1999zk, Shifman:2004dr, Hanany:2004ea} (see \cite{Shifman:2007ce} for review, we shall refer to it as \textit{4d/2d duality}). The four dimensional theory here sits at the root of its baryonic Higgs branch; therefore electric and flavor charges can be combined in a single set of quantum numbers. Together with magnetic charges they form two sets of quantum numbers which parameterize masses of four dimensional dyons. On the two dimensional side one has kinks interpolating between different supersymmetric vacua, their masses also depend on two sets of charges -- N\"{o}ether and topological ones. The statement of the 4d/2d duality in its original formulation \cite{Dorey:1999zk} is that the two sets perfectly match with each other. Thus the BPS spectrum of an entire four dimensional theory can be studied using a relatively simple two dimensional (gauged linear) sigma model with four supercharges. It was later shown by Shifman and Yung \cite{Shifman:2007ce}  that the 4d/2d correspondence is not accidental, the underlying two-dimensional theory in fact should be treated as a low energy effective theory on the worldsheet of a BPS vortex.

The first step in the direction of merging the 4d/2d correspondence and the gauge/quantum integrabilty duality together was put forward by Dorey, Hollowood and Lee \cite{Dorey:2011pa}. The authors considered all three ingredients at once -- four dimensional $\ssN=2$ gauge theory (SQCD with $N_f=2N$), a two dimensional ${\mathcal N}=(2,2)$ gauge linear sigma model (GLSM) (a certain $U(K)$ gauge theory), and an integrable system. Critical ingredient of the duality is that the 4d gauge theory sits at a baryonic root of the Higgs branch, which in Omega background undergoes a deformation, and scalar field VEV gets shifted by an amount proportional to $\epsilon$. Provided the baryonic Higgs root condition is satisfied, the 4d theory is shown to be dual to a given GLSM whose twisted superpotential plays a role of the Yang-Yang function for a $SL(2,\mathbb{R})$ Heisenberg magnet. The four dimensional superpotential is shown to be equal to the effective 2d twisted superpotential of the form \eqref{eq:EffTwistedExactSuperpot} on shell. 

The initial motivation for this research was the identification of the different dualities known for quantum integrable systems in the framework of SUSY gauge theories and their brane realizations. However, first it is necessary to explain the geometrical meaning of degrees of freedom in relevant integrable systems
which is a subtle issue. It was clear for a while that these degrees of freedom are described in terms of brane embeddings into the internal space. We refer the reader to \cite{Gorsky:2000ds}, where geometrical aspects of dualities between integrable systems are reviewed. 

A proper brane content involves surface operators, or equivalently nonabelian strings with large tension. To begin with we revisit
the classification of the BPS solitons of different codimensions in the $\epsilon$ deformed theory. There are some surprises. It turns out that some BPS states in the deformed theory to the best of our knowledge were overlooked. We study carefully the central charges of the Omega-deformed SUSY algebra and argue that there are new stringy and domain wall type central charges. The key point is that tensions of such strings and domain walls are proportional to the graviphoton field and these defects are absent in the undeformed theory. We shall investigate the string-like object both for the pure supersymmetric Yang-Mills (SYM) theory and for the SQCD.   It will be argued that there are some singularities akin to those of cosmic strings. The corresponding BPS equations for such strings will be derived and finiteness of their tension will be discussed. Similarly the domain walls solitons will be found, which are more expected to appear, since due to the Omega deformation the theory has the discrete set of vacuum states, hence domain walls naturally emerge. The monopoles in Omega background which have been already discussed in the literature \cite{Ito:2011wv, Ito:2011ta} also require some attention for their proper interpretation.  

Turning to the dualities in quantum integrable systems we shall focus on two subjects. First we shall consider the quantum bispectral duality relating two different integrable systems. Classically eigenvalues of the Lax operator in one system get interchanged with coordinates in the second system. Quantum mechanically it means that the single wave function serves for two systems simultaneously when considered as the function of the spectral or coordinate variables. Although this question has not yet been elaborated to the full extent in the literature, by employing the quantum version of the duality we were able to explain the details of the 4d/2d duality in Omega background \cite{Dorey:2011pa}. Geometrically bispectrality corresponds to a rotation of the brane configuration which represents the 4d gauge theory in question. We consider the relation between Bethe ansatz equations (BAE) for dual integrable systems and briefly discuss their degenerate solutions corresponding to the analogues of Argyres-Douglas points. Using the duality between the families of spin chain models and the Calogero-Ruijsenaars systems we shall identify bispectral pairs in both families. As a consequence, we will show that the Alday-Gaiotto-Tachikawa (AGT) duality \cite{Alday:2009aq} in the NS limit can be proved using the chain of dualities involving the bispectral pair, the BPS vortex solutions we found in the earlier sections and the 4d/2d duality established in \cite{Dorey:2011pa}, \cite{Chen:2011sj} will also play important roles in completing the story.

The paper is organized as follows. In the next section we review how the $\ssN=2$ supersymmetry algebra is affected by the Omega background. We then compute the central charge for a BPS string in pure $\ssN=2$ Super Yang Mills theory and calculate its tension. In \secref{sec:NekPartFunc} we consider domain walls and monopoles in Omega deformed SYM. Then in \secref{sec:SQCD} we investigate BPS strings in well-studied example of the 4d/2d duality -- the supersymmetric QCD. \Secref{sec:BraneConstruction} is devoted to brane constructions and integrable systems associated with $\ssN=2$ gauge theories and dualities between them. In \secref{sec:AGTNS} we show that in the NS limit, the celebrated AGT correspondence \cite{Alday:2009aq} can be reduced to the so-called \textit{bispectral} duality between two integrable systems. Finally in \secref{sec:Conclusions} we conclude and speculate on further research topics.

\section{Flux Tubes in Pure Super Yang-Mills Theory}\label{sec:FluxTube}

In the standard lore of topological defects in supersymmetric theories, the BPS strings only exist when a gauge group is at least semi-simple, e.g. $U(N)$. A simple reason for this is based on existence of a nontrivial fundamental group of the resulting moduli space due to presence of a $U(1)$ factor. The latter causes a nonzero Fayet-Iliopoulos (FI) term which supports string solutions, and we shall refer them as \textit{FI strings}. In this section, we shall instead consider to a new kind of string-like objects which have not been discussed in the literature before, we shall refer to them as $\epsilon$-\textit{strings}. As we shall later see their tension is proportional to $\epsilon^2$ and construct the relevant classical field configurations. For simplicity we shall only focus on the gauge group $SU(2)$ in this section.

\paragraph{Action.}

Let us start with the $\ssN=2$ Super Yang-Mills theory in four dimensions in Omega background. To set the notations, the Lagrangian of the undeformed theory reads
\[\label{eq:LagrSYM}
\mathcal{L} = \sfrac{1}{4\pi}\mathfrak{Im}\,\tau\left[\text{Tr}\int d^4\theta\, \bar{\Phi}e^{V}\Phi\, e^{-V} + \text{Tr}\int d^2\theta\, (W^\alpha)^2\right]\,,
\]
where $\Phi=(\phi,\psi,F)$ is adjoint chiral superfield, $V=(\sigma, \lambda, D)$ is adjoint vector superfield, $W^\alpha$ is its field strength, and $\tau=\frac{4\pi i}{g^2}+\frac{\theta}{2\pi}$ is coupling constant.

\paragraph{Omega deformation.}

The Omega deformation of a four dimensional theory like \eqref{eq:LagrSYM} can be constructed from a six dimensional theory by compactifying the theory on a two-torus with twisted boundary conditions \cite{Nekrasov:2003rj, Shadchin:2005mx}. Torus action on $\Reals^4$ is given by two matrices $\Omega^m_{a\,n}$ where $ m,n =1,2,3,4$ and $a=5,6$ which act by rotations in 12  and 34 planes respectively. In the NS limit matrix $\Omega_6$ vanishes, therefore we shall denote $\Omega=\Omega_5$. Metric on the deformed torus reads
\[\label{eq:MetricTorusOmega}
G_{AB}dx^A dx^B= Adzd\bar{z} + (dx^m+\Omega^m dz+\bar{\Omega}^m d\bar{z})^2\,,
\]
where $z=x^5+ix^6,\, \bar{z}=x^5-ix^6$ and the vector field $\Omega^m=\Omega^m_n x^n$. In the notations of \cite{Ito:2011wv} $\Omega^m=(-i\epsilon x^2, i\epsilon x^1,0,0)$. In other words vector field $\Omega=i\epsilon\,\partial_\varphi$ is a rotation generator around $x_3$-axis. Here we denote $\rho=\sqrt{x_1^2+x_2^2}$.
The components of the metric in the limit $A\to 0$ read
\[
G_{mn}=\delta_{mn}\,,\quad G_{am}=\Omega_{am}\,,\quad G_{ab}=\delta_{ab}+\Omega_a^m\Omega_{bm}\,.
\]
Upon the dimension reduction, the fifth and sixth components of the gauge field form an adjoint scalar, which undergoes the following deformation due to the Omega background
\[\label{eq:ScalarShift}
\phi \mapsto \phi - i\Omega^m\nabla_m+\sfrac{i}{2}\Omega^{mn}S_{mn}\,,
\]
where $\Omega^m,\,\Omega^{mn}$ were introduced after formula \eqref{eq:MetricTorusOmega} and $S_{mn}$ is the spin operator for adjoint representation of the gauge group. The latter does not affect the bosonic part of the theory, however it does modify the fermionic part. This issue will be important when we consider the supersymmetry algebra of the theory momentarily. Transformation \eqref{eq:ScalarShift} itself is not a well-defined change of coordinates, but $\phi$ enters the Lagrangian in a special way, this shift brings us to a well defined Lagrangian of a modified theory \cite{Nekrasov:2010ka}.  Another deformation of the theory consists of shifting of the coupling constant, thereby we promote it to a superfield. In the $\ssN=2$ superfield language\footnote{See Shadchin's PhD thesis \cite{Shadchin:2005mx} for details} the shift reads as follows
\[
\tau \mapsto \tau - \bar{\theta}^m\bar{\theta}^n(\bar{\Omega}_{mn})^\dag\,,
\]
where $\bar{\theta}^m=(\bar{\sigma}^m)^{\dot{\alpha} I}\bar{\theta}_{\dot{\alpha} I}$ is the twisted Grassmann variable for the diagonal $\mathfrak{su}(2)_{R+\mathcal{R}}$ generators. In components the Lagrangian of the $\ssN=2$ SYM after the deformation takes the following form
\<\label{eq:N2SYMOmegNoWL}
\mathcal{L} \eq \sfrac{1}{4g^2} (F^a_{mn})^2+\sfrac{1}{g^2}|\nabla_m\phi^a-F^a_{mn}\bar{\Omega}^n|^2+\sfrac{1}{2g^2}|\phi\tau^a\bar{\phi}-i\nabla_m(\Omega^m\bar{\phi}^a-\bar{\Omega}^m\phi^a)+i\bar{\Omega}^m\Omega^n F_{mn}^a|^2\nl
+\sfrac{1}{g^2}\bar{\lambda}^{fa}\sigma^m\nabla_m \lambda_f^a -\sfrac{i}{g^2}\lambda^{af}\bar{\phi}\,\tau^a\lambda_f+\sfrac{i}{g^2}\bar{\lambda}_f^a\phi\,\tau^a\bar{\lambda}^f\nl
+\sfrac{1}{g^2}\lambda^{fa}(\bar{\Omega}^m\nabla_m-\half\bar{\Omega}^{mn}\sigma_{mn})\lambda_f^a- \sfrac{1}{g^2}\bar{\lambda}_f^a(\Omega^m\nabla_m-\half\Omega^{mn}\sigma_{mn})\bar{\lambda}^{fa}\,,
\>
where $f=1,2$ denotes the R-symmetry index, and spinor indices are suppressed.

\paragraph{SUSY transformations.}

Recall that $\ssN=2$ supersymmetry algebra in four dimensions has the following form
\<\label{eq:N2SUSYalgebra}
\{Q^I_\alpha,\bar{Q}_{J\,\dot{\alpha}}\}\eq 2P_{\alpha\dot{\alpha}}\delta^I_J+2Z_{\alpha\dot{\alpha}}\delta^I_J\,,\nln
\{Q^I_\alpha,Q^J_\beta\}\eq \epsilon_{\alpha\beta}\epsilon^{IJ} Z_{\text{mon}}+(Z_{\text{d.w.}})^{IJ}_{\alpha\beta}\,.
\>
There are three types on central charges: string, monopole and domain wall types. We shall focus on the former in this section leaving monopoles and domain walls to \secref{sec:NekPartFunc}.

The full global symmetry of the theory is $SU(2)_L\times SU(2)_R\times SU(2)_{\mathcal{R}}$ (left, right and the R-symmetry). It is broken by the Omega background in the NS limit to $SU(2)_L\times SU(2)_{R+\mathcal{R}}$ by paring the R-symmetry with the right handed $SU(2)$. The supercharges undergo the Donaldson-Witten twist \cite{Witten:1988ze}
\[ \bar{Q}=\delta_I^{\dot{\alpha}} \bar{Q}^I_{\dot{\alpha}}\,,\quad Q_m=(\bar{\sigma}_m)^{I \alpha} Q_{I \alpha}\,,\quad \bar{Q}_{mn}=(\bar{\sigma}_{mn})^{\dot{\alpha}}_I \bar{Q}^I_{\dot{\alpha}}\,.
\]
These transformations can be inverted as follows
\[\label{eq:InvertedDWtransf}
Q_{\alpha}^I = \half (\sigma^m)^I_\alpha Q_m\,,\quad \bar{Q}_{\dot{\alpha} J} = \half\epsilon_{\dot{\alpha} J}\bar{Q} + \half (\bar{\sigma}_{mn})_{\dot{\alpha}J}\bar{Q}^{mn}\,.
\]

It turns out that a generic Omega background breaks all supersymmetries of the theory \eqref{eq:N2SYMOmegNoWL} except the BRST charge $\bar{Q}$. Moreover, it can be shown that the Lagrangian \eqref{eq:N2SYMOmegNoWL} is a $\bar{Q}$-exact expression \cite{Nekrasov:2002qd}, which makes it possible to compute the partition function of the theory by localization methods. 

It is more or less clear that the obstacle to supersymmetry is due to the spin operator terms $\half\Omega^{mn}\sigma_{mn}$ in the fermionic sector. The theory thus has to be further deformed to gain more supersymmetry. To understand what we need to do, let us look at the supersymmetry transformations of \eqref{eq:N2SYMOmegNoWL} with the problematic spin operators omitted, and see what additional terms do we need to introduce. One has \cite{Ito:2011wv} the following under variations
\<\label{eq:SusyTransform}
\delta\phi \eq \zeta^I_{\alpha}(\lambda^{\alpha}_I-\Omega^m(\sigma_m)^{\alpha\dot{\alpha}}\bar{\lambda}_{I\dot{\alpha}})+\bar{\zeta}^I_{\dot{\alpha}}\Omega^m(\bar{\sigma}_m)^{\alpha\dot{\alpha}}\lambda_{I\alpha}\,,\nln
\delta \lambda_{I\alpha} \eq \zeta_{I\beta}((\sigma^{mn})_\alpha^\beta F_{mn}+i[\phi,\bar{\phi}]\delta_\alpha^\beta+\nabla_m(\bar{\Omega}^m\phi-\Omega^m\bar\phi)\delta_\alpha^\beta)\nl
+\bar{\zeta}_{I\dot{\beta}}(\sigma^m)_\alpha^{\dot{\beta}}(\nabla_m\phi-F_{mn}\Omega^n)\,.
\>
However, since the action is not supersymmetric, the above transformations do not leave the Lagrangian invariant. As suggested in \cite{Nekrasov:2003rj} and later extended in \cite{Ito:2011wv}, one has to turn on R-symmetry Wilson lines properly to restore partial supersymmetries. Thus, in the NS limit one has to add
\[
-\bar{\mathcal{A}}_I^J\lambda^I\lambda_J-\mathcal{A}_I^J\bar{\lambda}^I\bar{\lambda}_J\,,\]%
where
\[
\mathcal{A}_I^J=-\half\bar{\Omega}_{mn}(\bar{\sigma}^{mn})^I_J\,,\]%
to the Lagrangian. One can also treat the above terms as emerging from a superpotential which we add to the theory, and we shall speculate more on this in \secref{sec:NekPartFunc}. In the NS limit, when $\epsilon_2=0$, the supersymmetry of the theory is enhanced to ${\mathcal N}=(2,2)$ \cite{Ito:2011wv} and is generated by the following supercharges
\[
Q_1, Q_2, \bar{Q}_{13}, \bar{Q}_{14}\,.
\]
Using the inverted transformation \eqref{eq:InvertedDWtransf} we conclude that in the original formulation of SUSY algebra \eqref{eq:N2SUSYalgebra} the following generators are included into ${\mathcal N}=(2,2)$ subalgebra
\[\label{eq:SUSYGens22}
Q_{12},\,Q_{21},\,\bar{Q}_{\dot{1}2},\,\bar{Q}_{\dot{2}1}\,.
\]
In the remaining part of the section we shall investigate $1/2$ BPS object -- a string which is annihilated by the above charges.

\paragraph{String central charge and string tension.}

The supercurrent for the Omega deformed SYM theory was computed in \cite{Ito:2011wv}. Its Euclidean time component has the following form (assuming static configuration, $B_3\neq 0$, others components of $F_{mn}$ vanish)
\<\label{eq:SuperCurrent}
J^4_{I\alpha} \eq \frac{1}{g^2}\left((-i[\phi,\bar{\phi}]+(\phi\bar{\Omega}^n-\bar{\phi}\Omega^n)\nabla_n)\sigma^4_{\alpha\dot{\alpha}}+\tilde{F}_{4n}\sigma^n_{\alpha\dot{\alpha}}\right)\bar{\lambda}^{\dot{\alpha}}_I\nl
+\frac{2\sqrt{2}}{g^2}(\sigma^{4n})^\beta_\alpha(-\nabla_n\phi+F_{np}\bar{\Omega}^p)\lambda_{I\beta}
\>
Let us find the string central charge current. Performing standard variation of the above supercurrent we get\footnote{Technically there is another contribution from the R-current \cite{Gorsky:1999hk}, which contributes to the $(\half,\half)$ central charge..}

\<\label{eq:StringChargeDensity}
\delta_{\zeta^I_{\alpha}}\bar{J}^{4 J}_{\dot{\alpha}}=2\sigma^4_{\alpha\dot{\alpha}}\delta_I^J\mathcal{P}_4+\partial_m\left( (\phi^a\bar{\Omega}^m-\bar{\phi}^a\Omega^m)B^a_3\right) \sigma^3_{\alpha\dot{\alpha}}\delta_I^J\,,
\>
where $\mathcal{P}_4$ is Hamiltonian of the system. Note that there is an additional contribution to the above string charge current which is bilinear in fermions of the form $\partial_m(\Omega^{m}\bar{\lambda} \lambda)$. For classical analysis, where all fermionic fields can be put to zero, this contribution can be omitted. We see that there is a correction which represents the string central charge, specifically the correction takes the following form
\[
\zeta_3=\sfrac{1}{2g^2}\partial_m\left((\phi^a\bar{\Omega}^m-\bar{\phi}^a\Omega^m)B^a_3\right) \sigma^3_{\alpha\dot{\alpha}}\delta^{IJ}\,,
\]
where $\rho^2=x_1^2+x_2^2$ is the transversal coordinate to the string. If $\epsilon$ is real then
\[
\zeta_3=\sfrac{1}{g^2}\partial_\varphi (\mathfrak{R}e\,\epsilon\bar{\phi}^aB^a_3)\,.
\]
The central charge is given by
\<
Z_{\text{string}}\eq\int d^3 x\, \zeta_3=\frac{1}{g^2}\int dz \int d\rho\, \rho \int\limits_0^{2\pi} d\varphi\,\partial_\varphi (\mathfrak{R}e(\epsilon\bar{\phi}^a)B^a_3)\nln
\eq\frac{1}{g^2}\int dz \int d\rho\, \rho B_3^a \,\mathfrak{R}e(\epsilon\bar{\phi}^a)\Big\vert_{0}^{2\pi} \,.
\>

We can immediately see that multi-valuedness of $\phi$ as a function of the azimuthal angle is required in order to make the central charge nonzero. The tension of the string solution under consideration (let's call them {\it $\epsilon$-strings}) is therefore given by
\[\label{eq:StringTension}
T = \frac{1}{g^2}\int\limits_0^\infty d\rho\, \rho B_3^a \,\mathfrak{R}e(\epsilon\bar{\phi}^a)\Big\vert_{0}^{2\pi} \,.
\]
Assuming that
\[\label{eq:AngleAnsarzPhi}
\phi(\rho,\varphi) = \phi(\rho)e^{i\alpha\varphi}\,,
\]
where $\alpha$ is an constant, we arrive to
\[\label{eq:StringTension}
T = \frac{1}{g^2}\int\limits_0^\infty d\rho \,\rho  \,\mathfrak{R}e\left(\epsilon B_3^a\bar{\phi}^a(e^{-2\pi i \alpha} -1)\right) \,.
\]
The above expression for the tension of $\epsilon$-string only makes sense if it is finite. In order to establish that one has to solve BPS equations \label{eq:BPSeqnsFull} in order to find the profile functions for $\phi$ and $B_3$ as function of the radial coordinate $\rho$.

\paragraph{BPS equations.}

Let us now find the BPS equations which describe such a string. Once supersymmetry algebra is understood \eqref{eq:SUSYGens22}, we can focus on the bosonic part of the action
\[\label{eq:ActionN2OmegaBos}
\mathcal{L} = \sfrac{1}{4g^2} F_{mn}^2+\sfrac{1}{g^2}|\nabla_m\phi-F_{mn}\bar{\Omega}^n|^2+\sfrac{1}{2g^2}|\phi\tau^a\bar{\phi}-i\nabla_m(\Omega^m\bar{\phi}^a-\bar{\Omega}^m\phi^a)|^2\,.
\]
Note that in the NS limit $\bar{\Omega}^m\Omega^n F_{mn}^a$ identically vanishes. Let us now do the Bogomolny completion, as the supersymmetry algebra suggests
\<
\mathcal{L}\eq \sfrac{1}{2g^2} |B^a_3+\phi\tau^a\bar{\phi}-i\nabla_m(\Omega^m\bar{\phi}^a-\bar{\Omega}^m\phi^a)|^2 + \sfrac{1}{2g^2} |\nabla_1\phi^a+i\nabla_2\phi^a-(\Omega_2-i\Omega_1)B^a_3 |^2 \nl
+ \sfrac{1}{2g^2}\partial_m(B^a_3(\Omega^m\bar{\phi}^a-\bar{\Omega}^m\phi^a))\geq \sfrac{1}{2g^2}\partial_m(B^a_3(\Omega^m\bar{\phi}^a-\bar{\Omega}^m\phi^a))\,.
\>
The above inequality is saturated provided that the following BPS equations are satisfied
\<\label{eq:BPSeqnsFull}
B^a_3+\bar{\phi}\tau^a\phi-i\nabla_m(\Omega^m\bar{\phi}^a-\bar{\Omega}^m\phi^a) \eq 0\,,\nln
\nabla_1\phi^a+i\nabla_2\phi^a-(\Omega_2-i\Omega_1)B^a_3 \eq 0\,.
\>
One can also check that the above BPS equations are consistent with the ${\mathcal N}=(2,2)$ supersymmetry algebra. Indeed, by looking at the gluino variation in \eqref{eq:SusyTransform} we need to set to zero all expressions which enter the right hand side together with $\zeta_{11},\,\zeta_{22}$ and their complex conjugates. Contributions proportional to $\zeta_{12}\,,\zeta_{21}$ and conjugated terms vanish automatically due to the BPS condition. By doing so one arrives at  equations \eqref{eq:BPSeqnsFull}.

Sometimes it is more convenient to switch to the complex coordinates
\[
w=x+iy\,,\quad \bar{w}=x-iy\,,\]%
than the BPS equations \eqref{eq:BPSeqnsFull} take the following form
\<\label{eq:BPSeqnsSYM}
\partial_{\bar{w}}A^a_w-\partial_w A^a_{\bar{w}}+\epsilon^{abc}\phi^b\bar{\phi}^c-i(w\nabla_{\bar{w}}-\bar{w}\nabla_w)(\epsilon\bar{\phi}^a-\bar{\epsilon}\phi^a)\eq 0\,,\nln
\partial_{\bar{w}}\phi^a-\epsilon^{abc}A_{\bar{w}}^b\phi^c+\epsilon w(\partial_{\bar{w}}A^a_w-\partial_w A^a_{\bar{w}})\eq 0\,.
\>
The above equations can be used in study of the effective two dimensional theory living on the $\epsilon$-string.

Note that the BPS equations for $\epsilon$-string \eqref{eq:BPSeqnsFull} can be regarded as Omega deformed Hitchin equations. Indeed, using the complex notation of \eqref{eq:BPSeqnsSYM} and starting from the Hitchin equations
\<
[\bar{\nabla},\nabla]+[\bar\phi,\phi]\eq 0\,,\nln
\bar{\nabla}\phi \eq 0\,,
\>
where $\nabla$ and $\bar{\nabla}$ are gauge covariant derivatives with respect to $w$ and $\bar{w}$ respectively, by means of \eqref{eq:ScalarShift}, we arrive to \eqref{eq:BPSeqnsSYM}.

\paragraph{Solution of BPS equations and vortex tension.}

Let us proceed with the solution of \eqref{eq:BPSeqnsFull}. We will look for a background solution when all fields are aligned along the Cartan subalgebra of the gauge algebra. Thus color superscript will always be $a=3$, in the rest of the section we shall omit it. One has the following
\<\label{eq:BPSeqnsCartan}
B_3-i\partial_m(\Omega^m\bar{\phi}-\bar{\Omega}^m\phi) \eq 0\,,\nln
\partial_1\phi+i\partial_2\phi-(\Omega_2-i\Omega_1)B_3 \eq 0\,.
\>
Decomposing $\phi = \phi_1+i\phi_2,\, \epsilon = e_1+ie_2$ into real and imaginary parts we obtain
\[\label{eq:B3harm}
B_3 = 2 \partial_\varphi(e_1 \phi_1+e_2\phi_2)\,,
\]
and two first order equations on $\phi_1$ and $\phi_2$. After some simple manipulations one gets
\<
\Delta \phi_1 + 4e_2\partial_\varphi(e_1 \phi_1+e_2\phi_2)\eq 0\,,\nln
\Delta \phi_2 - 4e_1\partial_\varphi(e_1 \phi_1+e_2\phi_2)\eq 0\,,
\>
which, after adding these equations with proper coefficients, implies that $e_1 \phi_1+e_2\phi_2$ is a harmonic function. However, at this point boundary conditions of the solution remain unclear as we need to have $\phi(2\pi)\neq \phi(0)$ in order to gain finite tension \eqref{eq:StringTension}. In order to make the problem mathematically precise we can make the following trick. The phase difference of $e^{2\pi i \alpha}$ will be identified with the deficit angle of a cone which is obtained by gluing $\varphi=0$ ray with $\varphi=2\pi$ one.

Solution of Laplace equation on a cone is formally given by a series of positive and negative powers of $\rho$ with angle dependent coefficients. The latter are normally expressed in terms of ellipsoidal harmonics. Since we are interested in normalizable solutions we only leave negative powers of the radial coordinate in the series. The solution will however be divergent at the origin. The dependence on phase $\alpha$ is then hidden in the harmonic coefficients. We shall refrain from giving more details here since we are not investigating any dynamics on $\epsilon$-strings in this paper. Rather we provide the evidence of existence of $\epsilon$-strings and finiteness of their tension.

Let us now look at the vortex tension \eqref{eq:StringTension}.  Using \eqref{eq:B3harm} we conclude that
\[\label{eq:Tint}
T = \frac{1}{2g^2}\int d^2 x\, \text{Tr}\, B^2 = \frac{1}{g^2}\int\limits_0^{2\pi} d\varphi \int\limits_0^\infty d\rho\, \rho\,\partial_\varphi (\mathfrak{R}e(\epsilon\bar{\phi}^a)B^a_3)\,.
\]
As we argued above the integral over the radial coordinate diverges, however it does not make the tension infinite. To see this let us regularize the radial integral on its lower limit by putting a cutoff at some small value of $\rho=\rho_0$. After integrating the full angular derivative we see that the contribution to the integral at the lower limit cancel each other as $\rho_0\to 0$. Thus we are left with the contribution from large $\rho$. As we are not specifying the full solution of the BPS equations we shall not evaluate integral \eqref{eq:Tint} here. However, from dimensional ground we anticipate
\[
T = A(\alpha) |\epsilon|^2\,,
\]
where $A$ is a constant. Therefore we find that  the tension of $\epsilon$-strings is quadratic in $\epsilon$, and it is only nonzero for fractional winding numbers.

To conclude this section let us make a few general comments concerning $\epsilon$-strings. First,
at large values of the graviphoton field the string tension is large and we can safely
consider it as semiclassical object and the string could serve as the new type of the surface
operator. In this case one can use the standard technique
to get the worldvolume theory. We shall discuss the worldvolume theory elsewhere. The second
point to be mentioned is some analogue with the string in the noncommutative gauge
theory found by Gross and Nekrasov \cite{Gross:2000wc}. They have discovered that the
Dirac string attached to the monopole in the noncommutative theory  becomes observable and its tension is proportional to the noncommutativity parameter. Since
using the chain of dualities \cite{Hellerman:2011mv, Reffert:2011dp, Hellerman:2012zf}
the $\epsilon$ parameter can be traded to the
noncommutativity in the internal space one could look for more close relation
between two types of strings. Finally, one could ask for the brane realization
of the $\epsilon$-string. Certainly it can not be identified as D2 brane similar
to the FI string since we can not reproduce the tension with such brane realization.
Hence the most natural candidate is the properly embedded D4 brane. We hope to discuss
the details of the brane realization of $\epsilon$-strings elsewhere.

\section{Monopoles and Domain Walls}\label{sec:NekPartFunc}

We have discussed color flux tubes in Omega background in \secref{sec:FluxTube}. Here we shall address two other types of topological defects we often encounter in supersymmetric theories -- monopoles and domain walls.

\paragraph{Central charge.}

Recall from \eqref{eq:N2SUSYalgebra} that domain walls and monopoles saturate holomorphic central charges in supersymmetry algebra.  Symmetric combination of these charges give domain wall piece, whereas an antisymmetric one contributes to monopoles. Analogously to a string charge density we have evaluated in \eqref{eq:StringChargeDensity}, we can proceed with the monopole and domain wall. The former calculation has been performed in \cite{Ito:2011wv}, and the latter gives
\[\label{eq:SuperCurrDW}
\delta_{\zeta_{I\beta}}J^4_{J\alpha} = \sfrac{1}{g^2}\delta_{IJ}(\sigma^4)_{\alpha\dot{\alpha}}(\sigma^3)_\beta^{\dot{\alpha}} \dpod{3}(\phi^a\nabla_n(\phi^a\bar{\Omega}^n-\bar{\phi}^a\Omega^n))\,.
\]
We have included here only the bosonic contribution to the supercurrent which is relevant for classical analysis. The full expression will also contain bilinear term in fermions $\partial_m(\Omega^{m}\lambda \lambda) + H.c.$ which will be manifest for quantum calculations \cite{Dvali:1996xe}.

Structurally \eqref{eq:SuperCurrDW} is very reminiscent of the string current \eqref{eq:StringChargeDensity}. As in the string charge case, a field dependent FI term appears
\[\label{eq:EffectiveFIdw}
\xi^a = \sfrac{1}{g^2}\nabla_m(\Omega^m\bar{\phi}^a-\bar{\Omega}^m\phi^a)\,,
\]
leading to the following expression of the domain wall tension
\[\label{eq:DomainWallTension}
T = \sfrac{1}{g^2}\xi^a(\phi^a_{+\infty}-\phi^a_{-\infty})\,,
\]
which can be viewed as a non-Abelian generalization of the standard calculation in theories with superpotentials. Let us now have a closer look on the BPS monopoles and domain walls appearing in $\ssN=2$ SYM theory, again for simplicity we shall consider $SU(2)$ gauge group here.

\paragraph{BPS equations and monopole solution.}

Ito et al \cite{Ito:2011wv, Ito:2011ta} have investigated BPS monopole solution of $\ssN=2$ SYM with gauge group $SU(2)$ in Omega background in the NS limit. It reads
\[\label{eq:BPSwOmega}
B_i^a - \nabla_i \phi^a +  i\epsilon_{ijk}\Omega^j B^{ak} = 0\,,
\]
or in components
\<\label{eq:BPSeqnsComplts}
B_3^a-\nabla_3\phi^a-\epsilon\, x^m B_m^a \eq 0\,,\nln
B_m^a-\nabla_m\phi^a+\epsilon\,  x_m B_3^a \eq 0\,,
\>
where $m=1,2$. The solution of these equations is given in \cite{Ito:2011wv}. The authors' conclusion is that the monopole's mass is not changed, the magnetic field strength has the same form as the one in the undeformed case for $\epsilon=0$. However, there is a correction of the scalar field profile. In the singular gauge (when only $\phi^3\neq 0$) the solution reads
\[\label{eq:Phi3ProfileIto}
\phi^3(\rho,z)=\sqrt{v^2 H(\rho ,z)^2+\epsilon ^2+\frac{2 v z
\epsilon  H(\rho ,z)}{\sqrt{\rho ^2+z^2}}+G(\rho,z)^2}\,,
\]
where function
\<G(\rho,z)^2\eq\frac{2 \rho ^2 v z \epsilon  F(\rho ,z) H(\rho ,z)}{\rho ^2+z^2}+\frac{\rho ^2 z^2 \epsilon ^2 F(\rho ,z)^2}{\rho ^2+z^2}-\frac{2 \rho ^2 \epsilon ^2 F(\rho ,z)}{\rho ^2+z^2}\nl
   -\frac{2 \rho ^2 v z \epsilon  F(\rho ,z) H(\rho,z)}{\left(\rho ^2+z^2\right)^{3/2}}+\frac{\rho ^4 \epsilon ^2 F(\rho ,z)^2}{\left(\rho ^2+z^2\right)^2}
\>
vanishes at $z\to\pm\infty$.  Functions $F$ and $H$ are taken from the 't-Hooft-Polyakov monopole solution \cite{Prasad:1975kr, Bogomolny:1975de}
\[ H(\rho ,z) = \coth{v\sqrt{\rho ^2+z^2}}+\frac{1}{v\sqrt{\rho ^2+z^2}}\,,\quad F(\rho,z)=1-\frac{v\sqrt{\rho ^2+z^2}}{\sinh{v\sqrt{\rho ^2+z^2}}}\,.
\]
Note that scalar field $\phi$ does not go to its vacuum value $v$ any longer as it does for $\epsilon=0$, but rather interpolates between $\phi_{+\infty}=v+\epsilon$ to $\phi_{-\infty}=v-\epsilon$ at plus and minus $z$-infinity respectively. This suggests us that maybe 1/2 BPS monopole is not a proper interpretation of the above solution and more structures can be involved.

Before we go further,  let us mention an useful symmetry of equations \eqref{eq:BPSeqnsComplts}. In the above analysis axial symmetry was assumed such that both $\phi$ and $B$ fields depended only on $\rho$ and $z$. We can also introduce azimuthal angle $\varphi$ in the game by giving the scalar field a phase
\[\label{eq:phirescaling}
\phi\mapsto\phi\, e^{i\alpha\varphi}\,.
\]
In order to preserve \eqref{eq:BPSeqnsComplts} the magnetic field strength also acquires a phase and its azimuthal component $B_\varphi^a$ gets generated due to $\nabla_m\phi$ term in the second equation. Provided that such a configuration is chosen, the FI field reads
\[ \xi^a = \sfrac{i}{g^2}\alpha(\bar{\epsilon}\phi^a+\epsilon\bar{\phi}^a)\,.
\]
Then the corresponding domain wall's tension \eqref{eq:DomainWallTension} becomes
\[ T=\sfrac{2}{g^2}\epsilon \text{Tr} \phi^2\,,
\]
for imaginary $\phi$ and real $\epsilon$. Since \eqref{eq:Phi3ProfileIto} still solves equations \eqref{eq:BPSeqnsComplts} we can compute the tension for the solution in hand. One gets
\[\label{eq:TensionDWeps}
T=\sfrac{2}{g^2}\epsilon\left((v+\epsilon)^2-(v-\epsilon)^2\right)=\sfrac{8}{g^2}v\epsilon^2\,.
\]
Let us now see how to construct monopoles and domain walls in $SU(2)$ SYM theory in the NS Omega background.

\paragraph{Monopole on a domain wall.}

We shall perform a slightly different Bogomolny completion of the action \eqref{eq:ActionN2OmegaBos} than the authors of \cite{Ito:2011wv, Ito:2011ta}
\<\label{eq:BPScomplAltern}
\lagr \eq \sfrac{1}{2g^2}\left|B^a_3-i\nabla_m(\Omega^m\bar{\phi}^a-\bar{\Omega}^m\phi^a)+\nabla_3\phi^a+i\epsilon_{3jk}\Omega^j B^{ak}\right|^2\nl
+\sfrac{1}{2g^2}\left|B_1^a+i B_2^a+(\nabla_1+i\nabla_2)\phi^a-(\Omega_2-i\Omega_1)B^a_3\right|^2\nl
+\sfrac{1}{g^2}\partial_m\left(B_3^a(\Omega^m\bar{\phi}^a-\bar{\Omega}^m\phi^a)\right)+\sfrac{1}{g^2}\partial_{3}(\phi^a\nabla_m(\Omega^m\bar{\phi}^a-\bar{\Omega}^m\phi^a))-\sfrac{1}{g^2}\partial_i(B_i^a\phi^a)\,.
\>
Three terms in the third line above correspond to strings, domain walls and monopoles respectively. Existence of the first two types of solitons solely relies on the non-trivial field dependent FI term \eqref{eq:EffectiveFIdw}
\[ 
\xi^a = \sfrac{1}{g^2}\nabla_m(\Omega^m\bar{\phi}^a-\bar{\Omega}^m\phi^a)\,.
\]
BPS equations follow from \eqref{eq:BPScomplAltern} immediately
\<\label{eq:BPSMononDW}
B^a_3+\nabla_3\phi^a+i\epsilon_{3jk}\Omega^j B^{ak}-i\nabla_m(\Omega^m\bar{\phi}^a-\bar{\Omega}^m\phi^a)\eq 0 \,,\nln
B_1^a+i B_2^a+(\nabla_1+i\nabla_2)\phi^a-(\Omega_2-i\Omega_1)B^a_3\eq 0\,.
\>
We can now observe that if $\epsilon$ and $\phi^a$ are real valued, as they were chosen to be in \cite{Ito:2011wv, Ito:2011ta}, then the latter equation above splits into two, one for the real part, one for the imaginary part of its l.h.s. Also the field dependent FI term vanishes. We immediately identify them as the last two equations of \eqref{eq:BPSeqnsComplts}. However, if a different ansatz is chosen, when either $\epsilon$ or $\phi^a$ or both have imaginary part, the FI parameter \eqref{eq:EffectiveFIdw} kicks in, and equations \eqref{eq:BPSMononDW} no longer decouple. Their solution is probably more complicated than reported in \cite{Ito:2011ta}, and will be reported elsewhere.

\paragraph{Boojums.}

\begin{figure}
\begin{center}
\includegraphics[height=5cm, width=4.55cm]{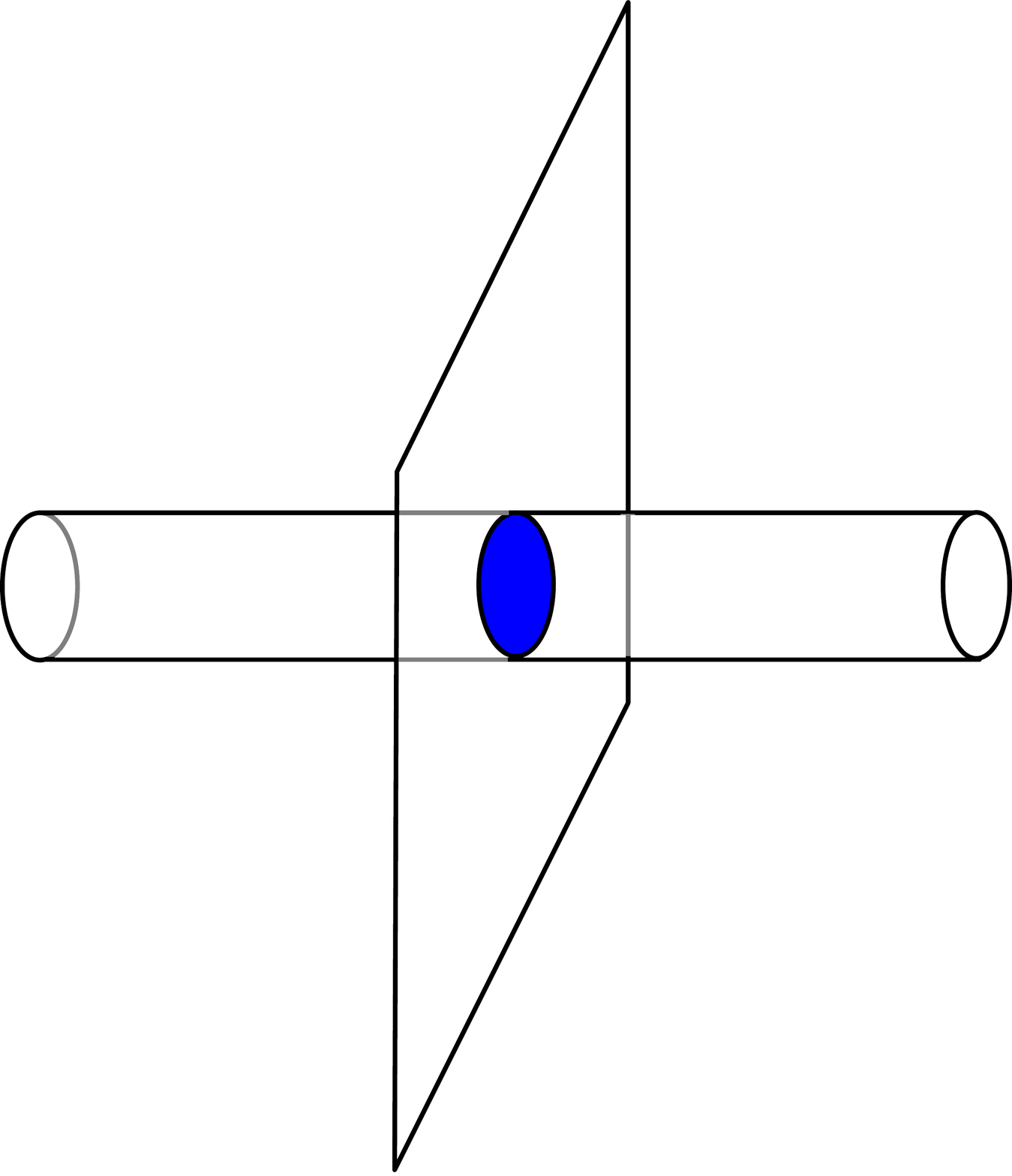}\qquad\qquad\qquad \includegraphics[height=5cm, width=6cm]{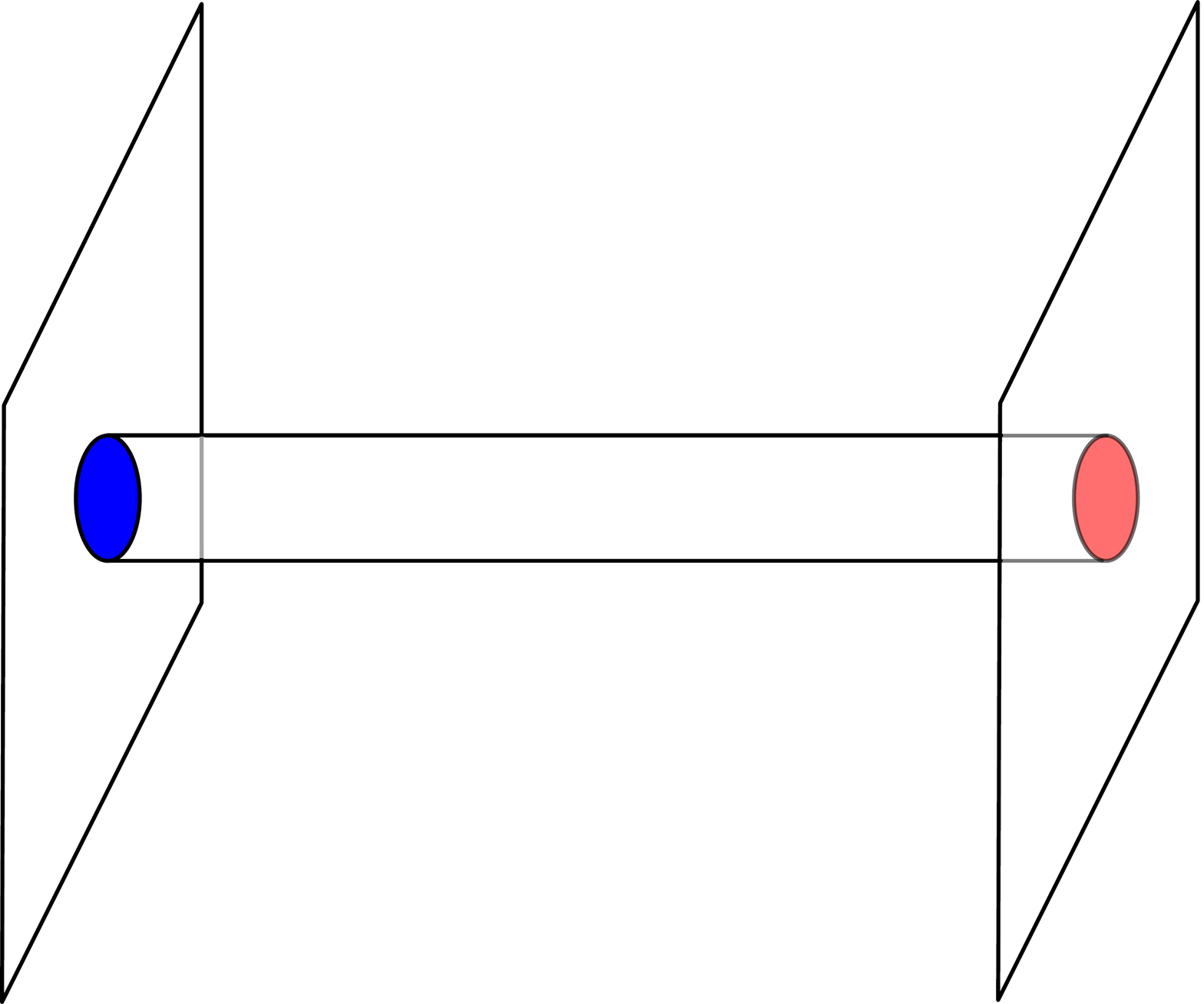}
\caption{Left: Boojum as a monopole-string-domain wall junction. The string is infinite and is stretched along the $z$-axis. Right: $\epsilon$-string ending at monopoles located on two parallel domain walls in $xy$-plane can be viewed as the superposition of two boojums. The string does not continue through the domain walls to the outer area, since the scalar field, main building block of the $\epsilon$-string, vanishes outside of the domain walls.}
\label{fig:epsStringMon}
\end{center}
\end{figure}

At this point let us also mention that the setup we have just described also admits strings provided that the parameter $\alpha$ in \eqref{eq:phirescaling} is non integer, otherwise the string central charge vanishes, indeed it follows from \eqref{eq:StringTension}. Thus the solution above describes a BPS monopole on a domain wall, and, if $\alpha$ is not integer, a more complicated \textit{boojum} construction \cite{Shifman:2007ce} which is a junction of a string, domain wall and a monopole \figref{fig:epsStringMon}; it is a $1/4$ BPS configuration. Remarkably, all three structures coexist together and emerge together from Omega deformation. It appears to be impossible, as far as our analysis suggests, to find, say, only domain wall without a monopole, or vice versa -- they always come in pairs. Strings, however, as we discussed in \secref{sec:FluxTube}, can exist on their on provided that $\alpha$ is a non integer.

Interestingly, the object we have just described -- a boojum, can be placed on ends of $\epsilon$-strings, the same way as monopoles in the Higgs phase of $\ssN=2$ theory can have non-Abelian flux tubes emerging from them \cite{Shifman:2007ce}. Such a string is depicted in \figref{fig:epsStringMon}
%

\paragraph{Relationship to coupled 4d/2d systems.}

\begin{figure}
\begin{center}
\includegraphics[height=5cm, width=7cm]{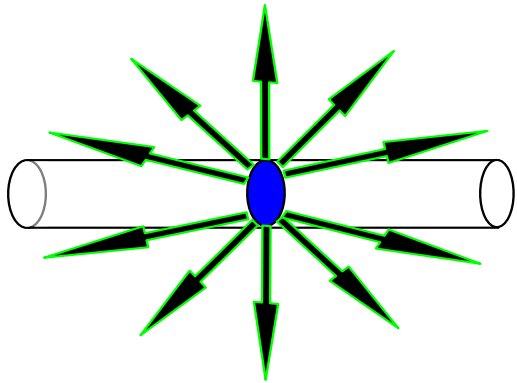}
\caption{'t-Hooft-Polyakov monopole on a surface defect.}
\label{fig:2d2dmon}
\end{center}
\end{figure}

In \cite{Gukov:2006jk, Gaiotto:2011tf} coupled 4d/2d systems were studied. An example of such a system is a four dimensional gauge theory with a surface operator insertion. The 4d theory is considered to be in the Coulomb branch, a 2d theory lives on the surface defect and both systems are coupled. Remarkably, both 2d (kinks) and 4d (monopoles, dyons) BPS states can be found in such systems and the authors of \cite{Gaiotto:2011tf} managed to derive the full 2d/4d wall crossing formula. Bound states of monopoles on surface defects are present in the theory, since the 4d theory is at the Coulomb branch, its magnetic field has a spherically symmetric pattern, unlike a Higgs monopole whose field lines are trapped to a vortex. These two pictures -- Higgsed monopole an a vortex and a Polyakov-'t-Hooft monopole on a surface defect \figref{fig:2d2dmon} may represent two different limiting configurations of a more generic setup, which involves more sophisticated 2d/4d dynamics. Keeping the calculations we have done in this section, we may hope that 4d theories in Omega background may be reasonable candidates for such a theory. It would be interesting to investigate the solution of BPS equations \eqref{eq:BPSMononDW} more closely and study different values of the deformation parameter $\epsilon$. At large $\epsilon$ the surface operator limit emerges and Gaiotto et al story  \cite{Gaiotto:2011tf} may also arise.
%

\section{$\ssN=2$ SQCD in Omega Background}\label{sec:SQCD}

The 4d/2d duality was initially formulated for four dimensional $\ssN=2$ supersymmetric QCD  with gauge group $U(N)$ with $N\leq N_f\leq 2N$ flavors  \cite{Dorey:1999zk}, it was found that its BPS spectrum on the so-called ``Baryonic Higgs branch'' precisely coincides with the BPS spectrum of certain mass-deformed two dimensional sigma model.
Physical explanation of the duality \cite{Hanany:2004ea, Shifman:2004dr} (see also \cite{Shifman:2009zz}) relies on the existence of BPS flux tubes (FI strings)  in this theory, and the mass deformed two dimensional sigma model is precisely the world volume theory of the flux tube.
This duality was further extended beyond the matching of BPS spectra, to an exact matching between the F-term potentials in Omega background  \cite{Dorey:2011pa, Chen:2011sj}, which we shall review next.
The vortex and duality results presented here also play an important role in verifying AGT correspondence \cite{Alday:2009aq} via the bi-spectral duality to be  discussed in the next two sections (See Figure 9 for a ``roadmap'').  
We shall also demonstrate how BPS vortex in Omega deformed theory which plays crucial role in establishing the results in \cite{Dorey:2011pa, Chen:2011sj}, preserving only ${\mathcal N}=(2,2)$ part of supersymmetry (4 out of 8 supercharges). Similarly to the pure SYM discussed in the previous section, the BPS vortex in question will be invariant under the unbroken part of the SUSY algebra, therefore it will still remain to be a BPS configuration.
Finally, we also sketch out a direct field theoretic derivation of the vortex world volume theory, complementing the D-brane based derivation earlier.

\subsection{Chen-Dorey-Hollowood-Lee duality}

Here we shall review main aspects of the duality unveiled in \cite{Dorey:2011pa, Chen:2011sj}. The authors have proposed and proved \footnote{We change the notations of \cite{Dorey:2011pa} to make them consistent with our notations.} that in NS limit,  four dimensional $U(N)$ SQCD with $N$ fundamental hypermultiplets of masses $m_1,\dots m_N$ together with $N$ antifundamental hypermultiplets of masses $\tilde{m}_1,\dots \tilde{m}_N$ and coupling constant $\tau$ is dual to the two dimensional $U(K)$ GLSM with $N$ chiral fundamentals of twisted masses $M_1,\dots,M_n$ together with $N$ chiral antifundamentals of twisted masses $\tilde{M}_1,\dots,\tilde{M}_n$ and coupling constant $\hat{\tau}$. The above statement holds provided that
the four dimensional theory is considered in the Higgs branch defined by the condition
\[\label{eq:DHLHiggs}
\phi_a = m_a - n_a\epsilon\,,
\]
for some $\mathbb{Z}_N$ vector $n_a$, rank of the gauge group of the 2d GLSM is given by
\[\label{eq:GGrank}
K = \sum\limits_{a=1}^N (n_a-1)=\sum\limits_{a=1}^N \hat{n}_a\,,
\]
masses of the 4d theory and twisted masses of the 2d theory are related to each other in the following way
\[\label{eq:massescorr}
M_a =m_a-\sfrac{3}{2}\epsilon\,,\quad \widetilde M_a =\widetilde m_a+\sfrac{1}{2}\epsilon\,,
\]
and coupling constants obey
\[\label{eq:couplingmatch}
\hat{\tau}=\tau+\half(N+1)\,.
\]
Quantitatively the CDHL duality states that in NS limit, the chiral rings of the 4d and 2d theories are isomorphic and we can relate effective twisted superpotential \eqref{eq:EffTwistedExactSuperpot} from the 4d gauge theory with the effective twisted superpotential of the corresponding 2d GLSM as follows
\[\label{DHLduality}
\widetilde{\mathcal{W}}(\phi_a=m_a-n_a \epsilon)-\widetilde{\mathcal{W}}(\phi_a=m_a-\epsilon) = \widetilde{\mathcal{W}}^{2d}_{\text{eff}}(\hat{n}_a)\,.
\]
This means that the sets of stationary points (vacua) of the two superpotentials are isomorphic and the above equality holds in the corresponding vacua. The equality \eqref{DHLduality} has been proven by computing the Nekrasov partition function on the Higgs branch of the theory \eqref{eq:DHLHiggs}, and deriving the effective twisted superpotential then matching it with the 2d superpotential on shell \cite{Dorey:2011pa, Chen:2011sj}. Let us mention a related contribution \cite{Bonelli:2011fq}, where moduli space of vortices was shown to be a submanifold of the instanton moduli space of the Omega deformed 4d theory.

Below we shall provide some further supports for this duality based on the study of non-Abelian BPS vortices in the Omega deformed four dimensional theory. This will be done by identifying the classical 2d theory living on the vortex along the course of the Shifman-Yung program \cite{Shifman:2007ce}.

\subsection{Constructing non-Abelian vortices $N_f=N$}

Let us now construct the action of the $\ssN=2$ SQCD in the NS Omega background in four dimensions. Although in \cite{Dorey:2011pa, Chen:2011sj} the superconformal $N_f=2N$ case was considered, from the viewpoint of the non-Abelian vortices it is more instructive to start with the left boundary of the stability window $N_f=N$.

\paragraph{Action.}

Let us begin again with the undeformed SQCD Lagrangian in four dimensions with $N=N_f$
\<\label{eq:SQCDundeformed}
\mathcal{L} \eq \frac{1}{4\pi}\text{Tr}\,\Bigg[\mathfrak{Im}\,\tau\left( \int d^4\theta\,\bar{\Phi}\,e^{V}\Phi\,e^{-V} + \int d^2\theta\, (W^\alpha)^2\right)  \nl
 +\int d^4\theta\left(\bar{Q}_ie^{V}Q^i + \widetilde{Q}_ie^{-V}\overline{\widetilde{Q^i}}\right)+\int d^2\theta\left(\widetilde{Q}_i\Phi Q^i+m_i\widetilde{Q}_iQ_i+ \text{H.c.}\right)\Bigg]\,,
\>
where $V$ is an adjoint $\ssN=1$ vector superfield, $\Phi=(\phi,\lambda, D)$ is an adjoint $SU(N)\subset U(N)$ chiral superfield, quark superfields $Q_i=(q_i,\psi_i, F_i)$ and $\tilde{Q}_i=(\tilde q_i,\tilde \psi_i,\tilde F_i)$ are transformed in $\textbf{N}$ and $\bar{\textbf{N}}$ of the $SU(N)$ and as $\textbf{N}_f$ and $\bar{\textbf{N}}_f$ under global flavor $SU(N_f)$ respectively; $m_i$ are quark masses and $\tau=\frac{4\pi i}{g^2}+\frac{\theta}{2\pi}$ is coupling constant.

\paragraph{Omega Deformation.}

Our task now is to construct the Omega deformed theory. As in the pure SYM case the deformation can naturally be understood in terms of the six dimensional $\ssN=1$ theory \cite{Nekrasov:2003rj}. It is convenient to use dual frame description $G_{AB}=e^{(c)}_Ae^{(c)}_B$. The components of sixbeins read
\[\label{eq:vielbeins}
e^{(m)}_n=\delta^m_n\,,\quad e^{(m)}_a = \Omega_a^m\,,\quad e^{(a)}_m=0\,,\quad e^{(a)}_b=\delta^a_b\,.
\]
Using the above equation we can rewrite the kinetic term for squarks
\[
e^{(B)}_a \nabla_{(B)}  q = \nabla_a q -i \Omega_a^m\nabla_m q\,.
\]
Thus we have
\[
|\nabla_A q|^2 = |\nabla_m q|^2 + |(\phi-i\Omega^m\nabla_m)q|^2\,,\]%
and analogously the kinetic term for anti squarks. The bosonic part of the action after quark masses $m_i$ and $\widetilde m_i$ are included reads
\<\label{eq:SQCDOmegaCompts}
\mathcal{L} \eq \sfrac{1}{4g^2} F_{mn}^2+\sfrac{1}{g^2}|\nabla_m\phi-F_{mn}\bar{\Omega}^n|^2+\sfrac{1}{2g^2}|\phi\tau^a\bar{\phi}-i\nabla_m(\Omega^m\bar{\phi}^a-\bar{\Omega}^m\phi^a)+g^2(\bar{q}\tau^a q-\widetilde{q}\tau^a \bar{\widetilde{q}})|^2 \nl
+\half|\nabla_m q|^2+\half|\nabla_m\widetilde{q}|^2+\half|(\phi-m_i-i\Omega^m\nabla_m)q_i|^2+\half|(\phi-\widetilde{m}_i-i\Omega^m\nabla_m)\widetilde{q}_i|^2\nl
+2g^2|\widetilde{q}\tau^a q|^2+\sfrac{g^2}{2}|\widetilde{q}_i{q}_i-N\xi_{FI}|^2+\sfrac{g^2}{8}(|q|^2-|\widetilde{q}|^2)^2\,,
\>
where we have included Fayet-Iliopoulos term $\xi_{FI}$. This theory has $U(N)_c\times SU(N)_f$ global color and flavor symmetry group.

\paragraph{Supersymmetry transformations.}

In what follows it is convenient to package squarks and anti-squarks into a single vector $q^{if}=(q^i,\,\,\tilde{\bar{q}}^i)$, where $f=1$ for squarks and $f=2$ for antisquarks.
Supersymmetry acts on the fields in the following way\footnote{We use chiral notation here.}
\<\delta \phi^a \eq -\sqrt{2}\zeta^{\alpha f}(\lambda^a_{\alpha f}-\Omega^{\alpha\dot{\alpha}}\bar{\lambda}^a_{\dot{\alpha} f})+\bar{\zeta}^f_{\dot{\alpha}}\Omega^{\alpha\dot{\alpha}}\lambda^a_{f\alpha}\,,\nln
\delta \lambda^{a\,f}_\alpha \eq -\zeta^{\beta f}F^a_{\alpha\beta}+i\zeta^g_\alpha \text{D}^{a\,f}_g+i\sqrt{2}\bar{\zeta}^{\dot{\alpha}f}\left(\nabla_{\alpha\dot{\alpha}}\phi^a-F^a_{\alpha\dot{\alpha}\beta\dot{\beta}}\Omega^{\beta\dot{\beta}}\right)\,,\nln
\delta q^{if} \eq \sqrt{2}\zeta^{\alpha f}\psi^i_\alpha + i\sqrt{2}\bar{\zeta}^f_{\dot{\alpha}}\bar{\tilde{\psi}}^{i\dot{\alpha}}\,,\nln
\delta \psi^i_\alpha \eq -i\sqrt{2}\bar{\zeta}^{\dot{\alpha}}_f\nabla_{\alpha\dot{\alpha}}q^{if}+2i\zeta^f_\alpha \left(\bar{\phi}_a(\tau^a)_j^i q^j_f-i\Omega^{\beta\dot{\beta}}\nabla_{\beta\dot{\beta}}q_f^i\right)\,,
\>
where $i=1,\dots,N$ runs through fundamental representation, $a=1,\dots, N^2$ runs through the adjoint representation of $U(N)$, $f,g =1,2$ denote $SU(2)$ R-symmetry index, and the D-term contribution in the first line above has the following form
\[ 
\text{D}^{a\,f}_g = -\bar{\phi}\tau^a\phi\delta_g^f-g^2\left(\bar{q}_g\tau^a q^f-\Xi^{a\,f}_g\right)\,,
\]
where the generalized FI-term  reads
\[\label{eq:GeneralizedFI}
\Xi^{a\,f}_g = \sfrac{i}{g^2} \nabla_{\alpha\dot{\alpha}}(\bar{\Omega}^{\alpha\dot{\alpha}}\phi^a-\Omega^{\alpha\dot{\alpha}}\bar\phi^a)\delta^f_g+\xi^f_{FI\,g}\delta^a_{N^2}\,.
\]
The first term we have already seen in the previous section, similar story here -- it is generated by the Omega background. The second contribution to $\Xi$ is the standard FI term. Here, as in \cite{Vainshtein:2000hu}, we formally kept the FI parameter $\xi^f_{FI\,g}$ as a triplet. Usually only diagonal part is left over, as it simplifies the calculations, however, it is absolutely unnecessary. As we can see, generalized FI parameter \eqref{eq:GeneralizedFI} is a sum of the two terms, the field dependent FI term, which appears due to Omega deformation, and the conventional $U(1)$ FI term, which is normally considered in supersymmetric QCD. As we know \cite{Hanany:1997hr}, presence of the latter does not affect the supersymmetry of the theory, however, it's broken to the $(2,2)$ SUSY due to the former. Similar to the pure SYM case, considered in the previous section, generators \eqref{eq:SUSYGens22} form the supersymmetry algebra. We shall use this information to extract the BPS equations later.

\paragraph{Classical vacua.}

Vacua of the theory can be chosen similarly to the undeformed SQCD. Indeed, after identifying
\[\label{eq:CFlockedVac}
\phi^a=m^a\,\quad \tilde{q}_{ai} = \bar{q}_{ai}=\sqrt{\xi_{FI}}\delta_{ia}\,,
\]
we find that the potential in \eqref{eq:SQCDOmegaCompts} vanishes. Alternatively one could have put $\tilde q^i=0$ and work only with squarks $q^i$ (see \cite{Shifman:2007ce} for details). This vacuum is invariant under the color-flavor rotations
\[ G_{c+f}:\quad \phi \mapsto U^{-1}_c \phi\, U_f\,,\quad q^i_a \mapsto (U^{-1}_c)_a^b \,q_b^j\, U_{f\,j}\,,
\]
where $U_c = U_f \in G_{c+f}$. In the most generic case, when masses $m^a$ are arbitrary, one has
\[ G_{c+f} = S(U(n_1)\times\dots\times U(n_k))\,,\quad \sum\limits_{j=1}^k n_j = N\,,
\]
where $S(\dots)$ stands for the stabilizer. If all masses are different then $G_{c+f}=S\left(U(1)^N\right)=U(1)^{N-1}$\,. The pattern of the symmetry breaking depends on the relationship between the masses $m_i$ and the FI parameter $\xi_{FI}$. In what follows we shall assume $\xi\gg m_i^2$ for any $i$ and all the masses to be of the same order, thus in our case we have the following breaking
\[\label{eq:symmetrybreaking}
U(N)_c\times SU(N)_f \xrightarrow{\sqrt{\xi}} U(1)\times SU(N)_{c+f} \xrightarrow{m_i} U(1)\times G_{c+f}\,.
\]
Note that $U(1)$ factor in the above formula will be very important in constructing vortices, which we shall now do. A vortex configuration will further break the above symmetry in a nontrivial way -- the above mentioned $U(1)$ will be coupled to the generators of the Cartan subalgebra of $G_{c+f}$. By using the extra $U(1)$ symmetry we can also change the second equality relation in \eqref{eq:CFlockedVac}, and this is what exactly is done in the DHL paper. For the convenience of the calculations of \cite{Dorey:2011pa} squarks have charge $-3/2$ and anti-squarks have charge $+1/2$ with respect to this symmetry. In the current paper it is more convenient to keep
the condition \eqref{eq:CFlockedVac} on the vortex solution as well.

\paragraph{Vortex configuration.}

In order to find non-Abelian BPS strings we need to organize winding around the $z$-axis. Thus we allow one of the flavors, say $q^N$, to depend on the azimuthal angle $e^{i\hat{n}\varphi}q^N$, where $\hat{n}$ is an integer. Algebraically it corresponds to breaking the symmetry of \eqref{eq:symmetrybreaking} down to $U(1)_{\text{diag}}\times SU(N-1)$, where the first $U(1)_{\text{diag}}$ factor is the diagonal subalgebra of the $U(1)$ from \eqref{eq:symmetrybreaking} and the $N-1$'st Cartan generator of the $G_{c+f}$. Then for the two terms in the second line of \eqref{eq:SQCDOmegaCompts} read
\[ V\supset\left|(\phi^i_j-m_N\delta^i_j+\hat{n}\epsilon \delta^i_j+\epsilon\rho (A_\varphi)^i_j)q^{Nj}\right|^2\,.
\]
It vanishes provided that the expression in the parentheses above is equal to zero. So we put
\[ \phi^N_N=m_N-\hat{n}\epsilon-\epsilon\rho (A_\varphi)^N_N\,.
\]
It can certainly be generalized to the case where more squark fields have angular dependences
\[\label{eq:PhiVacuumValue}
\phi^a=m^a-\hat{n}^a\epsilon-i\Omega^m A^a_m\,,
\]
where $\hat{n}^a$ is an integer valued vector, which is intended to count the number of flux quanta which flow through the vortex. We see that the above classical vacuum equation related the adjoint scalar and the gauge field.

\paragraph{Vortex BPS equations.}

While studying a $1/2$-BPS object we work with the half of supersymmetry algebra which acts trivially on it. In the case at hand this algebra is generated by \eqref{eq:SUSYGens22}. Remarkably it coincides with the BPS subalgebra of the non-Abelian vortex considered by Shifman and Yung \cite{Shifman:2007ce}. Thus even in the Omega deformed background in the NS limit, the vortex configuration we are considering in this section will remain $1/2$-BPS.

Performing Bogomol'ny completion of the action \eqref{eq:SQCDOmegaCompts} we get the following energy density
\<\label{eq:QCDOmegaBPSCompl}
\lagr \eq \sfrac{1}{2g^2}\left|(B^a_3)^2+g^2(\bar{q}\tau^a q-\Xi^a)\right|^2+\sfrac{i}{g^2}\left|(\nabla_1+i\nabla_2)\phi^a-(\Omega_2-i\Omega_1)B^a_3\right|^2\nl
+|(\nabla_1+i\nabla_2)q|^2+N\,\xi_{FI} B^N_3+\sfrac{1}{g^2}\partial_m(\Omega^m\bar{\phi}^a-\bar{\Omega}^m\phi^a)B_3^a\,.
\>
Here we assumed that the adjoint scalar and gauge field are only aligned along the Cartan subalgebra of the gauge Lie algebra. The last two terms in the second line of the above expression are total derivatives, but due to a different reason: the former is the Abelian field strength, which gives circulation of the gauge field after removing one integration, the latter involves $\partial_\varphi$ derivative and is of the same kind as \eqref{eq:StringChargeDensity}. We can see that the Lagrangian \eqref{eq:SQCDOmegaCompts} under the constraint  \eqref{eq:PhiVacuumValue} and color-flavor locked condition $\bar{q}^i=\widetilde{q}^i$ takes almost exactly the same form as for the undeformed case considered by Shifman and Yung \cite{Shifman:2007ce}. It means that the BPS construction for the vortex will also be almost exactly the same. The only difference is that adjoint scalar $\phi^a$ will have a nontrivial profile defined by the magnetic field and the Omega background. The corresponding BPS equations read
\<\label{eq:BPSeqnsSQCDFull}
B^a_3+g^2(\bar{q}_i\tau^a q^i-\Xi^a)\eq0\,,\nln
(\nabla_1+i\nabla_2)q^i\eq 0\,,\nln
(\nabla_1+i\nabla_2)\phi^a-(\Omega_2-i\Omega_1)B^a_3\eq 0\,,
\>
where, again as in \eqref{eq:GeneralizedFI}, the color index $a=1,\dots,N^2$ runs through all $U(N)$ generators. For convenience we can split up  $U(1)$ and $SU(N)$ parts and rewrite the first equation above using the definition of the generalized FI term \eqref{eq:GeneralizedFI}
\<\label{eq:Bzprofile}
B_3+g^2(|q|^2-\xi_{FI}-\xi^N_{nFI})\eq0\,,\nln
B^a_3+g^2(\bar{q}_i\tau^a q^i- \xi_{nFI}^a)\eq 0\,,
\>
where we have denoted
\[ \xi^a_{nFI} = \sfrac{i}{g^2} \nabla_{\alpha\dot{\alpha}}(\bar{\Omega}^{\alpha\dot{\alpha}}\phi^a-\Omega^{\alpha\dot{\alpha}}\bar\phi^a)=\sfrac{1}{g^2}\dpod{\varphi}(\bar{\epsilon}\phi^a+\epsilon\bar{\phi}^a)\,,\quad a=1,\dots, N\,,
\]
the non-Abeilan FI field. In its absence equations \eqref{eq:Bzprofile} and second equation in \eqref{eq:BPSeqnsSQCDFull} exactly reproduce the BPS set considered in \cite{Shifman:2007ce}; hence the solutions for the profile functions can be extracted from there directly. Thus the modification to the BPS vortex equations in the Omega background consist of introducing $\xi^a_{nFI}$ (which for some configurations can vanish) and the nontrivial profile for the adjoint scalar dictated by the third equation in
\eqref{eq:BPSeqnsSQCDFull}.

\paragraph{Asymptotic behavior of solutions.}

\begin{figure}
\begin{center}
\includegraphics[height=5cm, width=7cm]{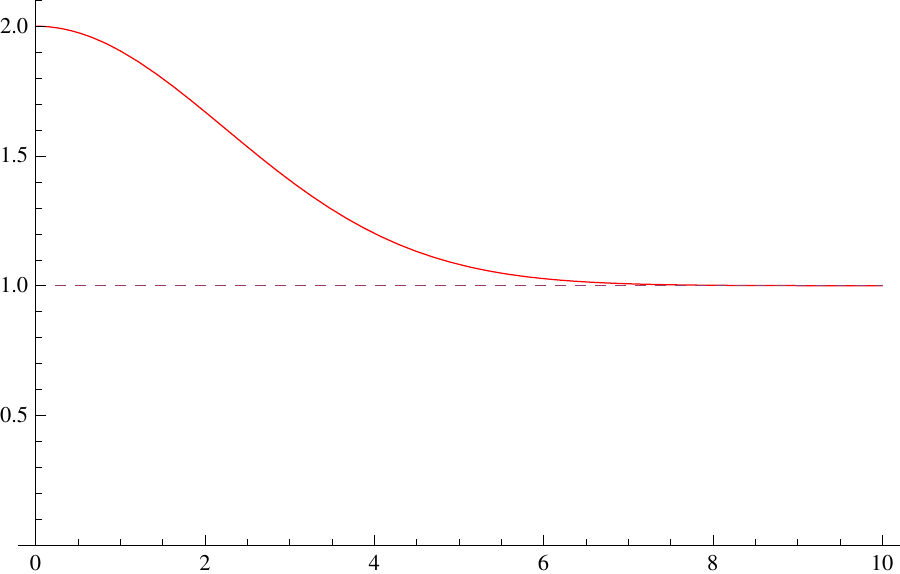}
\caption{Scalar field interpolating between two different minimum values of the potential}
\label{fig:phieps}
\end{center}
\end{figure}

Let us for the moment assume that $\phi^a$ is invariant under rotations around the $z$-axis, in other words $\xi^a_{nFI}$ vanishes. From the analysis of the previous section we conclude that it happens when $\phi$ does not depend on the azimuthal angle $\varphi$. Then, we know the solution for the magnetic field in all color directions, since it is exactly the same as in \cite{Shifman:2007ce}. In particular, far away from the vortex, the gauge field exhibits $1/\rho$ behavior. This makes the quantization condition \eqref{eq:PhiVacuumValue} physical and well defined. Indeed, it tells us that the adjoint scalar at large $\rho$ approaches its vacuum value
\[\label{eq:PhiVac}
\phi^a_{vac} = m^a - \epsilon(n^a+ k^a)\,,
\]
where $k^a$ is a $\mathbb{Z}_N$-valued vector of winding numbers of along the different Cartan color directions. We still need to figure out what $k^a$ is in terms of $n^a$. The reasoning for that comes from the following physical requirement -- string tension (energy per unit length) should be finite. Indeed as in the undeformed case, the conclusion comes from the requirement that $|\nabla_m q|^2$ terms are finite.
\[\label{eq:squarkkin}
\int\limits_0^{+\infty} d\rho\,\rho\, |\nabla_m q^i|^2\,.
\]
This was archived by a proper asymptotic behavior of the azimuthal component of gauge field $A_\varphi$ such that the integrand above could decay fast enough.
Indeed, let's say $q\sim e^{in\varphi}q(\rho)$, thus the integral becomes
\[ \int\limits_0^{+\infty} d\rho\,\frac{1}{\rho} \left|(in-iA_{\varphi}\rho\right)q^i|^2\,,
\]
so $A_\varphi\to n/\rho$ at large $\rho$. In other words, $A_\varphi$ should be proportional to the number of flux quanta which flow through the vortex. So, given \eqref{eq:PhiVac} we can easily figure out that $k^a=-n^a$ and $\phi$ tends to its undeformed value $m^a$ at large radial distances. Thus we conclude that the adjoint scalar interpolates between
\[\label{eq:Vac1vor}
\phi^a=m^a-n^a\epsilon
\]
at $\rho=0$ and
\[\label{eq:Vac2vor}
\phi^a=m^a
\]
at $\rho=\infty$, see \figref{fig:phieps}.

Now we understand why in the left hand side of \eqref{DHLduality}  involves the difference of the superpotential in two points -- they are two different minima of the effective potential for $\phi$ in four dimensional theory, and the vortex can now be viewed as a \textit{kink} which interpolates between these minima! Evidently \eqref{DHLduality} is only applicable for nonzero $\epsilon$. Note that the superpotential in the left hand side of \eqref{DHLduality} is evaluated at values of $\phi$ which are shifted by a unit of $\epsilon$. As we shall explain below it happens because of an additional $U(1)$ twist of (anti)squark fields.

Let us mention that a completely different situation occurs if $\phi$ acquires nontrivial topology in the spirit of the previous section. Then BPS equations \eqref{eq:BPSeqnsSQCDFull} do not decouple any longer. We expect a significant change in the asymptotic behavior of the solutions in that case. Investigations in this direction will be reported elsewhere.


\subsection{Constructing non-Abelian vortices $N<N_f\leq 2N$}

Let us now address semilocal vortices \cite{Shifman:2006kd}. In order to be more generic we shall keep $\widetilde N = N_f-N$ generic inside the conformal window. The vacuum condition is generalized as follows
\[ \phi = m\,,\quad q^i_a =
\begin{cases}
\sqrt{\xi_{FI}}\delta^i_a\,, & 1<i<N \\
 0\,, &i\geq N \geq 0 \,.
\end{cases}
\]
Symmetry breaking pattern is similar to \eqref{eq:symmetrybreaking}, only there is a residual global symmetry left due to the additional quark fields
\[ G_{c+f} = SU(N)_{c+f}\times SU(\widetilde N)\,.
\]
Most recent review of the semilocal vortex constructions can be found in \cite{Shifman:2011xc}. The structure of the BPS equations \eqref{eq:BPSeqnsSQCDFull} will not change, only the flavor index will now range $i=1,\dots,N+\widetilde N$. On the level of the low energy effective action the vortex theory will be modified by adding $\tilde N$ so-called \textit{size} moduli. Kinetic terms of size moduli bring logarithmic divergence to the energy density of the effective theory, therefore one has to introduce an infrared cutoff.

\paragraph{GLSM description.}

As usual, a GLSM description of the theory is more effective for computations. The 2d theory which is dual to the 4d SQCD in the NS Omega background with $N_f=N+\widetilde N$ quarks is given by the following Lagrangian provided that \eqref{eq:DHLHiggs}-\eqref{eq:couplingmatch} hold
\<\label{eq:HTGLSMAdj}
\lagr\eq \text{Tr} \int d^4\theta\,\left[\frac{1}{2e^2}|\Sigma|^2+\bar{\Phi}\,e^{\frac{V}{2}} \Phi\, e^{-\frac{V}{2}} + \sum\limits_{i=1}^N \bar{X}_i e^V\, X^i + \sum\limits_{i=1}^{\widetilde N} \bar{Y}_i e^{-V}\, Y^i\right]\nl
+\text{Tr} \int d^2\tilde\theta\, \tau\Sigma + H.c.\,, 
\>
where the trace is taken over the adjoint representation of $U(K)$ gauge group, $\Phi$ is adjoint chiral multiplet, and $\Sigma$ is field strength for 2d vector superfield $V$. The second line in the above Lagrangian represents the twisted F-terms of the theory. There are $N+\widetilde N+1$ twisted mass parameters turned on including $N+\widetilde N$ masses for $X$ and $Y$ fields together with the twisted mass for the adjoint scalar $\Phi$, which according to \cite{Dorey:2011pa, Chen:2011sj} equals to $\epsilon$.  In the limit $e\to \infty$ the gauge field becomes non dynamical, and we can integrate it out. In this limit we can recover the geometry of the NLSM's target space, which naturally appears in the derivation of the low energy theory.

In order to get the effective twisted superpotential in the right hand side of \eqref{DHLduality} we need to integrate out $X$'s, $Y$'s and $\Phi$'s in \eqref{eq:HTGLSMAdj}.
When $N_f=2N_c$ the theory is superconformal, the coupling does not run and no dynamical scale is generated.
\<\label{eq:Eff2dGLSM}
\widetilde{\mathcal{W}}_{\text{eff}}^{2d}(\lambda) \eq \epsilon \sum\limits_{a=1}^K\sum\limits_{i=1}^N f\left(\frac{\lambda_a- M_i}{\epsilon}\right)-\epsilon \sum\limits_{a=1}^K\sum\limits_{i=1}^N f\left(\frac{\lambda_a-\widetilde M_i}{\epsilon}\right)\nl
 + \epsilon \sum\limits_{a,b=1}^K f\left(\frac{\lambda_a-\lambda_b-\epsilon}{\epsilon}\right)+2\pi i \hat{\tau}\sum\limits_{a=1}^K\lambda_a\,,
\>
where $f(x)=x(\log x -1)$. Note the change of the coupling constant to $\hat{\tau}$ compared to \eqref{eq:HTGLSMAdj}. Minimizing the above superpotential we arrive to the ground state equations
\[\label{eq:XXXWsp}
\prod\limits_{l=1}^{N}\frac{\lambda_j-M_l}{\lambda_j-\widetilde M_l}=e^{2\pi i\hat{\tau}}\prod\limits_{k\neq j}^K\frac{\lambda_j-\lambda_k-\epsilon}{\lambda_j-\lambda_k+\epsilon}\,,
\]
which coincide with Bethe ansatz equations for the twisted anisotropic Heisenberg $SL(2,\Reals)$ magnet. This observation quantifies the so-called \textit{Bethe/gauge correspondence} for the $\ssN=2$ SQCD.

Theories with $\widetilde N<N$ can be obtained from the conformal theory by sending some masses to infinity and renormalizing the coupling constant. Dynamically generated scale $\Lambda_{QCD}$ will then appear.

\paragraph{A note on the $\ssN=2^*$ theory.}

Recently a similar to DHL and CDHL study of the $\ssN=2^*$ theory in five dimensions appeared in the literature \cite{Chen:2012we}, and a duality with a three dimensional integrable system was discussed. Although the calculations involving the Nekrasov partition function look very similar to \cite{Dorey:2011pa, Chen:2011sj}, there is a technical difference: the Higgs branch condition of the 5d theory looks similar to \eqref{eq:DHLHiggs}, however, there is no shift by $N$ in the rank of the 3d gauge group like in \eqref{eq:GGrank}. Clearly, is occurs because fundamental matter in \cite{Dorey:2011pa, Chen:2011sj} and adjoint matter in  \cite{Chen:2012we} contribute differently to the Nekrasov partition function; nevertheless physical understanding of the second duality remains to be uncovered. Since there are no BPS vortices in $\ssN=2^*$ theory, one cannot apply the method we used in the current section.

\section{Brane Constructions and Dualities in Integrable Systems}\label{sec:BraneConstruction}

Solutions to the Bethe ansatz equations mentioned above correspond to the ground states in the world volume theory on the non-abelian strings realized as D2 branes, hence it is desirable to translate the full powerful machinery of the integrable systems into the brane language. In this section we focus on the particular issue namely the realization of known dualities between quantum integrable systems using brane language. 

We shall first review the Hanany-Witten type IIA brane construction which yields the $\ssN=2$ SQCD and integrable systems related to it -- the XXX spin chain and Gaudin model together with the dualities these models are involved in. Employing the Gaudin/XXX duality, we will be able to give a vortex interpretation of the AGT duality in the next section, where the XXX model appears on the $\ssN=2$ theory side and the Gaudin model naturally arises in study of Liouville CFT. Here we shall make some preparations to that study. In addition to that, the Gaudin/XXX duality will be examined by studying the simplest examples of Argyres-Douglas type points and wall crossing phenomena in presence of the Omega background. At the end we shall discuss yet another duality between spin chains and Calogero-Moser systems for further study.

\subsection{Dualities from the Hanany-Witten brane construction}

\begin{figure}
\begin{center}
\includegraphics[height=11cm, width=14cm]{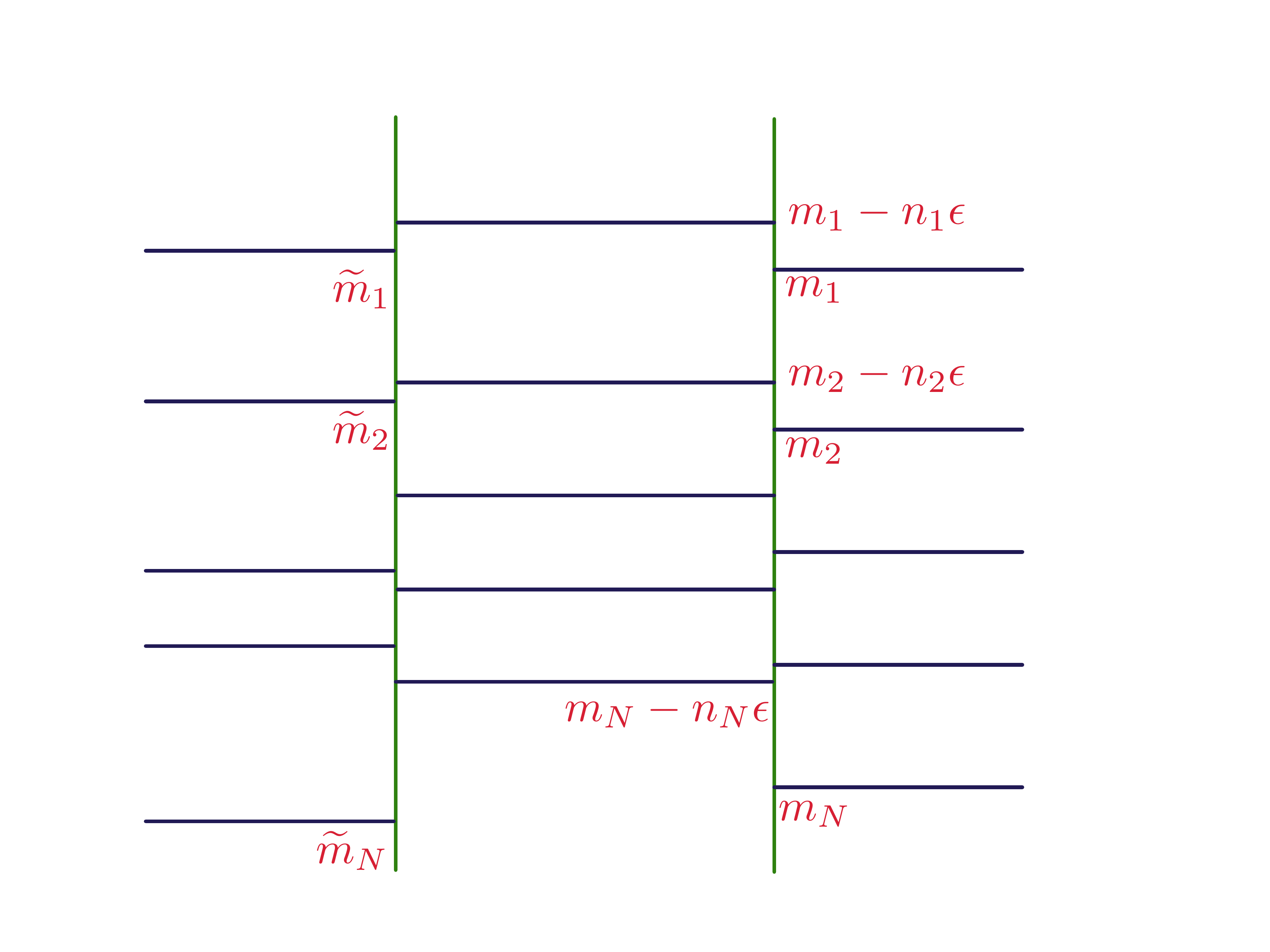} \includegraphics[height=11cm, width=14cm]{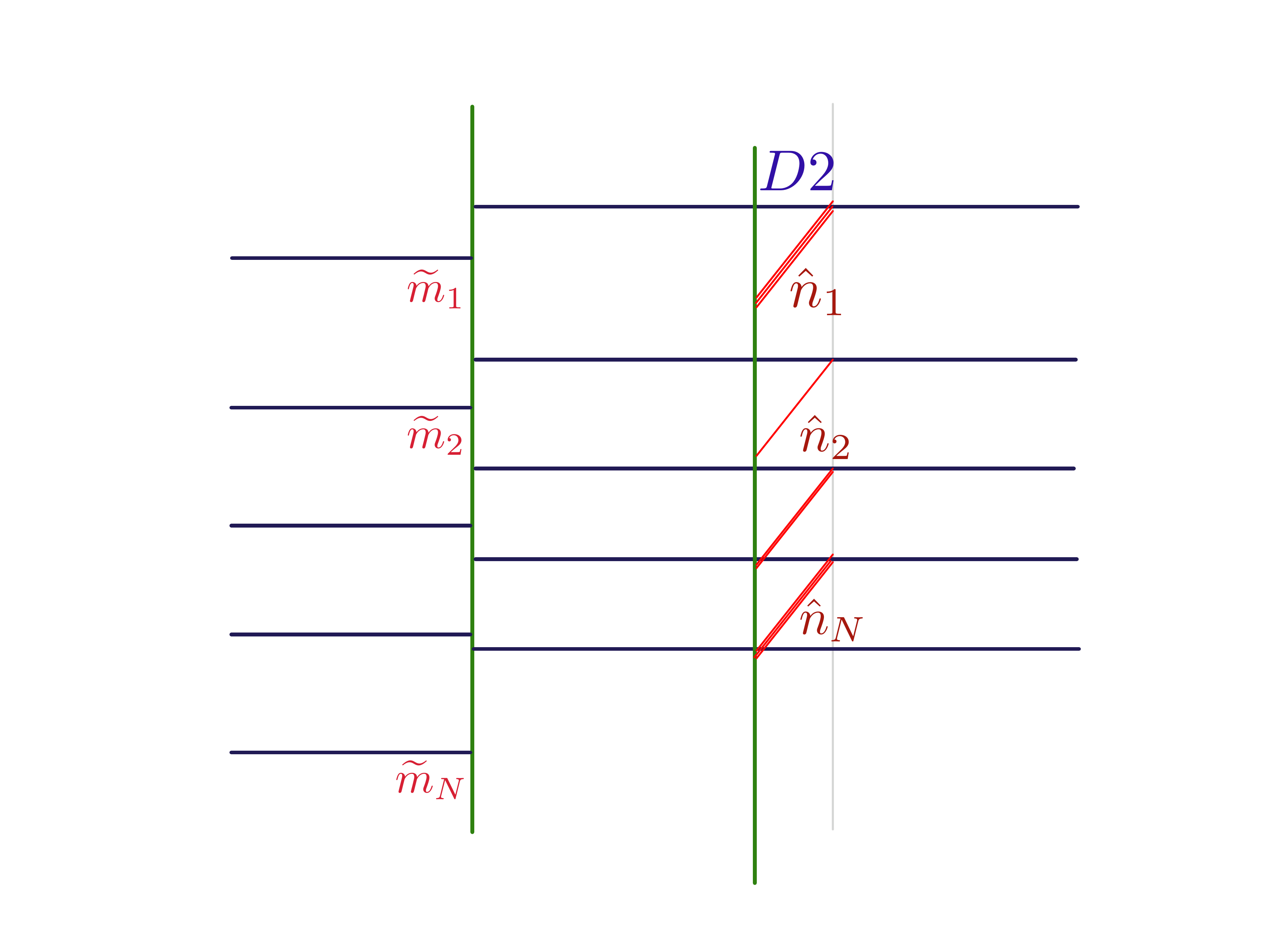}
\caption{Top: Type IIA brane picture. Positions of semi infinite D4 branes in 45 plane is given by $m_i + n_i \epsilon$, where $i=1,\dots, N$.  In the deformed configuration the Higgs phase of the theory is given by the condition $a_i = m_i+n_i\epsilon$.
Bottom: Vortex strings as D2 branes stretched in the 7th direction.}
\label{fig:hweps}
\end{center}
\end{figure}

Brane configuration for the $\ssN=2$ SQCD employs the Hanany-Witten construction \cite{Hanany:1996ie}. As it was shown in\cite{Dorey:2011pa} and further explained here in \secref{sec:SQCD}, in presence of Omega background the Higgs branch condition gets deformed \eqref{eq:DHLHiggs}. Hence the positions of the flavor D4 branes are shifted by $n_a\epsilon$ for each color (see bottom picture in \figref{fig:hweps}). It contains two NS5, N D4 branes which are stretched between the two NS5 branes and two sets of semi infinite D4's which are attached to NS5's. All D4 branes occupy 01236 directions, NS5's lie in 012345 directions.
\begin{center}
\begin{tabular}{|c|c|c|c|c|c|c|c|c|c|c|}
\hline
         & 0 & 1 &  2 & 3 & 4 & 5 & 6 & 7 & 8 & 9 \\
\hline \hline
NS5 & x &  x &  x & x & x &  x &    &     &    &    \\
\hline
D4    & x &  x & x  & x &    &     & x &    &     &    \\
\hline
D2    & x &     &     & x &    &     &    &  x &    &    \\
\hline
\end{tabular}
\end{center}
Under geometric transition the brane configuration described in \cite{Dorey:2011pa, Chen:2011sj} interpolate between the 4d theory and the 2d theory. The latter can be obtained by moving the right NS5 brane in the 7th direction and emerging D2 branes (037) which are stretched between this NS5 and D4's (see right picture in \figref{fig:hweps}). The value of $x_7$ gives tension of D2 strings which is equal to $\epsilon$ in our construction.

The rank of the gauge group of the two dimensional GLSM is given by summing up all the D2 branes $K=\sum_i \hat{n}_i$, where, we remind, $\hat{n}_i=n_i-1$. The low energy dynamics of the two dimensional theory is given by the effective twisted superpotential and the following ground state equations
\[\label{eq:2dXXXBAE}
\prod\limits_{l=1}^{N}\frac{\lambda_j-M_l}{\lambda_j-\widetilde M_l}=q\prod\limits_{k\neq j}^K\frac{\lambda_j-\lambda_k-\epsilon}{\lambda_j-\lambda_k+\epsilon}\,,
\]
which is the Bethe ansatz equations for the anisotropic $SL(2)$ spin chain. Note that for generic 2d masses $M_a$ and $\widetilde{M}_a$ at each spins at each site $a=1,\dots, N$ have different representations. Indeed, in order to match each term in the left hand side of \eqref{eq:2dXXXBAE} with phases of anisotropic chain
\[
\prod\limits_{l=1}^{N}\frac{\lambda_i-\theta_a+S_a\epsilon}{\lambda_i-\theta_a-S_a\epsilon}\,,
\]
where $\nu_a$ are anisotropies and $S_a$ are spins\footnote{For example, $S_a=-1/2$ gives the $SL(2)$ chain}, one identifies \cite{Nekrasov:2009ui}
\[\label{eq:MassesNoether}
M_a = \theta_a - S_a\epsilon\,,\quad \widetilde{M}_a = \theta_a + S_a\epsilon\,.
\]
%


\subsection{The Gaudin/XXX duality}\label{sec:bispec}

It is known that the Gaudin model \cite{refId0} enjoys several dualities\footnote{Some details about the Gaudin model are given in \appref{sec:GaudLiouville}.}. First we recall the duality introduced at the classical level in \cite{springerlink:10.1007/BF00626526}. It relates the rational Gaudin model with SL(N) group at M sites and SL(M) group at N sites. The positions of marked points $z_i$ on the sphere corresponding to the inhomogenities
and the diagonal element of the twist matrix get interchanged. At the classical level the spectral curves and the action differentials are equivalent. At the quantum level the Bethe ansatz equations reflect this symmetry at the level of spectra.

Let us explain this symmetry in the brane picture. Let us first remind ourselves the similar symmetry in the Toda system discussed in \cite{Gorsky:1997jq}. It the Toda case this symmetry merely implies the equivalence of $2\times 2$ and $N\times N$ Lax operator representations which can be explained as the 90 degrees rotation of the viewpoint of the brane picture. In the first representation the gauge group is connected to NS5 branes, while in the second case it is defined by the number of D4 branes in the IIA picture.

\begin{figure}
\begin{center}
\includegraphics[height=4cm, width=14cm]{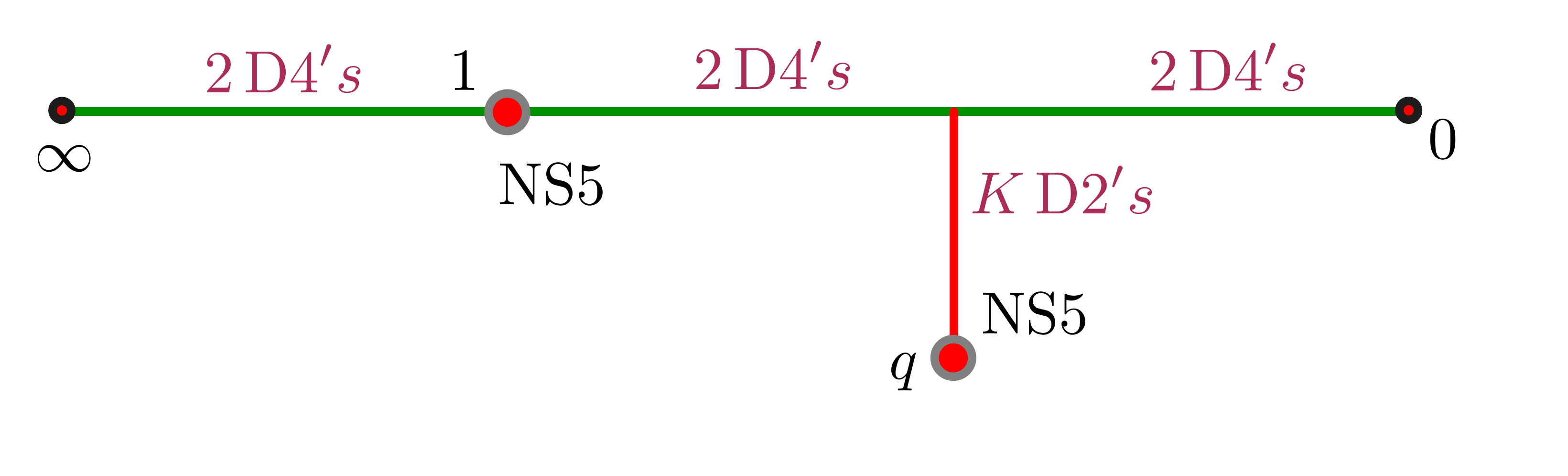}
\caption{$(6+i10,7)$ section of the HW brane construction (view from ``below'').}
\label{fig:bispec}
\end{center}
\end{figure}

If we add the fundamental matter and consider the conformal case there are additional data which have to be matched via the duality. In the $2\times 2$ representation
the $SL(2)$ twist matrix emerges which reflects the positions of NS5 branes in the 6-10 plane \figref{fig:bispec}.
The masses of the fundamentals provide the inhomogenities at the corresponding lattice sites. Upon the 90 degrees rotation similar to the Toda case the two sets of data get interchanged.

The duality between a pair of rational Gaudin models can be generalized to a similar duality between a trigonometric Gaudin
model and a XXX spin chain via the so-called $\mathfrak{gl}(M)/\mathfrak{gl}(N)$ duality \cite{MR2409414}.  For $M=N=2$ Bethe ansatz equations read as follows\footnote{We have adopted the notation and made some change of variable compared to \cite{MR2409414}.}
\[\label{eq:TrigomGaud}
\frac{\mathcal{M}_1-\mathcal{M}_2-\epsilon}{t_i} + \sum_{b=1}^2 \frac{\nu_b\epsilon}{t_i-z_b} -\sum_{\substack{j=1\\ j\neq i}}^{\kappa_2} \frac{2\epsilon}{t_i-t_j}=0, \,\quad i=1,\dots,\kappa_2\,,
\]
for trigonometric Gaudin, and
\[\label{eq:BAElargeEps}
\prod\limits_{a=1}^{2}\frac{\lambda_i+\mathcal{M}_a}{\lambda_i+\mathcal{M}_a+\kappa_a \epsilon}=\frac{z_2}{z_1} \prod_{\substack{
j = 1\\ j\neq i}}^{\nu_2}\frac{\lambda_i-\lambda_j-\epsilon}{\lambda_i-\lambda_j+\epsilon}\,,\quad i=1,\dots, \nu_2\,,
\]
for the $SL(2)$ XXX chain. The Mukhin-Tarasov-Varchenko (MTV) duality \cite{MR2409414} states that \eqref{eq:TrigomGaud} as set of equations with respect to $t_1,\dots t_{\kappa_2}$ has isomorphic space of orbits of solutions with the one of \eqref{eq:BAElargeEps} as set w.r.t. $\lambda_1,\dots,\lambda_{\nu_2}$ provided that %
\[\label{eq:Levelbalance}
\kappa_1+\kappa_2=\nu_1+\nu_2\,.
\]
Parameters $\mathcal{M}_{1,2}$ and $z_{1,2}$ are generic. We can now recognize \eqref{eq:2dXXXBAE} in \eqref{eq:BAElargeEps} with
\[
M_a=-\mathcal{M}_a,\quad \widetilde{M}_a =-\mathcal{M}_a-\kappa_a \epsilon,\quad K=\nu_2,\quad N=2,\quad z_1=1,\quad z_2=q\,,
\]
and parameters $\kappa_{1,2}$ and $\nu_1$ will be specified later. Also it will be more useful for us to use the 4d masses instead of the 2d ones. We can then rewrite set of MTV dual equations \eqref{eq:TrigomGaud,eq:BAElargeEps} as follows
\<\label{eq:MTVdualref}
\frac{-m_1+m_2-\epsilon}{t_i}+\frac{\nu_1\epsilon}{t_i-z_1} + \frac{K\epsilon}{t_i-z_2} \eq\sum_{\substack{j=1\\ j\neq i}}^{\kappa_2} \frac{2\epsilon}{t_i-t_j}\,,\nln
\prod\limits_{a=1}^{2}\frac{\lambda_i-m_a+\sfrac{3}{2}\epsilon}{\lambda_i-\widetilde{m}_a-\half\epsilon}\eq\frac{z_2}{z_1} \prod_{\substack{
j = 1\\ j\neq i}}^{K}\frac{\lambda_i-\lambda_j-\epsilon}{\lambda_i-\lambda_j+\epsilon}\,.
\>
Thus we can see that twists $z_1,z_2$, corresponding to the positions of the NS5 branes in 6-10 plane \figref{fig:bispec}, and masses of the fundamentals $m_1, m_2$
interchange their roles upon the duality. We see that matching to the BAE corresponding to $U(2),\, N_f=4$ SQCD shows that the strange nonequal mass shifts to the fundamentals and antifundamentals \eqref{eq:massescorr} have now clear interpretation within the duality. Namely, the number of the Gaudin Bethe roots yields the asymmetry between the fundamental and antifundamental masses. Also Gaudin spins match with the number of Bethe roots at the XXX side. Later in the next section we shall use these spins in order to make the AGT duality manifest.

Let us emphasize that the Hamiltonian of the Gaudin model is nothing but the r.h.s. of the Knizhnik-Zamolodchikov (KZ) equation \cite{Knizhnik198483} on the sphere with $L+3$ marked points $z_i$ \cite{Babujian_Flume_1993}
\[\label{eq:KZeq}
b^2\frac{d \Psi(z_i)}{d z_i}=\mathcal{H}_{Gaud}\Psi(z_i)\,, \quad i=1,\dots,L\,,
\]
where $b$ is some constant. In the next section, when we will discuss Liouville theory on the same Riemann surface, we shall specify its value.

One could also introduce the so called dynamical operators with respect to boundary conditions \cite{MR2409414}. Under the bispectral duality transformations the Gaudin KZ operator and the dynamical operators get interchanged as well. The number of marked points in the $N\times N $ representation of the Lax operator corresponds to the number of NS5 branes involved in the gauge theory brane construction.

\subsection{Bispectral duality and Argyres-Douglas points}

Classically, the bispectral duality just states that two systems
have almost the same (differ by the simple factor) spectral
curves and action differentials. At the quantum level the situation
is more subtle.
Since naively the bispectral duality connects the systems 
with different degrees of freedom, one should be able to analyze
the phenomena of merging of two degrees of freedom into the
single one. Below the simplest example shall be considered.

Let us now consider \eqref{eq:TrigomGaud} again, this time we identify $\mathcal{M}_1=-\mathcal{M}_2=l\epsilon$. Then one has
\begin{equation}
\label{BAE_Gaudin}
\frac{2l-1}{t_i}+\frac{\nu_1}{t_i-z_1}+\frac{\nu_2}{t_i-z_2}- \sum_{\substack{
b = 1\\ b\neq a}}^{\kappa_2}\frac2{t_i-t_j}=0,\quad  i=1,\dots,\kappa_2,
\end{equation}
for the Gaudin model and
\begin{equation}
\label{BAE_XXX}
\frac{s_a-l-\epsilon}{s_a-l-\epsilon-\kappa_1\epsilon}\frac{s_a+l-\epsilon}{s_a+l-\epsilon-\kappa_2\epsilon}\frac{z_1}{z_2} \prod_{\substack{
b = 1\\ b\neq a}}^{\nu_2}\frac{s_a-s_b-\epsilon}{s_a-s_b+\epsilon}=1,\quad a=1,\dots,\nu_2
\end{equation}
for the XXX model. Integers $\kappa_a, \nu_a$ satisfy the relation $\kappa_1+\kappa_2=\nu_1+\nu_2$. One of the results in \cite{MR2409414} is the precise correspondence of the orbits of solutions to the Bethe equations under the group of permutations of variables (permutations of $t_1,...,t_{\kappa_2}$ for the Gaudin model and of $s_1,...,s_{\nu_2}$ for the XXX model). At first glance such correspondence seems to be quite weak as it does not establish a direct connection between the roots of both systems and does not allow one to simplify one set of equations having known the solution of the other. But it preserves one important feature of the XXX model, namely the degeneration locus which could be called a little bit loosely ``quantum Argyres-Douglas (AD) points" \cite{Argyres:1995jj}.

The ``classical" Argyres-Douglas manifold is the locus in the moduli space of the theory 
where different vacua merge together. To get such merging typically one starts with ${\mathcal N}=2$ theory 
and perturbs it down to ${\mathcal N}=1$ and tunes the external parameters like couplings and masses.
In the ``quantum case" we have one more parameter from Omega deformation
$\epsilon$ and the AD manifold involves this additional coordinate
in the parameter space. The rest is the same and AD manifold
corresponds to the colliding vacua. In this subsection we shall normalize $\epsilon=1$.

Since the solution to the BA equation correspond to the  vacuum state
``quantum AD point" corresponds to the appearance of multiple roots. We shall consider a simple case $\kappa_1=\nu_2=2,\,\kappa_2=\nu_1=1$ as an example and find the AD manifolds for both models. Bethe ansatz equations the XXX chain \eqref{BAE_XXX} then read
\<\label{BAE_example}
\frac{s_1-l-1}{s_1-l-3}\frac{s_1+l-1}{s_1+l-2}\frac{s_1-s_2-1}{s_1-s_2+1}\eq q\,,\nln
\frac{s_2-l-1}{s_2-l-3}\frac{s_2+l-1}{s_2+l-2}\frac{s_2-s_1-1}{s_2-s_1+1}\eq q\,,
\>
and the Gaudin system is described by a single Bethe equation. In general, the XXX BAE contain a number of degenerate solutions, such that for some $i,j$ the roots coincide $s_i=s_j$. The vacua of the theory correspond to the non-degenerate solutions of the system \cite{Dorey:2011pa}. Obviously every solution to the degenerate system is a solution to the full BAE system. This property can be used to lower the degree of the Bethe equations. In our case the degenerate solution is $s_1=s_2$, so from \eqref{BAE_example} we obtain
\begin{equation}
\label{BAE_degenerate}
\frac{s-l-1}{s-l-3}\frac{s+l-1}{s+l-2}=-q.
\end{equation}
We can now solving the first equation of (\ref{BAE_example}) with respect to $s_2$ and substitute this solution into the second equation to obtain a polynomial roots of which solve the XXX BAE system. In order to eliminate the degenerate roots we merely need to divide this polynomial by (\ref{BAE_degenerate}). To find the AD manifold we calculate the discriminant of the reduced polynomial
\<\label{Discr1}
D_1(l,q)\eq 4 (1-q)^6 q^2 \left(4 l^2 q^2-8 l^2 q+4 l^2-4 l q+4 l+8 q+1\right)^2\nl
\big(4 l^2 q^4-32 l^2 q^3+56 l^2 q^2-32 l^2 q+4 l^2+4 l q^4-36 l q^3\nl
+28 l q^2+4 l q+q^4-18 q^3+17 q^2-8 q\big).
\>
However, the set $D_1=0$ still contains extra roots. Although the equation was divided by the degenerate one, the discriminant still captures the cases when the roots of the reduced system coincide with the roots of the degenerate system. In order to exclude such cases discriminant $D_1$ \eqref{Discr1} must be divided by the resultant of the reduced and degenerate polynomials. This resultant turns out to be precisely the polynomial in the last parentheses in (\ref{Discr1}). Thus the resulting AD set is described by the zero locus of the following polynomial (we do not discuss trivial cases when $q=0$ and $q=1$)
\begin{equation}
\label{Discr}
D(l,q)=4 l^2 q^2-8 l^2 q+4 l^2-4 l q+4 l+8 q+1\,,
\end{equation}
which is precisely the discriminant of the Gaudin equation written in a polynomial form! 

The above example describes the way to calculate the quantum AD set for the XXX BAE. First, we identify the degenerate subset and divide all equations by the corresponding polynomials. Then we find the discriminant of the reduced polynomial and divide it by all possible resultants with the degenerate polynomials. The same procedure can be done for the Gaudin system. The resulting polynomial describes the set of the quantum AD points and coincides for XXX and Gaudin systems. Certainly we have considered the simplest example and this
issue deserves a separate study.

\subsection{Walls of marginal stability}

One more issue we shall briefly discuss concerns the interpretation of the walls of marginal stability in the Omega deformed theory in the language of the quantum integrable system. The wall can be described in terms of the superpotential however the Yang-Yang function in the quantum integrable system has the interpretation of the twisted superpotential as well. Hence we could formulate the problem of finding the walls of marginal stability in terms of the YY function for the spin chain or Bethe ansatz equations.

The $\epsilon$-deformed theory has exact effective twisted superpotential \eqref{eq:Eff2dGLSM}. XXX Bethe ansatz equations \eqref{eq:XXXWsp} specify the positions of the vacua. Kinks in the two-dimensional theory which interpolate between these vacua possess two kinds of charges: Noether charges $M_l, \widetilde M_l$ or 2d twisted masses \eqref{eq:MassesNoether} and the topological charges which are given by the difference of the vacuum values of the superpotential \eqref{eq:Eff2dGLSM} evaluated at the vacua the kink is hopping between. A wall of marginal stability is the solution to the following kinematical condition of the decay of the kink into its constituents
\begin{equation}
\label{MS}
\mathfrak{Im}\frac{Z_{\text{top}}}{Z_{\text{N\"other}}}=\frac{\widetilde{\mathcal{W}}(\lambda_{vac}^{(1)})-\widetilde{\mathcal{W}}(\lambda_{vac}^{(2)})}{M_1-M_2}=0\,.
\end{equation}

We follow the procedure described in \cite{Bolokhov:2011mp, Bolokhov:2012dv}. The idea of the analysis is that the vacua and the vacuum values of the superpotential should be continuous functions of all parameters across the wall of marginal stability. We investigate the system depending on a single complex parameter $\theta$ and make all the deformation parameters to be equally spaced on a circle
\begin{equation}
\theta_l=\theta \exp\left(\frac {2\pi il}N\right)\,,
\end{equation}
where $\theta_l$ are the spin chain impurities and contribute to 2d masses \eqref{eq:MassesNoether}. Note that the above $\mathbb{Z}_L$ choice of $\theta$'s is a simplification as it certainly cannot be done for generic masses. However, as it argued in \cite{Bolokhov:2011mp, Bolokhov:2012dv}, it is generic enough for the study of wall crossing phenomena in 2d.

We can now interpret the vacuum solutions to the BAE equations $\left\{\left.\lambda_{vac}^{(i)}\right|_{\theta}\right\}$ not as separate functions, but rather as branches of some continuous function $\lambda_{vac}(\theta)$ on the complex plane. The number of branches $\ell$ is of course finite and coincide with the degree of the BAE system as a system of polynomial equations. The superpotential is the continuous function depending on $\lambda_{vac}(\theta)$. It has infinitely many branches but the sequence of branches has a certain periodicity of order $\ell$.

The supersymmetric $\mathbb{P}^{N-1}$ theory considered in \cite{Bolokhov:2011mp, Bolokhov:2012dv, Dorey:2012dq} has infinitely many walls of marginal stability. The vacuum as the function of the twisted mass parameter $M_0$
\[
M_l=M_0\exp{\left(\frac{2\pi il}N\right)}
\]
has $N$ branches and the branches of the superpotential differ by the quantity proportional to $m_0$
\begin{equation}
\widetilde{\mathcal{W}}_{vac}^{(n)}(M_0)=\exp\left(\frac{2\pi in}{N}\right)\left(\widetilde{\mathcal{W}}_{vac}^{(0)}(M_0)-2\pi i nM_0\right).
\end{equation}
The condition for the kink interpolating between $n$th and $(n+1)$st vacua to merge with a state of N\"other charge $m_l-m_0$ on the wall of marginal stability reads
\begin{equation}
\label{MS_BSY}
\mathfrak{Re} \left(\frac {\widetilde{\mathcal{W}}_{vac}^{(0)}}{M_l}-\frac{2\pi in\exp\left(\frac {2\pi i}N\right)}{\exp\left(\frac {2\pi i l}N\right)-1}\right)=0\,.
\end{equation}
The above equation describes $N$ walls which are roughly logarithmic spirals with infinitely many branches.

\begin{figure}[h]
\begin{center}
\includegraphics{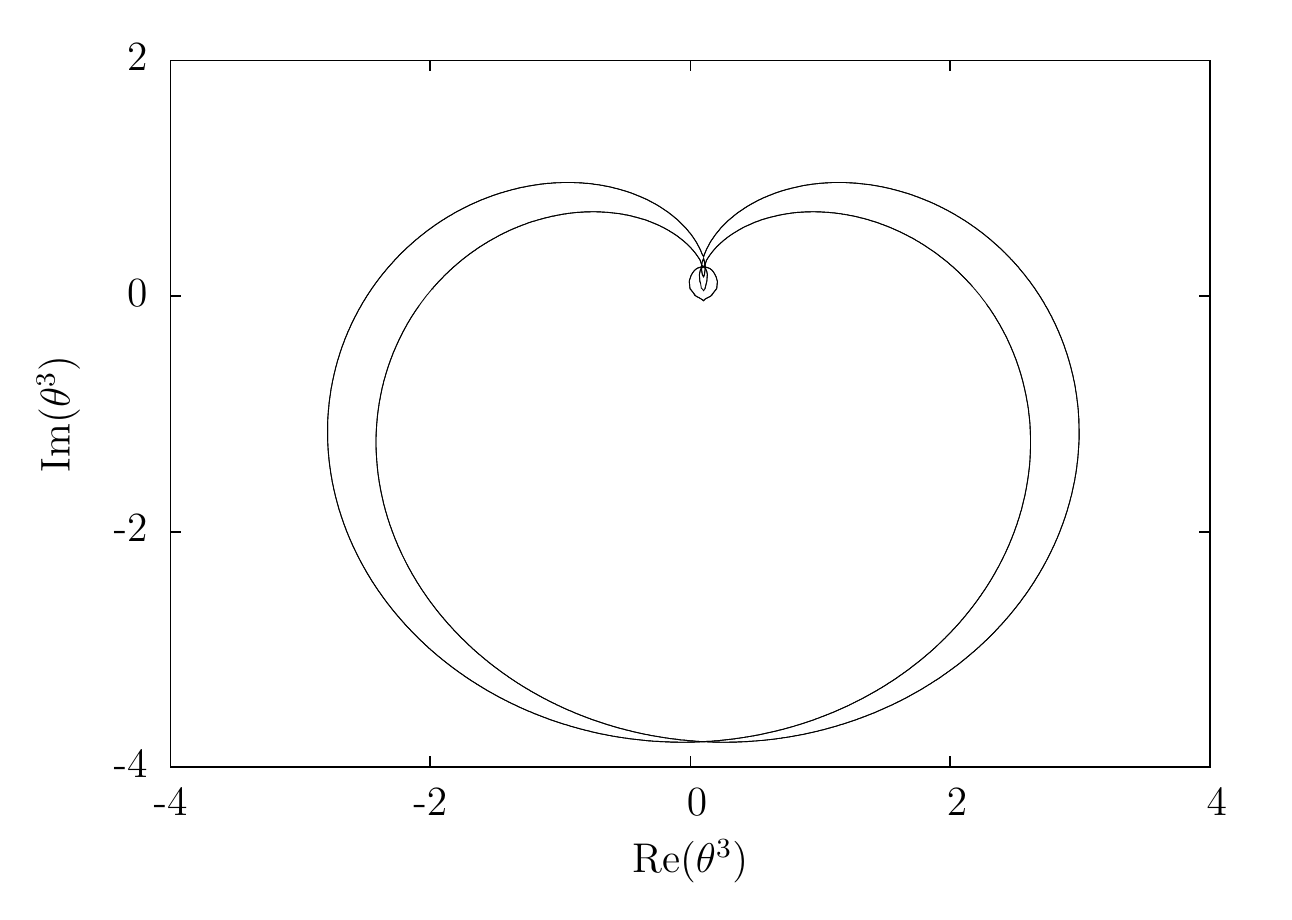}
\caption{The walls of marginal stability for the XXX superpotential on the $\theta^3$ plane. The parameters are: $q=-1,\, S=\frac 12,\,\epsilon=1$. The AD point is at $\theta^3=\frac i4$.}
\label{b3}
\end{center}
\end{figure}

\begin{figure}[h]
\begin{center}
\includegraphics{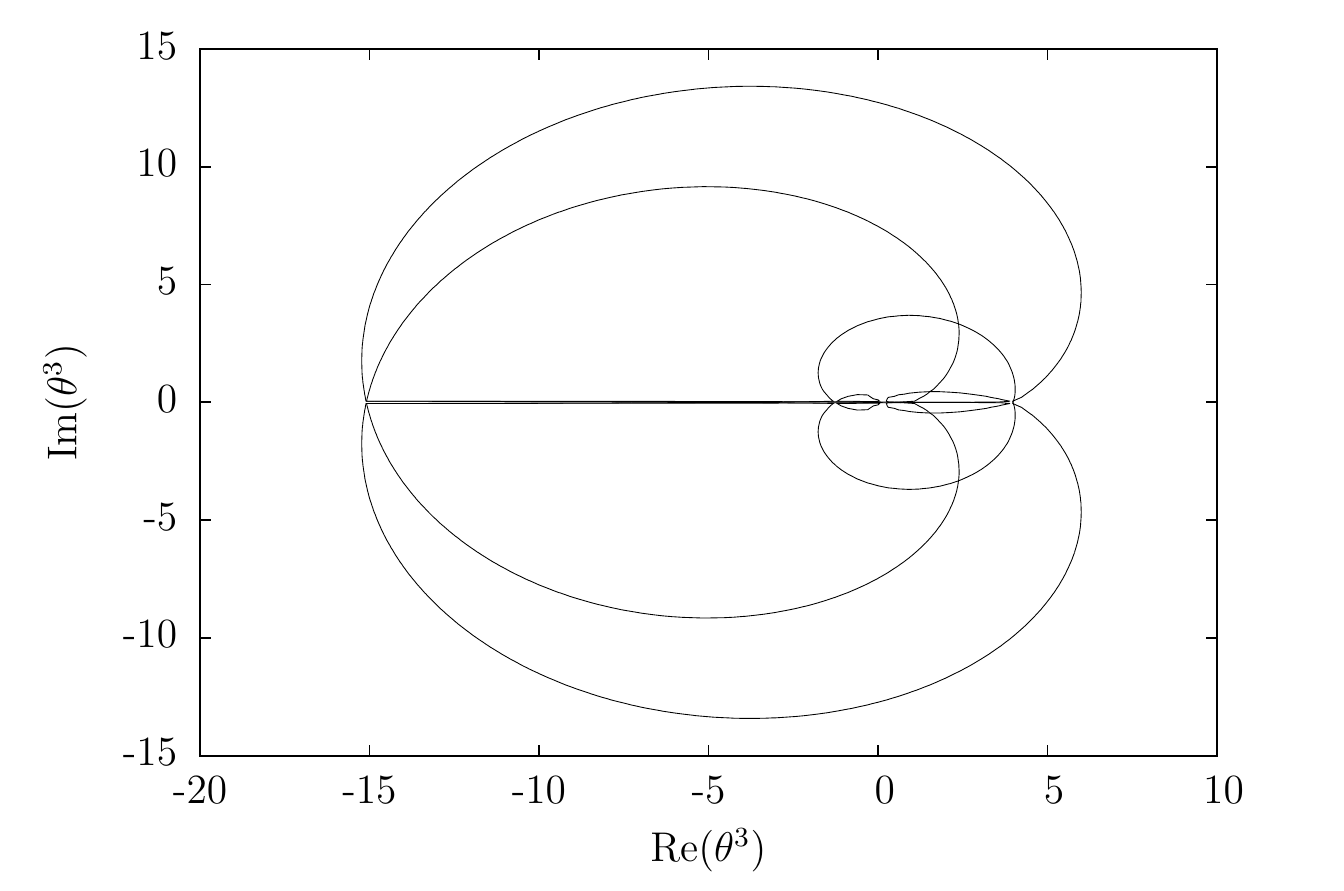}
\caption{The walls of marginal stability for the Gaudin superpotential on the $\theta^3$ plane. The parameters are: $G=1,\, S=\frac 12$. The AD point is at $\theta^3=4$.}
\label{g3}
\end{center}
\end{figure}

In the case of the Omega deformed theory the situation changes a little bit. 
In order to depict the walls of marginal stability graphically in general case, one needs to solve the system of BAE. We take the case $N=3,\,K=1$ in \eqref{eq:2dXXXBAE} as the simplest example. The BAE system is
\begin{equation}
\frac{(\lambda+S\epsilon)^3-\theta^3}{(\lambda-S\epsilon)^3-\theta^3}=q.
\end{equation}
The solution has three branches. The difference between $n$th and $(n+3)$rd branches of the superpotential equals $4\pi i$. The walls of marginal stability are depicted in \figref{b3}.
In the limit $\epsilon \to 0$ the XXX system passes into the Gaudin system. The walls of marginal stability can be plotted for the Gaudin superpotential too. They possess the main features of the walls of marginal stability for the XXX system. The superpotential \eqref{eq:Eff2dGLSM} at small $\epsilon$ becomes
\begin{equation}
\label{W_G}
\widetilde{\mathcal{W}}_{\text{Gaudin}}(\lambda_j)=G\sum_{j=1}^K\lambda_j-2\sum_{j=1}^K\sum_{l=1}^Ns_l\log\left(\lambda_j-\theta_l\right)-\sum_{j\neq k}^N\log\left(\lambda_j-\lambda_k\right)\,,
\end{equation}
where $G$ is external field. In the case $N=3,\, K=1$ the solution to the corresponding Gaudin equation has three branches. The difference between $n$th and $(n+3)$rd branches of the superpotential remains $4\pi i$ as in the XXX case. The corresponding walls of marginal stability are drawn in \figref{g3}.
%

\subsection{On the spin chain/Calogero duality}

One may wonder if the bispectral duality between two different types of the spin chains has any relation to a similar duality discussed in the context of the Calogero--Ruijsenaars family of the integrable systems. The answer turns out to be positive. The prototype of this duality has been discovered in \cite{1996alg.geom.12001G}, where the correspondence between the zero locus of the Toda Hamiltonians in the phase space and the manifold associated with the quantum model has been
discovered. This correspondence has been generalized for the rational Calogero model \cite{MR0280103, MR0375869, Sutherland:1971ks} at fixed coupling which turns out to be dual in the same sense to the rational Gaudin model \cite{2009arXiv0906}. 

The identification of the parameters goes as follows. The inhomogenities in the Gaudin model $z_i$ are identified with the coordinates of Calogero model $x_i$, while the eigenvalues of the Gaudin Hamiltonians $H_i$ are identified with the Calogero momenta. The inhomogenities are Poisson dual to the positions of the marked points, hence their identification with the Poisson pair in the Calogero model is natural. The duality above can be extended from the zero locus at the Calogero side to the arbitrary values of the classical Calogero Hamiltonians \cite{2012arXiv1201.3990M}
\[
\mathcal{H}_{\text{Cal},k}=\sigma_k(\lambda)\,,
\]
where $\sigma_k$ is the k-th symmetric power of the Lax connection eigenvalues. It turns out that the Lax eigenvalues at the Calogero side are mapped onto the eigenvalues of the twist matrix at the spin chain side.

\begin{figure}
\begin{center}
\includegraphics[height=5cm, width=12cm]{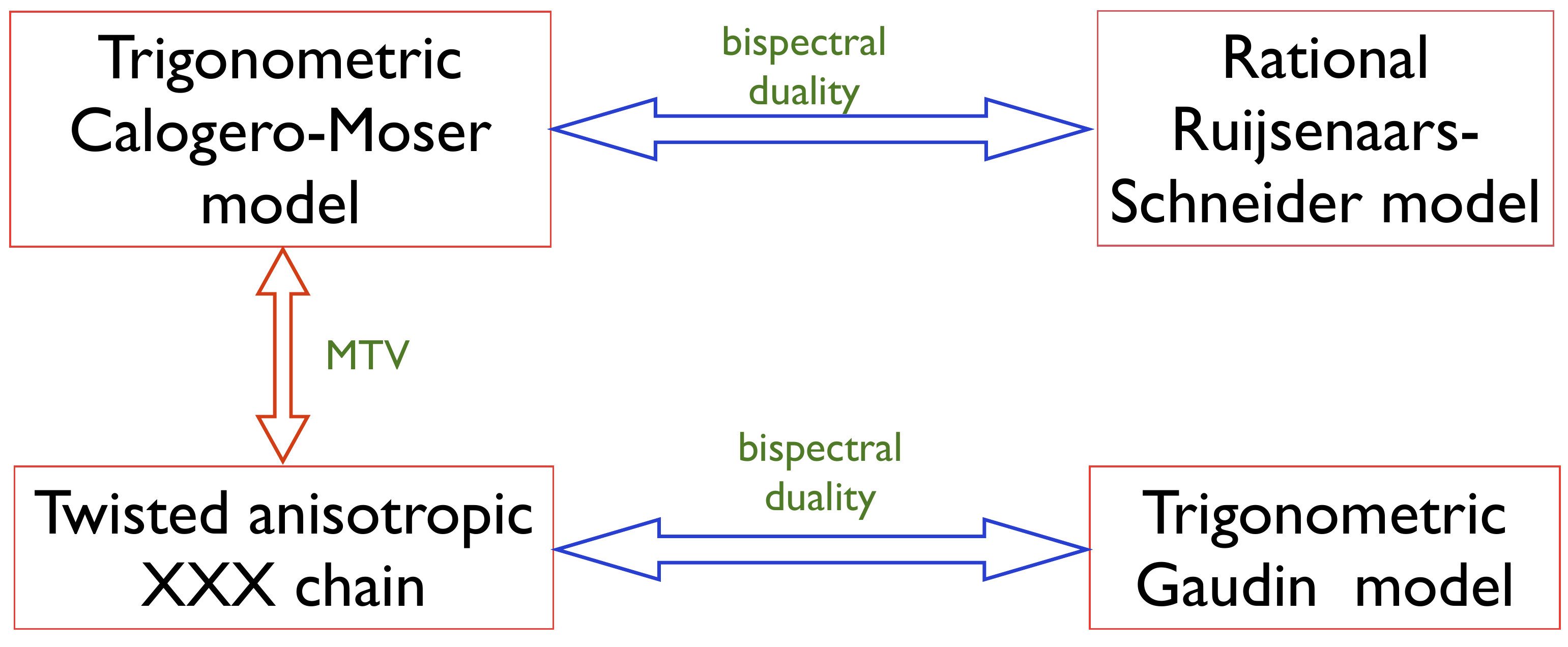}
\caption{A pair of bispectral dualities mapped onto each other} 
\label{fig:MTVbispec}
\end{center}
\end{figure}

As we have discussed above the rational Gaudin model enjoys the marked points/twist duality   and in some sense is selfdual. Its bispectral dual -- the rational Calogero model is selfdual as well in the same sense. The bispectral duality can be generalized to the trigonometric and relativistic cases \cite{MR1322943, MR1329481, MR887995, MR929148}. Thus the trigonometric Calogero-Moser model is known to be dual to the rational Ruijsenaars-Schneider model \cite{MR851627, MR887995}\footnote{See also \cite{2009JPhA...42r5202F,2010JMP....51j3511F,2011CMaPh.301...55F,2012NuPhB.860..464F}}. The quantum version of this duality has been elaborated in \cite{chalykh:5139}. Recall that in \secref{sec:bispec} we discussed another bispectrally dual pair between the trigonometric Gaudin and the XXX models \cite{MR2729945}. We can now see that the bispectrality at the Calogero--Ruijsenaars side matches perfectly with the duality at the Gaudin--XXX side. Note that this duality has the clear interpretation in terms of the Chern-Simons theory with inserted Wilson lines and its Yang-Mills
degenerations \cite{Fock:1999ae}. The relationship between the two dualities is summarized in \figref{fig:MTVbispec}

It would be interesting to make the next step and consider the selfdual trigonometric Ruijsenaars model at the Calogero side of the correspondence. Its dual on the spin chain side is expected to be the XXZ chain which, according to the correspondence considered above, should enjoy some kind of bispectral selfduality. Another important issue concerns the generalization of these dualities to the elliptic integrable models. Not much is known about the self-dual elliptic model yet (see, however, \cite{Fock:1999ae, Braden:1999aj, Gorsky:2000ds, Braden:2001yc,Braden:2003gv})

To conclude this section note  that curiously we arrive to the claim that since the twists at the spin side corresponds to the spectrum of the dual Calogero model, therefore the SQCD gauge coupling, which corresponds to the twist, could be quantized! This point deserves the further analysis.

\section{The AGT Correspondence in the NS Limit}\label{sec:AGTNS}

The Alday-Gaiotto-Tachikawa duality \cite{Alday:2009aq} relates a conformal block of the Liouville CFT on a Riemann surface of genus $g$ and $n$ punctures with the instanton part of the Nekrasov partition function of a quiver gauge theory naturally associated with this Riemann surface. Furthermore it concludes that the $n$-point correlator in the Liouville theory on this  Riemann surface can be evaluated as an integral of the square of the absolute value of the full Nekrasov partition function.

Having done the above analysis on spectral duality between integrable systems,  we arrive at an interesting observation which envisages the AGT correspondence for the Liouville theory with large central charge on $S^2$ with four punctures and $U(2)$ SQCD with four flavors. It has already been addressed in the literature earlier \cite{Mironov:2009qn} where a Liouville conformal block at large $c$ simply becomes a hypergeometric function of the conformal dimensions and the instanton number \cite{springerlink:10.1007/BF01214585} was used. Then the authors figured out that only chiral terms in the Nekrasov partition function will contribute, either $\epsilon_1$ or $\epsilon_2$ have to vanishes; it enabled them to identify each multi-instanton contribution with corresponding terms of the hypergeometric function's expansion. The proof is rather formal and it will be desirable to have a more physical rationale to it. The current section is intended to fulfill this goal.

\begin{figure}
\begin{center}
\includegraphics[height=10.5cm, width=16.5cm]{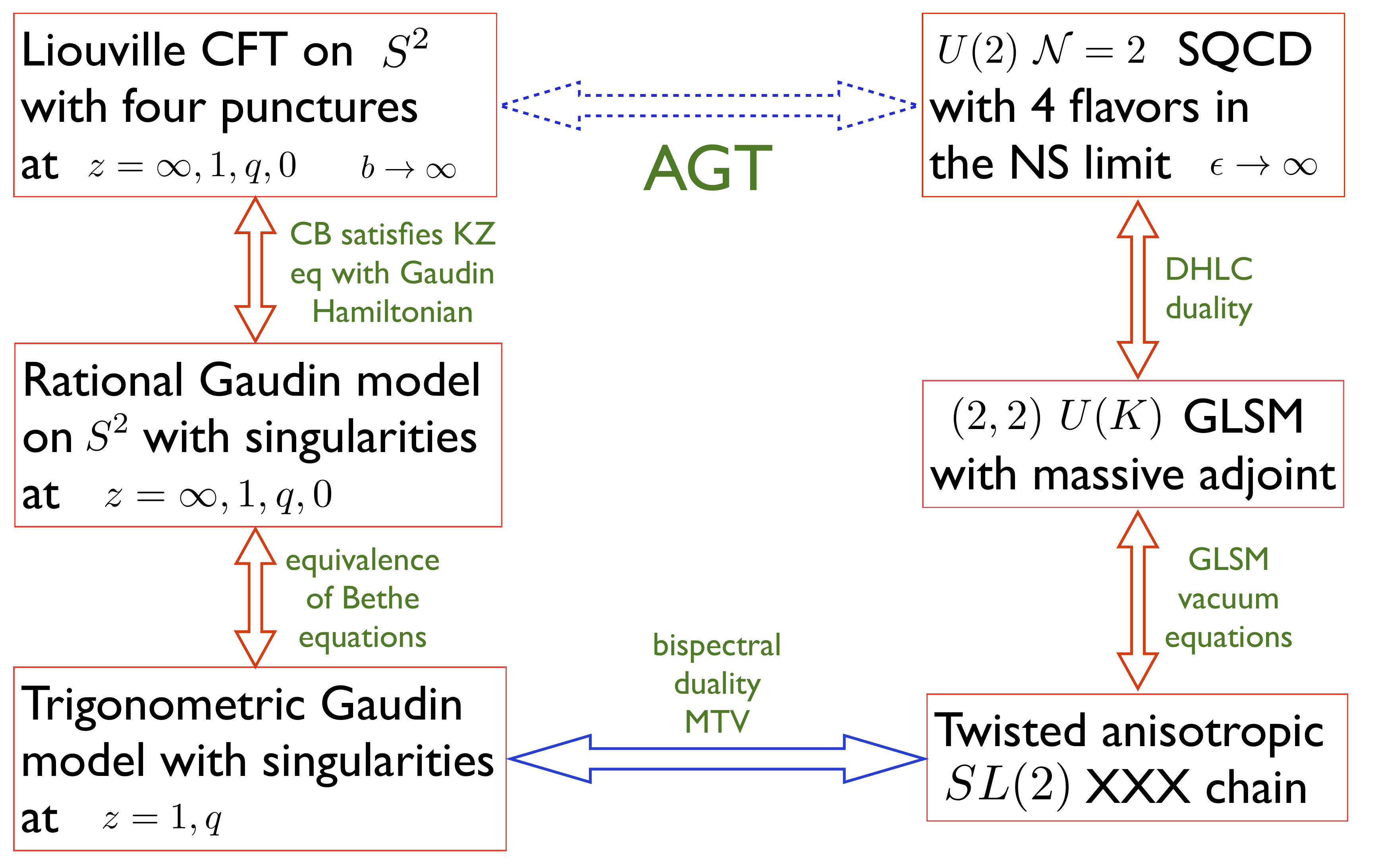}
\caption{Roadmap of the AGT duality in the NS limit. The main statement (top horizontal line) is obtained by the chain of dualities between various integrable systems.}
\label{fig:roadmap}
\end{center}
\end{figure}

In order to see how the AGT relation comes about, in what follows we shall relate both the gauge theory and the Liouville theory to a pair of integrable systems which enjoy a certain spectral duality between them. The roadmap we shall use to guide us through this section is presented in \figref{fig:roadmap}.
Starting from the 4d gauge theory on the top right of \figref{fig:roadmap} we shall use the results of \secref{sec:SQCD} and \cite{Chen:2011sj} in order to relate the 4d theory with the corresponding 2d GLSM. As it was shown in \cite{Chen:2011sj}, the equivalence was established \textit{to all orders in the instanton parameter}. Thus, instead of working with each summand of the instanton partition function, as all known proofs of the AGT \cite{Mironov:2009qn, Fateev:2009aw, 2012arXiv1202.2756S}  did so far, we shall look at the entire expression and the effective twisted superpotential \eqref{eq:EffTwistedExactSuperpot} which follows from it. Following this line of thought we provide a physical rationale for the AGT correspondence in the NS limit, namely, modulo certain dualities between two given integrable systems, it is reduced to the 4d/2d duality in the NS limit \cite{Chen:2011sj}. The latter exists due to non-Abelian semilocal BPS vortices which we have discussed in \secref{sec:SQCD}, also vortices exist provided that the corresponding FI parameter is turned on, hence we uncover why an extra $U(1)$ factor on the gauge theory side is important.

 The quantum duality, together with its brane interpretation, was discussed earlier in \secref{sec:bispec}. Now we shall start with the left column of \figref{fig:roadmap} by reminding ourselves how the Gaudin model is related to Liouville conformal blocks, and later on, by means of the bispectral duality, we shall connect the story to the Heisenberg $SL(2)$ chain and to the 4d gauge theory.\footnote{Some remarks on the role of bispectral duality between the trigonometric Gaudin model and the XXX chain on the classical level can be found in \cite{Mironov:2012uh, Mironov:2012ba}}

\subsection{Liouville theory and rational Gaudin model}

Recall that the Liouville theory has central charge
\[
c=1+6Q^2\,, \qquad Q = b+\frac{1}{b}\,,
\]
and in the classical limit $b\to \infty$ so $Q\to\infty$ as well. Let us now consider a conformal block $\mathcal{F}_{\alpha_0\,\,\,\alpha\,\,\,\,\alpha_1}^{\,\,\mu_0\,\,\mu_1}(q)$ of the Virasoro algebra with central charge $c\to \infty$ with the four primary operators of dimensions
\[\label{eq:ChiralPrimConfDim}
\Delta_1 = \alpha_0(Q-\alpha_0)\,,\quad \Delta_2 = \mu_0(Q-\mu_0)\,,\quad \Delta_3 = \mu_1(Q-\mu_1)\,,\quad \Delta_4 = \alpha_1(Q-\alpha_1)\,,
\]
inserted at points $\infty,1,q,0$ respectively on the $S^2$ with an intermediate $s$-channel state of dimension $\Delta =\alpha(Q-\alpha)$. In the above formula
\[\label{eq:alphas}
\alpha_0 = \half Q + \widetilde{\mu}_0\,,\quad \alpha = \half Q +a\,,\quad\alpha_1 = \half Q + \widetilde{\mu}_1\,,
\]
where $a$ is the $SU(2)$ Coulomb branch coordinate. In the above formulae the mass parameters represent the following linear combinations of the SQCD quark masses $m_{1,2,3,4}$
\[
\mu_0=\half(m_1+m_2),\quad \widetilde{\mu}_0=\half(m_1-m_2),\quad \mu_1=\half(m_3+m_4),\quad \widetilde{\mu}_1=\half(m_3-m_4)\,.
\]
There is an obvious notational conflict with \cite{Alday:2009aq}, where $\mu$'s and $m$'s are interchanged compared to our paper. We had to switch the notations in order to be consistent with \secref{sec:SQCD}, were $m$'s are used for the quark masses. As far as the rest of the notations are concerned, they will be in agreement with \cite{Alday:2009aq}. Note that in \secref{sec:SQCD} we treated all the four flavors as fundamental hypermultiplets, however, in \cite{Alday:2009aq} as well as in \cite{Dorey:2011pa} two of them, with masses $m_3$ and $m_4$ are considered to be fundamental and two others, with masses $m_1$ and $m_2$ to be antifundamental. For the purposes of \secref{sec:SQCD} this turned out to be a mild difference and we were able to relate the 4d and 2d theories by studying the vortex effective theory. Also from the GLSM perspective it was natural to distinguish fundamental and antifundamental fields. In this section we have to be more careful about this issue as contributions from the fundamental and anti-fundamental multiplets to the Nekrasov partition at finite $\epsilon$ are different.

Note that all conformal dimensions \eqref{eq:ChiralPrimConfDim} diverge at least linearly with $b$, however, as we shall later see, in order to match the Liouville CFT with the four dimensional theory in this limit, the dimensions will diverge quadratically and proper regularization is needed. Teschner  in \cite{Teschner:2010je} have identified effective twisted superpotential \eqref{eq:EffTwistedExactSuperpot}\footnote{According to the NS dictionary this is also a Yang-Yang function} with the NS limit of a Liouville conformal block on the sphere as well as the proper regularization of the conformal dimensions. Conformal block $\Psi(z_i)$ as the function of punctures' locations was found to satisfy the KZ equation \eqref{eq:KZeq} for the dual WZNW model with level $k$ and $b^2 = -(k+2)^{-1}$
\[\label{eq:KZClassical}
-\frac{1}{k+2}\frac{d \Psi(z_i)}{d z_i}=\mathcal{H}_{Gaud}\Psi(z_i)\,, \quad i=1,\dots,L\,,
\]
where $\mathcal{H}_{\text{Gaud}}$ is the Hamiltonian of the rational Gaudin model\footnote{See also \cite{Bonelli:2011na} where the matrix model approach to Hitchin systems in connection with the Liouville CFT was constructed.}. Thus the large $b$ limit corresponds to taking $k\to -2$. The conformal dimensions of chiral primary operators get rescaled and become
\[\label{eq:rescaleddims}
\delta_i = -\frac{\Delta_i}{b^2}\,,
\]
as $b \to\infty$\,. For $S^2$ with four punctures at $\infty,1,q$ and $0$ respectively from \eqref{eq:ChiralPrimConfDim, eq:alphas} and \eqref{eq:rescaleddims} we obtain
\<\label{eq:RescaledConfDims}
\delta_1 \eq \left(\frac{\widetilde{\mu}_0}{b}-\frac{1}{2}\right)\left(\frac{\widetilde{\mu}_0}{b}+\frac{1}{2}\right)\,,\nln
\delta_2 \eq \left(\frac{\mu_0}{b}-1\right)\frac{\mu_0}{b}\,,\nln
\delta_3 \eq \left(\frac{\mu_1}{b}-1\right)\frac{\mu_1}{b}\,,\nln
\delta_4 \eq \left(\frac{\widetilde{\mu}_1}{b}-\frac{1}{2}\right)\left(\frac{\widetilde{\mu}_1}{b}+\frac{1}{2}\right)\,,
\>
as $b\to \infty$. Our next step is to allow the mass parameters $\mu_a$ and $\widetilde{\mu}_a$ scale with $b$ upon identification with the 4d theory. 

Equivalently one can also probe Liouville conformal blocks with surface operator insertions \cite{Alday:2009fs, Drukker:2009id}, those conformal blocks satisfy Gaudin eigenvalue problem in the NS limit\footnote{See also \cite{Dimofte:2010tz} where a systematic study of 4d gauge theories with surface operators in Omega-background was done.}.

\subsection{$\ssN=2$ SQCD in the NS Omega background}

On the 4d gauge theory side, we compute the Nekrasov partition function for the 4d $\ssN=2$ SQCD with mass parameters $\mu_0,\widetilde{\mu}_0,\mu_1,\widetilde{\mu}_1$ whose instanton part is 
\[\label{eq:PartFuncLiouv}
\mathcal{Z}_{\text{inst}}(a,\mu_0,\widetilde{\mu}_0,\mu_1,\widetilde{\mu}_1)  = (1-q)^{2\mu_0(Q-\mu_1)}\mathcal{F}_{\alpha_0\,\,\,\alpha\,\,\,\alpha_1}^{\,\,\mu_0\,\,\mu_1}(q)\,,
\]
where $\alpha = \half Q-a$ and $a$ is the $SU(2)$ Coulomb modulus. For a generic Omega background the AGT dictionary says the deformation parameters are related to the 2d Liouville theory  via
\[
b^2 = \frac{\epsilon_1}{\epsilon_2}\,,\quad \hbar^2 = \epsilon_1 \epsilon_2\,.
\]
The usual NS limit $\epsilon_2\to 0$ corresponds to $b\to\infty$ and $\epsilon_1$ is kept fixed, then the Liouville theory becomes classical as $\hbar \to 0$. 
In this section, we will be rather interested in the quantum regime of the Liouville theory so we shall allow  $\epsilon_1\to\infty$ such that $\hbar$ is kept fixed. It is clear that by a proper tuning of $\epsilon_1$ and $\epsilon_2$ one can obtain any desired value of the Planck constant.

As we have already discussed above, in the NS limit a more appropriate object to study is not the Nekrasov partition function but the effective twisted superpotential \eqref{eq:EffTwistedExactSuperpot}. As it was shown in \cite{Dorey:2011pa} that this superpotential also emerges from the $(2,2)$ GLSM which we have described in \secref{sec:SQCD}.

The DHL paper has done a perturbative calculation in the instanton number $q$ in order to establish their 4d/2d duality \eqref{DHLduality} and the proof  to all orders was further established in \cite{Chen:2011sj}. CDHL showed that in the NS limit the Nekrasov partition function can be represented as an integral over a finite set of variables and can be explicitly evaluated, and the saddle point condition is shown to be equivalent to the Bethe ansatz equations for the $SL(2)$ XXX chain. 
One may ask immediately why the vortices are relevant, indeed they only exist in a Higgs branch of the four dimensional theory, whereas the AGT statement relates Liouville momenta with Coulomb branch coordinates. In order to understand this, let us recall that at zero value of the FI term the Higgs branch touches the Coulomb branch, and as it was pointed out in \cite{Dorey:2011pa}, by making a proper limit in the relation\footnote{From now on we shall work with the $U(2)$ SQCD with 4 flavors.}
\[\label{eq:CoulombHiggs}
a_a = m_{2+a} - n_a\epsilon\,, \quad a=1,2\,,
\]
one may recover \textit{any} point of the Coulomb branch of the $U(2)$ SQCD. Indeed, as $\epsilon\to 0$ the Higgs lattice becomes more and more dense filling the Coulomb branch in that limit. However, for what we are doing here, the opposite $\epsilon\to\infty$ limit is relevant, as it is required by the connection to the Liouville theory. Still we want to be able to cover any point on the Coulomb branch, so one has to scale the fundamental masses $m_a$ with $\epsilon$ as well in order to keep combination \eqref{eq:CoulombHiggs} finite. So at any given Liouville momentum we only need to sit at a certain point on a Coulomb branch and the Higgs branch root has all information we need about that point. Recall that the anti-fundamental masses and, correspondingly $\mu_0$ and $\widetilde{\mu}_0$ are not affected by \eqref{eq:CoulombHiggs} and therefore do not scale with $\epsilon$.

We now make an observation that the ground state equations for the $(2,2)$ GLSM \eqref{eq:2dXXXBAE} (or second equation in \eqref{eq:MTVdualref} where $\widetilde{m}_{1,2}$ are now denoted as $m_{1,2}$ (antifundamental) and $m_{1,2}$ became $m_{3,4}$ (fundamental) respectively\footnote{The obvious notational conflict occurs here. In spite of this we still keep the notations of this section and  \secref{sec:SQCD} as they are since both are natural in where they stand. We hope this issue will not confuse the reader.}.)
\[\label{eq:XXXBAE6}
\prod\limits_{a=1}^{2}\frac{\lambda_i-m_{2+a}+\sfrac{3}{2}\epsilon}{\lambda_i-m_a-\sfrac{1}{2}\epsilon}=q \prod_{\substack{
j = 1\\ j\neq i}}^{K}\frac{\lambda_i-\lambda_j-\epsilon}{\lambda_i-\lambda_j+\epsilon}\,,
\]
can be written as the second equation from the MTV dual pair \eqref{eq:MTVdualref}. In order to see this we need to employ \eqref{eq:CoulombHiggs} and substitute $m_3$ and $m_4$ into the numerators of the let hand side of \eqref{eq:XXXBAE6}. Then we take the limit of large $\epsilon$ keeping in mind that rapidities $\lambda_i$ also scale with $\epsilon$. Neither Coulomb moduli $a_a$ nor the antifundamental masses $m_{1,2}$ enjoy this scaling, so they will drop out from the equations. We then arrive to \eqref{eq:MTVdualref} where $z_2/z_1=q$ and
\[\label{eq:GaudinRootsnaMatch}
n_a=\kappa_a+2\,.
\]
%

\subsection{The duality}

Now let us start connecting the story with the Liouville. By means of the bispectral duality these equations are mapped onto \eqref{eq:TrigomGaud} yielding the trigonometric Gaudin model from the Heisenberg chain. Note that \eqref{eq:TrigomGaud} depends only on two points $z_1$ and $z_2$ corresponding to the locations of the NS5 branes in 6-10 plane in \figref{fig:hweps}. However, the Liouville conformal block depends on four operators sitting at $\infty, 1, q, 0$. Interestingly we can mention that trigonometric Gaudin Bethe equations \eqref{eq:TrigomGaud} when only $z_1$ and $z_2$ punctures are involved can be treated as a \textit{rational} $\mathfrak{sl}(2)$ Gaudin Bethe equations on $S^2$ with all four punctures included. Indeed,
\[\label{eq:sl2Gaudz4}
\sum_{b=0}^3 \frac{\nu_b\epsilon}{t_i-z_b} -\sum_{\substack{j=1\\ j\neq i}}^{\kappa_2} \frac{2\epsilon}{t_i-t_j}=0\,,
\]
where $z_{0,1,2,3}=\{\infty, 1, q, 0\}$, is equivalent to \eqref{eq:TrigomGaud} with
\[
\nu_2=K,\quad   \epsilon \nu_3=m_3-m_4-\epsilon=2\widetilde{\mu}_1 - \epsilon\,,
\label{eq:Spins1q}
\]
being spins of the $\mathfrak{sl}(2)$ representations sitting at points $q$ and $0$. Specification of $\nu_0$ is not important, as the corresponding contribution drops out from the equation since $z_0=\infty$. Also, as we have already mentioned in \eqref{eq:GaudinRootsnaMatch} there is an exact matching between the number of the Gaudin Bethe roots with the parameters $n^a$ of the Higgs branch. In other words, all sectors of the trigonometric Gaudin model Hilbert space parameterized by number of Bethe roots (excitations over the Bethe vacuum), by means of the bispectral duality, are mapped onto various points of the Higgs branch lattice $\{n_a\}$ of the four dimensional theory. Finally, the value of $\nu_1$ can be found from \eqref{eq:Levelbalance} and  \eqref{eq:GaudinRootsnaMatch}. We conclude that $\nu_1=-2$, which formally corresponds to the spin $-1$ representation for $z_1=1$

Note that one should also take out the $U(1)$ factor from the $U(2)$ gauge group, as it does not have an analogue in the Liouville theory. Imposing it on the $U(2)$ Coulomb moduli $a_1, a_2$ with the help of \eqref{eq:CoulombHiggs} we get
\[
m_3+m_4-(n_1+n_2)\epsilon=0\,,
\]
or, using the Liouville mass parameters, one gets
\[\label{eq:U1cond}
\frac{\mu_1}{\epsilon} = \frac{n_1+n_2}{2}\,.
\]
The $U(1)$ condition balances the count of the parameters on both sides of the correspondence as in order to match $\mathfrak{sl}(2)$ spin at $z_4=0$ we used only one antifundamental mass parameter (which is related to the fundamental one).

Again, from the gauge theory perspective we are interested in keeping Coulomb branch parameters in \eqref{eq:CoulombHiggs} finite while masses $\mu_0$ and $\mu_1$ and $\epsilon$ are sent to infinity. The rescaled conformal dimensions \eqref{eq:RescaledConfDims} upon identification $b=\epsilon$ and by using \eqref{eq:U1cond} therefore read
\[\label{eq:rescaleddims}
\delta_0=-\frac{1}{4}\,,\quad \delta_1 = 0\,, \quad \delta_2 = \frac{K}{2}\left(\frac{K}{2}+1\right)\,,\quad \delta_3 = \gamma_0(\gamma_0+1)\,,
\]
where
\[
\gamma_0=-\frac{\hat{n}_1}{2}+\frac{\hat{n}_2}{2}-\frac{1}{2}=\frac{\nu_3}{2}\,,
\]
and we recall $K=\hat{n}_1+\hat{n}_2$ is the number of the D2 branes stretched between the NS5 brane at $z_2=q$ and the D4 branes. We can also see that in such limit the $SU(2)$ Coulomb coordinate and the anti-fundamental masses have dropped from the formulae. In the last two terms of \eqref{eq:rescaleddims} we recognize  $\mathfrak{sl}(2)$ quadratic Casimir eigenvalues on representations of spins $\half K$ and $\sfrac{\nu_3}{2}$ respectively. We can see from \eqref{eq:Spins1q} and \eqref{eq:rescaleddims} that the matching occurs at these points. Vanishing eigenvalues $\delta_1$ confirms the fact that $\nu_1=-2$ corresponds to the spin $-1$ representation. Spin of the representation at $z_0=\infty$ is formally equal to $-\half$.

Here is the summary table of the correspondence between the objects we have discussed in this section in addition to the standard AGT dictionary
\begin{center}
\begin{tabular}{|c|c|}
\hline
Liouville conformal block at  $b\to\infty$   &    $U(2)\,, N_f=4$ SQCD  instanton  \\
on $S^2$ with four punctures                     &    partition function in the NS limit \\
\hline \hline
Rational Gaudin model from KZ                              &    $SL(2)$ spin chain from the ground state    \\
 equation on conformal blocks    &   equation  for the 2d GLSM dual to 4d theory  \\
\hline
Punctures' positions $z_2/z_1$  & Instanton number $q$ \\
\hline
$\mathfrak{sl}(2)$ spin at $z_2=q$  & $U(1)$ condition, number of D2 branes\\
                                                           &  emerging at NS5 brane at $z_2=q$ \\
\hline
Conformal dimensions of chiral operators  & Quadratic $\mathfrak{sl}(2)$ Casimir eigenvalues on \\
at points $z_2=q,\,z_3=0$    &   spin $\half\hat{n}_1+\half\hat{n}_2$ and   \\
                                                   &   $-\half\hat{n}_1+\half\hat{n}_2-\half$ representations   \\
\hline
Degenerate puncture at $z_2=q$  & Higgs branch condition \\
\hline
Gaudin Hilbert space sectors with   & Higgs branch lattice $\{n_a\}$ \\
different number  $\kappa_a$ of Bethe roots        &     \\
\hline
\end{tabular}
\end{center}
%

\subsection{Generalization to $SU(2)$ linear quivers}

One can easily generalize the above construction to the Liouville theory on $S^2$ with $L+3$ punctures. A natural quiver gauge theory associated to this Riemann surface has $L$ $SU(2)$ gauge nodes with Coulomb moduli $a_i$ successively connected together. Liouville conformal block of $L+3$ operators located at points
\[
\infty,\, 1,\, q_1,\, q_1 q_2,\, \dots,\, q_1 q_2\dots q_{L},\, 0\,,
\]
with the following scaling dimensions
\[\label{eq:confdimsL}
\alpha_0(Q-\alpha_0)\,,\quad \mu_0(Q-\mu_0)\,, \dots, \quad m_L(Q-m_L)\,, \quad \alpha_{L+1}(Q-\alpha_{L+1})\,,
\]
respectively glued by operators of dimensions $\alpha_i(Q-\alpha_i)$ in the intermediate $s$-channels.

\begin{figure}
\begin{center}
\includegraphics[height=4cm, width=15cm]{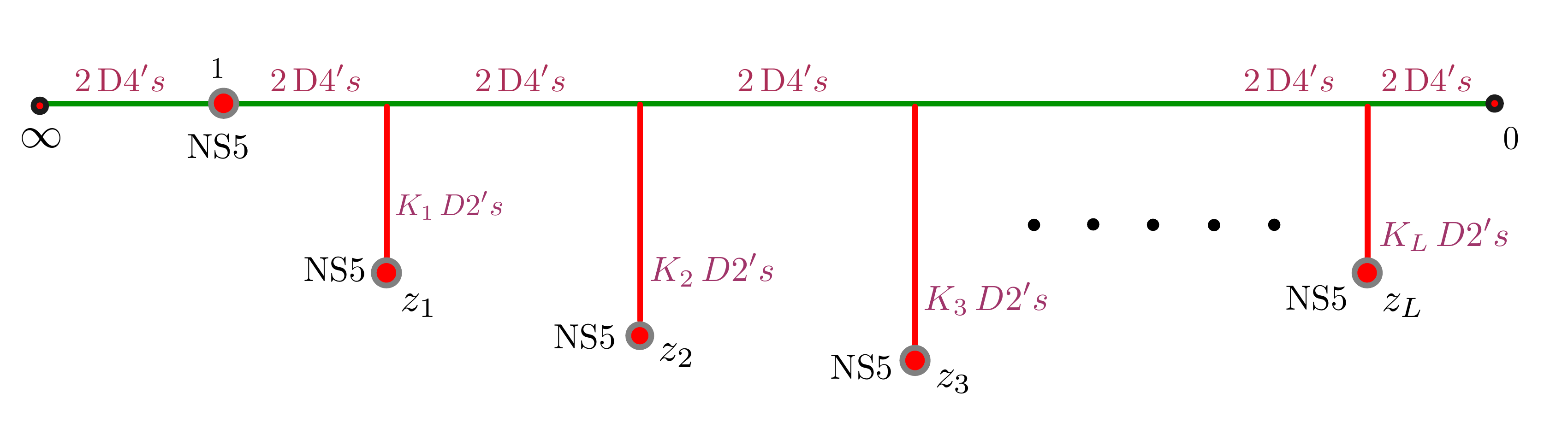}
\caption{$(6+i10)-7$ slice of the CDHL quiver construction. Vertical lines correspond to stacks of $D2$ branes.}
\label{fig:CDHLquiver}
\end{center}
\end{figure}

NS limit of such quiver theory has been elaborated in \cite{Chen:2011sj}. Brane interpretation of the bispectral duality is very useful in this case. The corresponding Hanany-Witten picture view from ``below'' (in $(6+i10)-7$ space) is shown in \figref{fig:CDHLquiver}.
Quiver theories have bifundamental matter with masses $\mu_k^{(p)}$, so the Higgs branch conditions get changed
\[
a^{(p)}_a = m_a^{(p)} + n^{(p)}_a \epsilon + \sum\limits_{k=1}^L \mu_k^{(p)}\,.
\]
Using the above relation we can express conformal dimensions \eqref{eq:confdimsL} in terms of Coulomb branch coordinates, quantization parameters and bifundamental masses. Performing the rescaling analogous to \eqref{eq:RescaledConfDims} we conclude that operators located at
\[
z_1=q_1,\quad z_2=q_1q_2,\quad z_3 = q_1 q_2 q_3,\quad\dots, \quad z_{L}=q_1\dots q_L
\]
have dimensions
\[
\frac{K_1}{2}\left(\frac{K_1}{2}+1\right),\quad \dots, \quad \frac{K_L}{2}\left(\frac{K_L}{2}+1\right)\,,
\]
where $K_i = \hat{n}^{(i)}_1+\hat{n}^{(i)}_2,\,i=1,\dots, L$, corresponding to the  $\mathfrak{sl}(2)$ Casimir eigenvalues on representations of spins $\half K_1\,\dots,\half K_L$. Spins sitting at each point $z_i$ (multiplied by 2) correspond to the total number of D2 branes stretched between the $i$'th NS5 brane and both D4 branes. Our construction has to be supplemented by $L$ $U(1)$ conditions similar to \eqref{eq:U1cond} for each gauge group.

The full treatment of the AGT duality for linear quivers requires the construction of the bispectral dual to the twisted anisotropic $SL(L,\mathbb{R})$ chain, which emerges from it \cite{Chen:2011sj}. We shall postpone this analysis for the future work.

\section{Conclusions and Outlook}\label{sec:Conclusions}

In this paper we have investigated BPS solitons -- strings, monopoles and domain walls in $\ssN=2$ four dimensional gauge theories in Omega background with the Nekrasov-Shatashvili limit imposed. We derived the central charges for these solitons from the supersymmetry algebra and observed that string and domain wall charges are proportional to the external graviphoton field present in Omega background. At large values of $\epsilon$, string and domain wall tensions are large which makes semiclassical considerations legitimate. Existence of a BPS string in pure SYM implies fractional windings for the adjoint scalar and hence presence of a conical singularity. We have presented arguments that its tension is nevertheless finite. Moreover we have argued that BPS monopoles found in \cite{Ito:2011ta} are actually located on domain walls interpolating between two different vacua of the theory. The next step along this road will be to analyze the complete set of the moduli space of the corresponding solutions and worldvolume theories of BPS solitons. Also one may expect more surprises in study of wall crossing phenomena for such BPS objects. We postpone the discussion on these issues for a separate publication.

There is an interesting question regarding the behavior of solitonic BPS states at small $\epsilon$. In the pure SYM strings and domain walls become highly quantum objects and could potentially condense. However it is not clear if such condensation of the extended defects should be taken into account. In SQCD at large value of the FI term in the large $\epsilon$ limit the standard nonabelian string is recovered. Note that small $\epsilon$ limit corresponds to semiclassical regime from the integrability viewpoint, hence the potential condensation of BPS solitons also deserves further investigation in the Hamilton-Jacobi framework.

An interplay between quantum integrability and Omega deformed gauged theories allowed us to apply some known dualities in the integrable systems literature. Degrees of freedom in an integrable systems are interpreted as coordinates of the corresponding brane positions since all dualities can be reformulated in brane language. The integrability being just the reflection of the symmetry of the brane geometry involved plays a role of a consistency condition for the whole construction. We have shown that different dualities connecting Gaudin, XXX and Calogero type models can be explained in terms of brane geometry both at classical and at quantum levels.

As a byproduct of the vortex construction and the dualities between integrable systems we were able to reconstruct the AGT correspondence in the NS limit. The Liouville CFT has infinite central charge and all scaling dimensions needed to be regularized. In the paper we discussed the Liouville theory on $S^2$ with $L+3$ punctures, which was shown to be dual to linear quiver gauge theories with corresponding matter content. Our construction certainly has to be extended to other known AGT dual pairs, e.g. a torus with multiple punctures, etc.

There are many questions which certainly deserve additional study. Surely, more complicated defects involving strings, domain walls and monopoles have to be explored. It would be interesting to investigate similar defects in five and six dimensions and with the complete Omega deformation beyond the NS limit.

\section*{Acknowledgements}

We are grateful to N. Bobev, S. Bolognesi, N. Dorey, V. Kumar, S. Lee, D. Morrison, J. Polchinski, M. Shifman, A. Vainshtein, W. Vinci, A. Yung, A. Zabrodin, and A. Zotov for illuminating and inspiring discussions. This research was supported in part by the National Science Foundation under Grant No. NSF PHY11-25915. 
HYC would like to thank the University of Cincinnati and University of Kentucky for the generous financial supports where part of the work was done. 
He is also supported in part by National Science Council and Center for Theoretical Sciences at National Taiwan University.
PK's work is supported in part by DOE grant DE-FG02-94ER40823. The work of AG is supported in part by grants RFBR-12-02-00284 and PICS-12-02-91052. AG thanks FTPI at University of Minnesota where the part of the work has done for the hospitality and support.

\appendix

\section{The Gaudin Model}\label{sec:GaudLiouville}

Here we discuss the Gaudin model -- the key tool in our AGT construction, its relations with the XXX spin chain. 

\paragraph{Gaudin model from XXX chain.}

The Gaudin model is the simplest example of the Hitchin system on a sphere with marked points \cite{Nekrasov:1995nq}. It is also known to be a large impurity limit of an anisotropic twisted XXX spin chain. This fact can be realized both in the transfer matrix at the classical limit and in the Bethe ansatz equations in the quantum case. We shall be interested in the quantum case and upon the proper limit Bethe ansatz equations for the Gaudin model can be obtained. Let us start with Bethe equations for anisotropic $XXX_\frac{S}{2}$ spin chain\footnote{We measure spectral parameters $\lambda_i$ in units of $i\epsilon$ here.} with twist $q=e^{2\pi i\hat{\tau}}$
\[
\prod\limits_{a=1}^N\frac{\lambda_i-\nu_a+\sfrac{\epsilon}{2} S_a}{\lambda_i-\nu_a-\sfrac{\epsilon}{2} S_a}=q\prod_{\substack{
j = 1\\ j\neq i}}^K\frac{\lambda_i-\lambda_j-\epsilon}{\lambda_i-\lambda_j+\epsilon}\,.
\]
By taking logarithms of both parts of the above equations, then rescaling
\[
\lambda_i\mapsto x\lambda_i,\,\quad \nu_a\mapsto x\nu_a,\,\,\quad \hat{\tau}\mapsto \frac{\hat{\tau}}{x}\,,
\]
and sending $x\to\infty$ we arrive at the following set of equations
\[
\frac{\log{q}}{\epsilon} - \sum\limits_{a=1}^N\frac{S_a}{\lambda_i-\nu_a} = \sum_{\substack{j = 1\\ j\neq i}}^K\frac{2}{\lambda_i-\lambda_j}\,,
\]
which are nothing but Bethe equations for the Gaudin model. The anisotropies $\nu_a$ at each site still play the role of the inhomogenities in the model, while the twist $q$ in the XXX chain play the role of the external field in the Gaudin system. As we can see the latter vanishes as $\epsilon\to \infty$.

\paragraph{Bethe ansatz equations for the Gaudin model.}

Let us now recall how the Bethe ansatz equations for the rational Gaudin model with the Lie algebra symmetry $\mathfrak{g}$ are derived. For our purposes we need merely $\mathfrak{g}=\mathfrak{sl}(2)$ and and $L$ points on the sphere. At each point we fix a representation $V(\nu_1),\dots,V(\nu_L)$ of $\mathfrak{sl}(2)$ algebra with some dominant weights $\nu_a,\, a =1,\dots, L$. According to the Bethe ansatz prescription \cite{refId0} we construct the following operator
\[
S(u)=\sum\limits_{a=1}^4\frac{\mathcal{H}_a}{u-z_a}+\sum\limits_{a=1}^4\frac{\Delta(\nu_a)}{(u-z_a)^2}\,,
\]
where $\mathcal{H}_a$ are Gaudin Hamiltonians at each site of the lattice
\[
\mathcal{H}_a=\sum\limits_{b\neq a}\sum\limits_{\alpha=1}^{\text{dim}(\mathfrak{g})}\frac{\mathfrak{J}_\alpha^{(b)}\mathfrak{J}^{\alpha\,(b)}}{z_a-z_b}\,,
\]
where $\mathfrak{J}_\alpha^{(b)}$ of the acts with $\mathfrak{J}_\alpha\in\mathfrak{sl}(2)$ on the $b$-th site of the spin chain and with identity on the others. $\Delta(\nu_a)$ are eigenvalue of the $\text{U}(\mathfrak{sl}(2))$ quadratic Casimir acting on $V(\nu_a)$.  For such a system Bethe ansatz equations for the sector with $\kappa_a$ Bethe roots read as follows
\[\label{eq:RationalGaudGen}
\sum_{b=1}^L \frac{\nu_b\epsilon}{t_i-z_b} -\sum_{\substack{j=1\\ j\neq i}}^{\kappa_a} \frac{2\epsilon}{t_i-t_j}=0, \,\quad i=1,\dots,\kappa_a\,.
\]
%



\section{Supersymmetry Algebra and Central Charges}\label{sec:Setup}

$\ssN=2$ supersymmetry algebra in four dimensions has the following form
\<\label{eq:N2SUSYalgebra}
\{Q^I_\alpha,\bar{Q}_{J\,\dot{\alpha}}\}\eq 2P_{\alpha\dot{\alpha}}\delta^I_J+2Z_{\alpha\dot{\alpha}}\delta^I_J\,,\nln
\{Q^I_\alpha,Q^J_\beta\}\eq \epsilon_{\alpha\beta}\epsilon^{IJ} Z_{\text{mon}}+(Z_{\text{d.w.}})^{IJ}_{\alpha\beta}\,.
\>
There are three types on central charges: string, monopole and domain wall types.
The full global symmetry of the theory is $SU(2)_L\times SU(2)_R\times SU(2)_{\mathcal{R}}\times SU(2)_c$. It is broken by the Omega background in the NS limit to $SU(2)_{R+\mathcal{R}}\times SU(2)_c$. Twisted supercharges
\[ \bar{Q}=\delta_I^{\dot{\alpha}} \bar{Q}^I_{\dot{\alpha}}\,,\quad Q_m=(\bar{\sigma}_m)^{I \alpha} Q_{I \alpha}\,,\quad \bar{Q}_{mn}=(\bar{\sigma}_{mn})^{\dot{\alpha}}_I \bar{Q}^I_{\dot{\alpha}}\,.
\]
The former operator above is also known as BRST operator. The transformations can be inverted as
\[\label{eq:InvertedDWtransf}
Q_{\alpha}^I = \half (\sigma^m)^I_\alpha Q_m\,,\quad \bar{Q}_{\dot{\alpha} J} = \half\epsilon_{\dot{\alpha} J}\bar{Q} + \half (\bar{\sigma}_{mn})_{\dot{\alpha}J}\bar{Q}^{mn}\,.
\]
Plugging these formulae into \eqref{eq:N2SUSYalgebra} we get the twisted version of the supersymmetry algebra. 

\section{Some Notations}\label{sec:Notations}

In this work we are intended to use Euclidean signature. We benefit from this while studying static configurations, where in the gauge $A_4=0$ the Lagrangian is nothing but the energy density. Below we list some definitions and conventions. Raising and lowering of spinor indices is performed by means of Levi-Civita symbol
\[ \epsilon^{12}=-\epsilon^{21}=-\epsilon_{12}=\epsilon_{21}=1\,,
\]
the same for the Levi-Civita symbol with dotted indices. The definition is consistent with
\[ \epsilon^{\alpha\beta}\epsilon_{\beta\gamma} = \delta^\alpha_\gamma\,.
\]
Scalars can then be obtained by contracting spinor indices, for example $\eta_\alpha \chi^\alpha = -\eta^\alpha\chi_\alpha$. Vector indices are contracted with Euclidean metric $\delta_{mn}$.

Sigma matrices
\<
(\sigma^m)_{\alpha\dot{\alpha}} \eq (-i\tau^1,\,-i\tau^2,\,-i\tau^3,\, 1)\,,\nln
(\bar{\sigma}^m)_{\dot{\alpha}\alpha} \eq (i\tau^1,\,i\tau^2,\,i\tau^3,\, 1) = ((\sigma^m)_{\alpha\dot{\alpha}})^\dag\,,
\>
where $\tau^{1,2,3}$ are standard Pauli matrices. Thus for $\sigma^m$ the undotted index goes first, whereas for $\bar{\sigma}^m$ it is the last.

Lorentz projectors
\<
(\sigma^{mn})_{\alpha\beta}\eq\quarter \left((\sigma^m)_{\alpha\dot{\alpha}}(\bar{\sigma}^n)_{~\beta}^{\dot{\alpha}}-(\sigma^n)_{\alpha\dot{\alpha}}(\bar{\sigma}^m)_{~\beta}^{\dot{\alpha}}\right)\,,\nln
(\bar{\sigma}^{mn})_{\dot{\alpha}\dot{\beta}}\eq\quarter \left((\bar{\sigma}^m)_{\dot{\alpha}\alpha}(\sigma^n)^{\alpha}_{~\dot{\beta}}-(\bar{\sigma}^n)_{\dot{\alpha}\alpha}(\sigma^m)_{~\dot{\beta}}^{\alpha}\right)\,.
\>

Chiral and antichiral electromagnetic field strength
\<
F_{\alpha\beta} \eq -\half F_{mn}(\sigma^m)_{\alpha\dot{\alpha}}(\sigma^n)_\beta^{~\dot{\alpha}}=-(\vec{E}+\vec{B})\cdot\vec{\sigma}\,\sigma^2\,,\nln
\bar{F}^{\dot{\alpha}\dot{\beta}} \eq \half F_{mn}(\bar{\sigma}^m)^{\dot{\alpha}\alpha}(\bar{\sigma}^n)_{~\alpha}^{\dot{\beta}}=(-\vec{E}+\vec{B})\cdot\vec{\sigma}\,\sigma^2\,.
\>
%

\bibliography{cpn1}

\begin{thebibliography}{10}
\ifx\href\asklfhas\newcommand{\href}[2]{#2}\fi
\ifx\arxivref\asklfhas\newcommand{\arxivref}[1]{\href{http://arxiv.org/abs/#1}%
{#1}}\fi
\ifx\doiref\asklfhas\newcommand{\doiref}[2]{\href{http://dx.doi.org/#1}{#2}}\fi
\raggedright
\small
\parskip 0pt

\bibitem{Nekrasov:2009rc}
N.~A.~Nekrasov and S.~L.~Shatashvili,
\textit{``{Quantization of Integrable Systems and Four Dimensional Gauge
  Theories}''},
\texttt{\arxivref{0908.4052}}.
%
\bibitem{Gorsky:1995zq}
A.~Gorsky, I.~Krichever, A.~Marshakov, A.~Mironov and A.~Morozov,
\textit{``{Integrability and Seiberg-Witten exact solution}''},
\textsf{\doiref{10.1016/0370-2693(95)00723-X}{Phys.Lett.~B355,~466~(1995)}},
\texttt{\arxivref{hep-th/9505035}}.
%
\bibitem{Donagi:1995cf}
R.~Donagi and E.~Witten,
\textit{``{Supersymmetric Yang-Mills theory and integrable systems}''},
\textsf{\doiref{10.1016/0550-3213(95)00609-5}{Nucl.Phys.~B460,~299~(1996)}},
\texttt{\arxivref{hep-th/9510101}}.
%
\bibitem{Gorsky:1997jq}
A.~Gorsky, S.~Gukov and A.~Mironov,
\textit{``{Multiscale N=2 SUSY field theories, integrable systems and their
  stringy / brane origin. 1.}''},
\textsf{\doiref{10.1016/S0550-3213(98)00055-8}{Nucl.Phys.~B517,~409~(1998)}},
\texttt{\arxivref{hep-th/9707120}}.
%
\bibitem{Nekrasov:2002qd}
N.~A.~Nekrasov,
\textit{``{Seiberg-Witten prepotential from instanton counting}''},
\textsf{Adv.Theor.Math.Phys.~7,~831~(2004)},
\texttt{\arxivref{hep-th/0206161}},
To Arkady Vainshtein on his 60th anniversary.
%
\bibitem{Nekrasov:2009ui}
N.~A.~Nekrasov and S.~L.~Shatashvili,
\textit{``{Quantum integrability and supersymmetric vacua}''},
\textsf{\doiref{10.1143/PTPS.177.105}{Prog.Theor.Phys.Suppl.~177,~105~(2009)}},
\texttt{\arxivref{0901.4748}},
21 pp., short version II, conference in honour of T.Eguchi's 60th anniversary.
%
\bibitem{Nekrasov:2009uh}
N.~A.~Nekrasov and S.~L.~Shatashvili,
\textit{``{Supersymmetric vacua and Bethe ansatz}''},
\textsf{\doiref{10.1016/j.nuclphysbps.2009.07.047}{Nucl.Phys.Proc.Suppl.~192-1%
93,~91~(2009)}},
\texttt{\arxivref{0901.4744}}.
%
\bibitem{Dorey:1999zk}
N.~Dorey, T.~J.~Hollowood and D.~Tong,
\textit{``{The BPS spectra of gauge theories in two and four dimensions}''},
\textsf{JHEP~9905,~006~(1999)},
\texttt{\arxivref{hep-th/9902134}}.
%
\bibitem{Shifman:2004dr}
M.~Shifman and A.~Yung,
\textit{``{Non-Abelian string junctions as confined monopoles}''},
\textsf{\doiref{10.1103/PhysRevD.70.045004}{Phys.~Rev.~D70,~045004~(2004)}},
\texttt{\arxivref{hep-th/0403149}}.
%
\bibitem{Hanany:2004ea}
A.~Hanany and D.~Tong,
\textit{``{Vortex strings and four-dimensional gauge dynamics}''},
\textsf{JHEP~0404,~066~(2004)},
\texttt{\arxivref{hep-th/0403158}}.
%
\bibitem{Shifman:2007ce}
M.~Shifman and A.~Yung,
\textit{``{Supersymmetric Solitons and How They Help Us Understand Non-Abelian
  Gauge Theories}''},
\textsf{\doiref{10.1103/RevModPhys.79.1139}{Rev.~Mod.~Phys.~79,~1139~(2007)}},
\texttt{\arxivref{hep-th/0703267}}.
%
\bibitem{Dorey:2011pa}
N.~Dorey, S.~Lee and T.~J.~Hollowood,
\textit{``{Quantization of Integrable Systems and a 2d/4d Duality}''},
\texttt{\arxivref{1103.5726}}.
%
\bibitem{Gorsky:2000ds}
A.~Gorsky and V.~Rubtsov,
\textit{``{Dualities in integrable systems: Geometrical aspects}''},
\texttt{\arxivref{hep-th/0103004}}.
%
\bibitem{Ito:2011wv}
K.~Ito, S.~Kamoshita and S.~Sasaki,
\textit{``{Deformed BPS Monopole in Omega-background}''},
\texttt{\arxivref{1110.1455}},
* Temporary entry *.
%
\bibitem{Ito:2011ta}
K.~Ito, S.~Kamoshita and S.~Sasaki,
\textit{``{BPS Monopole Equation in Omega-background}''},
\textsf{\doiref{10.1007/JHEP04(2011)023}{JHEP~1104,~023~(2011)}},
\texttt{\arxivref{1103.2589}}.
%
\bibitem{Alday:2009aq}
L.~F.~Alday, D.~Gaiotto and Y.~Tachikawa,
\textit{``{Liouville Correlation Functions from Four-dimensional Gauge
  Theories}''},
\textsf{\doiref{10.1007/s11005-010-0369-5}{Lett.Math.Phys.~91,~167~(2010)}},
\texttt{\arxivref{0906.3219}}.
%
\bibitem{Chen:2011sj}
H.-Y.~Chen, N.~Dorey, T.~J.~Hollowood and S.~Lee,
\textit{``{A New 2d/4d Duality via Integrability}''},
\textsf{\doiref{10.1007/JHEP09(2011)040}{JHEP~1109,~040~(2011)}},
\texttt{\arxivref{1104.3021}}.
%
\bibitem{Nekrasov:2003rj}
N.~Nekrasov and A.~Okounkov,
\textit{``{Seiberg-Witten theory and random partitions}''},
\texttt{\arxivref{hep-th/0306238}},
90 pp. plain TeX, 15 pictures Report-no: IHES-P/03/43, PUMD-2003, ITEP-36/03
  Subj-class: High Energy Physics - Theory: Mathematical Physics: Statistical
  Mechanics: Algebraic Geometry: Exactly Solvable and Integrable Systems:
  Probability Theory.
%
\bibitem{Shadchin:2005mx}
S.~Shadchin,
\textit{``{On certain aspects of string theory/gauge theory correspondence}''},
\texttt{\arxivref{hep-th/0502180}},
Ph.D. Thesis.
%
\bibitem{Nekrasov:2010ka}
N.~Nekrasov and E.~Witten,
\textit{``{The Omega Deformation, Branes, Integrability, and Liouville
  Theory}''},
\textsf{\doiref{10.1007/JHEP09(2010)092}{JHEP~1009,~092~(2010)}},
\texttt{\arxivref{1002.0888}}.
%
\bibitem{Witten:1988ze}
E.~Witten,
\textit{``{Topological Quantum Field Theory}''},
\textsf{\doiref{10.1007/BF01223371}{Commun.Math.Phys.~117,~353~(1988)}}.
%
\bibitem{Gorsky:1999hk}
A.~Gorsky and M.~A.~Shifman,
\textit{``{More on the tensorial central charges in N=1 supersymmetric gauge
  theories (BPS wall junctions and strings)}''},
\textsf{\doiref{10.1103/PhysRevD.61.085001}{Phys.Rev.~D61,~085001~(2000)}},
\texttt{\arxivref{hep-th/9909015}}.
%
\bibitem{Gross:2000wc}
D.~J.~Gross and N.~A.~Nekrasov,
\textit{``{Monopoles and strings in noncommutative gauge theory}''},
\textsf{JHEP~0007,~034~(2000)},
\texttt{\arxivref{hep-th/0005204}}.
%
\bibitem{Hellerman:2011mv}
S.~Hellerman, D.~Orlando and S.~Reffert,
\textit{``{String theory of the Omega deformation}''},
\textsf{\doiref{10.1007/JHEP01(2012)148}{JHEP~1201,~148~(2012)}},
\texttt{\arxivref{1106.0279}},
36 pages. References added, brane construction clarified, edited for style.
%
\bibitem{Reffert:2011dp}
S.~Reffert,
\textit{``{General Omega Deformations from Closed String Backgrounds}''},
\textsf{JHEP~1204,~059~(2012)},
\texttt{\arxivref{1108.0644}}.
%
\bibitem{Hellerman:2012zf}
S.~Hellerman, D.~Orlando and S.~Reffert,
\textit{``{The Omega Deformation From String and M-Theory}''},
\texttt{\arxivref{1204.4192}}.
%
\bibitem{Dvali:1996xe}
G.~Dvali and M.~A.~Shifman,
\textit{``{Domain walls in strongly coupled theories}''},
\textsf{\doiref{10.1016/S0370-2693(97)00131-7,
  10.1016/S0370-2693(97)00131-7}{Phys.Lett.~B396,~64~(1997)}},
\texttt{\arxivref{hep-th/9612128}},
Several typos corrected Report-no: CERN-TH/96-356, TPI-MINN-96/26-T.
%
\bibitem{Prasad:1975kr}
M.~Prasad and C.~M.~Sommerfield,
\textit{``{An Exact Classical Solution for the 't Hooft Monopole and the
  Julia-Zee Dyon}''},
\textsf{\doiref{10.1103/PhysRevLett.35.760}{Phys.Rev.Lett.~35,~760~(1975)}}.
%
\bibitem{Bogomolny:1975de}
E.~Bogomolny,
\textit{``{Stability of Classical Solutions}''},
\textsf{Sov.J.Nucl.Phys.~24,~449~(1976)}.
%
\bibitem{Gukov:2006jk}
S.~Gukov and E.~Witten,
\textit{``{Gauge Theory, Ramification, And The Geometric Langlands Program}''},
\texttt{\arxivref{hep-th/0612073}}.
%
\bibitem{Gaiotto:2011tf}
D.~Gaiotto, G.~W.~Moore and A.~Neitzke,
\textit{``{Wall-Crossing in Coupled 2d-4d Systems}''},
\texttt{\arxivref{1103.2598}}.
%
\bibitem{Shifman:2009zz}
M.~Shifman and A.~Yung,
\textit{``{Supersymmetric solitons}''}.
%
\bibitem{Bonelli:2011fq}
G.~Bonelli, A.~Tanzini and J.~Zhao,
\textit{``{Vertices, Vortices and Interacting Surface Operators}''},
\textsf{\doiref{10.1007/JHEP06(2012)178}{JHEP~1206,~178~(2012)}},
\texttt{\arxivref{1102.0184}}.
%
\bibitem{Vainshtein:2000hu}
A.~I.~Vainshtein and A.~Yung,
\textit{``{Type I superconductivity upon monopole condensation in
  Seiberg-Witten theory}''},
\textsf{\doiref{10.1016/S0550-3213(01)00394-7}{Nucl.Phys.~B614,~3~(2001)}},
\texttt{\arxivref{hep-th/0012250}}.
%
\bibitem{Hanany:1997hr}
A.~Hanany, M.~J.~Strassler and A.~Zaffaroni,
\textit{``{Confinement and strings in MQCD}''},
\textsf{\doiref{10.1016/S0550-3213(97)00651-2}{Nucl.Phys.~B513,~87~(1998)}},
\texttt{\arxivref{hep-th/9707244}}.
%
\bibitem{Shifman:2006kd}
M.~Shifman and A.~Yung,
\textit{``{Non-Abelian semilocal strings in N=2 supersymmetric QCD}''},
\textsf{\doiref{10.1103/PhysRevD.73.125012}{Phys.Rev.~D73,~125012~(2006)}},
\texttt{\arxivref{hep-th/0603134}}.
%
\bibitem{Shifman:2011xc}
M.~Shifman, W.~Vinci and A.~Yung,
\textit{``{Effective World-Sheet Theory for Non-Abelian Semilocal Strings in N
  = 2 Supersymmetric QCD}''},
\texttt{\arxivref{1104.2077}}.
%
\bibitem{Chen:2012we}
H.-Y.~Chen, T.~J.~Hollowood and P.~Zhao,
\textit{``{A 5d/3d duality from relativistic integrable system}''},
\texttt{\arxivref{1205.4230}}.
%
\bibitem{Hanany:1996ie}
A.~Hanany and E.~Witten,
\textit{``{Type IIB superstrings, BPS monopoles, and three-dimensional gauge
  dynamics}''},
\textsf{\doiref{10.1016/S0550-3213(97)00157-0}{Nucl.Phys.~B492,~152~(1997)}},
\texttt{\arxivref{hep-th/9611230}}.
%
\bibitem{refId0}
{Gaudin, M.},
\textit{``Diagonalisation d'une classe d'hamiltoniens de spin''},
\textsf{\doiref{10.1051/jphys:0197600370100108700}{J.~Phys.~France~37,~1087~(1%
976)}},
\href{http://dx.doi.org/10.1051/jphys:0197600370100108700}{\texttt{http://dx.d%
oi.org/10.1051/jphys:0197600370100108700}}.
%
\bibitem{springerlink:10.1007/BF00626526}
M.~Adams, J.~Harnad and J.~Hurtubise,
\textit{``Dual moment maps into loop algebras''},
\textsf{Letters~in~Mathematical~Physics~20,~299~(1990)},
10.1007/BF00626526,
\href{http://dx.doi.org/10.1007/BF00626526}{\texttt{http://dx.doi.org/10.1007/%
BF00626526}}.
%
\bibitem{MR2409414}
E.~Mukhin, V.~Tarasov and A.~Varchenko,
\textit{``Bispectral and (gl(N),gl(M)) dualities, discrete versus
  differential''},
\textsf{\doiref{10.1016/j.aim.2007.11.022}{Adv.~Math.~218,~216~(2008)}},
\href{http://dx.doi.org/10.1016/j.aim.2007.11.022}{\texttt{http://dx.doi.org/1%
0.1016/j.aim.2007.11.022}}.
%
\bibitem{Knizhnik198483}
V.~Knizhnik and A.~Zamolodchikov,
\textit{``Current algebra and Wess-Zumino model in two dimensions''},
\textsf{\doiref{10.1016/0550-3213(84)90374-2}{Nuclear~Physics~B~247,~83
  ~(1984)}},
\href{http://www.sciencedirect.com/science/article/pii/0550321384903742}{\text%
tt{http://www.sciencedirect.com/science/article/pii/0550321384903742}}.
%
\bibitem{Babujian_Flume_1993}
H.~M.~Babujian and R.~Flume,
\textit{``Off-Shell Bethe Ansatz Equation for Gaudin Magnets and Solutions of
  Knizhnik-Zamolodchikov Equations''},
\textsf{ModPhysLett~A9,~10~(1993)},
\href{http://arxiv.org/abs/hep-th/9310110}{\texttt{http://arxiv.org/abs/hep-th%
/9310110}}.
%
\bibitem{Argyres:1995jj}
P.~C.~Argyres and M.~R.~Douglas,
\textit{``{New phenomena in SU(3) supersymmetric gauge theory}''},
\textsf{\doiref{10.1016/0550-3213(95)00281-V}{Nucl.~Phys.~B448,~93~(1995)}},
\texttt{\arxivref{hep-th/9505062}}.
%
\bibitem{Bolokhov:2011mp}
P.~A.~Bolokhov, M.~Shifman and A.~Yung,
\textit{``{BPS Spectrum of Supersymmetric CP(N-1) Theory with ZN Twisted
  Masses}''},
\textsf{\doiref{10.1103/PhysRevD.84.085004}{Phys.Rev.~D84,~085004~(2011)}},
\texttt{\arxivref{1104.5241}}.
%
\bibitem{Bolokhov:2012dv}
P.~A.~Bolokhov, M.~Shifman and A.~Yung,
\textit{``{2D-4D Correspondence: Towers of Kinks versus Towers of Monopoles in
  N=2 Theories}''},
\textsf{Phys.Rev.~D85,~085028~(2012)},
\texttt{\arxivref{1202.5612}}.
%
\bibitem{Dorey:2012dq}
N.~Dorey and K.~Petunin,
\textit{``{On the BPS Spectrum at the Root of the Higgs Branch}''},
\textsf{JHEP~1205,~085~(2012)},
\texttt{\arxivref{1202.5595}}.
%
\bibitem{1996alg.geom.12001G}
A.~Givental,
\textit{``{Stationary Phase Integrals, Quantum Toda Lattices, Flag Manifolds
  and the Mirror Conjecture}''},
in: \textit{``eprint arXiv:alg-geom/9612001''},
12001p.
%
\bibitem{MR0280103}
F.~Calogero,
\textit{``Solution of the one-dimensional {$N$}-body problems with quadratic
  and/or inversely quadratic pair potentials''},
\textsf{J.~Mathematical~Phys.~12,~419~(1971)}.
%
\bibitem{MR0375869}
J.~Moser,
\textit{``Three integrable {H}amiltonian systems connected with isospectral
  deformations''},
\textsf{Advances~in~Math.~16,~197~(1975)}.
%
\bibitem{Sutherland:1971ks}
B.~Sutherland,
\textit{``{Exact results for a quantum many body problem in one-dimension.
  2.}''},
\textsf{\doiref{10.1103/PhysRevA.5.1372}{Phys.Rev.~A5,~1372~(1972)}}.
%
\bibitem{2009arXiv0906}
E.~Mukhin, V.~Tarasov and A.~Varchenko,
\textit{``{Bethe algebra of Gaudin model, Calogero-Moser space and Cherednik
  algebra}''},
\texttt{\arxivref{0906.5185}},
\href{http://adsabs.harvard.edu/abs/2009arXiv0906.5185M}{\texttt{http://adsabs%
.harvard.edu/abs/2009arXiv0906.5185M}}.
%
\bibitem{2012arXiv1201.3990M}
E.~Mukhin, V.~Tarasov and A.~Varchenko,
\textit{``KZ characteristic variety as the zero set of classical Calogero-Moser
  Hamiltonians''},
\texttt{\arxivref{1201.3990}},
\href{http://adsabs.harvard.edu/abs/2012arXiv1201.3990M}{\texttt{http://adsabs%
.harvard.edu/abs/2012arXiv1201.3990M}}.
%
\bibitem{MR1322943}
S.~N.~M.~Ruijsenaars,
\textit{``Action-angle maps and scattering theory for some finite-dimensional
  integrable systems. {II}. {S}olitons, antisolitons, and their bound
  states''},
\textsf{\doiref{10.2977/prims/1195164945}{Publ.~Res.~Inst.~Math.~Sci.~30,~865~%
(1994)}},
\href{http://dx.doi.org/10.2977/prims/1195164945}{\texttt{http://dx.doi.org/10%
.2977/prims/1195164945}}.
%
\bibitem{MR1329481}
S.~Ruijsenaars,
\textit{``Action-angle maps and scattering theory for some finite-dimensional
  integrable systems. {III}. {S}utherland type systems and their duals''},
\textsf{\doiref{10.2977/prims/1195164440}{Publ.~Res.~Inst.~Math.~Sci.~31,~247~%
(1995)}},
\href{http://dx.doi.org/10.2977/prims/1195164440}{\texttt{http://dx.doi.org/10%
.2977/prims/1195164440}}.
%
\bibitem{MR887995}
S.~N.~M.~Ruijsenaars,
\textit{``Complete integrability of relativistic {C}alogero-{M}oser systems and
  elliptic function identities''},
\textsf{Comm.~Math.~Phys.~110,~191~(1987)},
\href{http://projecteuclid.org/getRecord?id=euclid.cmp/1104159234}{\texttt{htt%
p://projecteuclid.org/getRecord?id=euclid.cmp/1104159234}}.
%
\bibitem{MR929148}
S.~N.~M.~Ruijsenaars,
\textit{``Action-angle maps and scattering theory for some finite-dimensional
  integrable systems. {I}. {T}he pure soliton case''},
\textsf{Comm.~Math.~Phys.~115,~127~(1988)},
\href{http://projecteuclid.org/getRecord?id=euclid.cmp/1104160851}{\texttt{htt%
p://projecteuclid.org/getRecord?id=euclid.cmp/1104160851}}.
%
\bibitem{MR851627}
S.~N.~M.~Ruijsenaars and H.~Schneider,
\textit{``A new class of integrable systems and its relation to solitons''},
\textsf{\doiref{10.1016/0003-4916(86)90097-7}{Ann.~Physics~170,~370~(1986)}},
\href{http://dx.doi.org/10.1016/0003-4916(86)90097-7}{\texttt{http://dx.doi.or%
g/10.1016/0003-4916(86)90097-7}}.
%
\bibitem{2009JPhA...42r5202F}
L.~{Feh{\'e}r} and C.~{Klim{\v c}{\'{\i}}k},
\textit{``{On the duality between the hyperbolic Sutherland and the rational
  Ruijsenaars-Schneider models}''},
\textsf{\doiref{10.1088/1751-8113/42/18/185202}{Journal~of~Physics~A~Mathemati%
cal~General~42,~185202~(2009)}},
\texttt{\arxivref{0901.1983}}.
%
\bibitem{2010JMP....51j3511F}
L.~{Feh{\'e}r} and V.~{Ayadi},
\textit{``{Trigonometric Sutherland systems and their Ruijsenaars duals from
  symplectic reduction}''},
\textsf{\doiref{10.1063/1.3492919}{Journal~of~Mathematical~Physics~51,~103511~%
(2010)}},
\texttt{\arxivref{1005.4531}}.
%
\bibitem{2011CMaPh.301...55F}
L.~{Feh{\'e}r} and C.~{Klim{\v c}{\'{\i}} k},
\textit{``{Poisson-Lie Interpretation of Trigonometric Ruijsenaars Duality}''},
\textsf{\doiref{10.1007/s00220-010-1140-6}{Communications~in~Mathematical~Phys%
ics~301,~55~(2011)}},
\texttt{\arxivref{0906.4198}}.
%
\bibitem{2012NuPhB.860..464F}
L.~{Feh{\'e}r} and C.~{Klim{\v c}{\'{\i}}k},
\textit{``{Self-duality of the compactified Ruijsenaars-Schneider system from
  quasi-Hamiltonian reduction}''},
\textsf{\doiref{10.1016/j.nuclphysb.2012.03.005}{Nuclear~Physics~B~860,~464~(2%
012)}},
\texttt{\arxivref{1101.1759}}.
%
\bibitem{chalykh:5139}
O.~A.~Chalykh,
\textit{``Bispectrality for the quantum Ruijsenaars model and its integrable
  deformation''},
\textsf{\doiref{10.1063/1.533399}{Journal~of~Mathematical~Physics~41,~5139~(20%
00)}},
\href{http://link.aip.org/link/?JMP/41/5139/1}{\texttt{http://link.aip.org/lin%
k/?JMP/41/5139/1}}.
%
\bibitem{MR2729945}
E.~Mukhin, V.~Tarasov and A.~Varchenko,
\textit{``Gaudin {H}amiltonians generate the {B}ethe algebra of a tensor power
  of the vector representation of (gl(N))''},
\textsf{\doiref{10.1090/S1061-0022-2011-01152-5}{Algebra~i~Analiz~22,~177~(201%
0)}},
\href{http://dx.doi.org/10.1090/S1061-0022-2011-01152-5}{\texttt{http://dx.doi%
.org/10.1090/S1061-0022-2011-01152-5}}.
%
\bibitem{Fock:1999ae}
V.~Fock, A.~Gorsky, N.~Nekrasov and V.~Rubtsov,
\textit{``{Duality in integrable systems and gauge theories}''},
\textsf{JHEP~0007,~028~(2000)},
\texttt{\arxivref{hep-th/9906235}}.
%
\bibitem{Braden:1999aj}
H.~Braden, A.~Marshakov, A.~Mironov and A.~Morozov,
\textit{``{On double elliptic integrable systems. 1. A Duality argument for the
  case of SU(2)}''},
\texttt{\arxivref{hep-th/9906240}}.
%
\bibitem{Braden:2001yc}
H.~Braden, A.~Gorsky, A.~Odessky and V.~Rubtsov,
\textit{``{Double elliptic dynamical systems from generalized Mukai-Sklyanin
  algebras}''},
\textsf{\doiref{10.1016/S0550-3213(02)00248-1}{Nucl.Phys.~B633,~414~(2002)}},
\texttt{\arxivref{hep-th/0111066}}.
%
\bibitem{Braden:2003gv}
H.~W.~Braden and T.~J.~Hollowood,
\textit{``{The Curve of compactified 6-D gauge theories and integrable
  systems}''},
\textsf{JHEP~0312,~023~(2003)},
\texttt{\arxivref{hep-th/0311024}}.
%
\bibitem{Mironov:2009qn}
A.~Mironov and A.~Morozov,
\textit{``{Proving AGT relations in the large-c limit}''},
\textsf{\doiref{10.1016/j.physletb.2009.10.074}{Phys.Lett.~B682,~118~(2009)}},
\texttt{\arxivref{0909.3531}}.
%
\bibitem{springerlink:10.1007/BF01214585}
A.~B.~Zamolodchikov,
\textit{``Conformal symmetry in two dimensions: An explicit recurrence formula
  for the conformal partial wave amplitude''},
\textsf{Communications~in~Mathematical~Physics~96,~419~(1984)},
10.1007/BF01214585,
\href{http://dx.doi.org/10.1007/BF01214585}{\texttt{http://dx.doi.org/10.1007/%
BF01214585}}.
%
\bibitem{Fateev:2009aw}
V.~Fateev and A.~Litvinov,
\textit{``{On AGT conjecture}''},
\textsf{JHEP~1002,~014~(2010)},
\texttt{\arxivref{0912.0504}}.
%
\bibitem{2012arXiv1202.2756S}
O.~{Schiffmann} and E.~{Vasserot},
\textit{``{Cherednik algebras, W algebras and the equivariant cohomology of the
  moduli space of instantons on A**2}''},
\texttt{\arxivref{1202.2756}}.
%
\bibitem{Mironov:2012uh}
A.~Mironov, A.~Morozov, Y.~Zenkevich and A.~Zotov,
\textit{``{Spectral Duality in Integrable Systems from AGT Conjecture}''},
\texttt{\arxivref{1204.0913}}.
%
\bibitem{Mironov:2012ba}
A.~Mironov, A.~Morozov, B.~Runov, Y.~Zenkevich and A.~Zotov,
\textit{``{Spectral Duality Between Heisenberg Chain and Gaudin Model}''},
\texttt{\arxivref{1206.6349}}.
%
\bibitem{Teschner:2010je}
J.~Teschner,
\textit{``{Quantization of the Hitchin moduli spaces, Liouville theory, and the
  geometric Langlands correspondence I}''},
\texttt{\arxivref{1005.2846}}.
%
\bibitem{Bonelli:2011na}
G.~Bonelli, K.~Maruyoshi and A.~Tanzini,
\textit{``{Quantum Hitchin Systems via beta-deformed Matrix Models}''},
\texttt{\arxivref{1104.4016}}.
%
\bibitem{Alday:2009fs}
L.~F.~Alday, D.~Gaiotto, S.~Gukov, Y.~Tachikawa and H.~Verlinde,
\textit{``{Loop and surface operators in N=2 gauge theory and Liouville modular
  geometry}''},
\textsf{\doiref{10.1007/JHEP01(2010)113}{JHEP~1001,~113~(2010)}},
\texttt{\arxivref{0909.0945}}.
%
\bibitem{Drukker:2009id}
N.~Drukker, J.~Gomis, T.~Okuda and J.~Teschner,
\textit{``{Gauge Theory Loop Operators and Liouville Theory}''},
\textsf{JHEP~1002,~057~(2010)},
\texttt{\arxivref{0909.1105}}.
%
\bibitem{Dimofte:2010tz}
T.~Dimofte, S.~Gukov and L.~Hollands,
\textit{``{Vortex Counting and Lagrangian 3-manifolds}''},
\textsf{\doiref{10.1007/s11005-011-0531-8}{Lett.Math.Phys.~98,~225~(2011)}},
\texttt{\arxivref{1006.0977}}.
%
\bibitem{Nekrasov:1995nq}
N.~Nekrasov,
\textit{``{Holomorphic bundles and many body systems}''},
\textsf{\doiref{10.1007/BF02099624}{Commun.Math.Phys.~180,~587~(1996)}},
\texttt{\arxivref{hep-th/9503157}}.
%
\end{thebibliography}
\bibliographystyle{nb}

\end{document}